\pdfoutput=1
\documentclass[11pt,a4paper]{article}
\usepackage{jheppub}
\usepackage[utf8]{inputenc}
\usepackage{feynmp}
\usepackage{slashed}
\usepackage{graphicx}
\usepackage{ifdraft}
\usepackage{epstopdf}
\usepackage{subfigure}
\usepackage{etoolbox}
\usepackage{amsfonts}
\usepackage{amssymb}
\usepackage{xcolor}
\usepackage{graphicx}
\usepackage{amsmath}
\usepackage{array,booktabs}
\usepackage{tikz}
\usetikzlibrary{automata, arrows, calc, decorations.markings, decorations.pathreplacing, intersections,
  patterns, positioning, shapes.geometric, shapes.misc,topaths}
\usepackage{tqft}
\usetikzlibrary{shapes}
\usepackage{empheq}
\usepackage{mathrsfs}
\usepackage{microtype}
\usepackage{amsmath,mathtools}
\newcommand{\mathtikz}[2][]{\begin{tikzpicture}[baseline=\the\dimexpr-\fontdimen22\textfont2\relax,#1]#2\end{tikzpicture}}
\usepackage{hyperref,hypcap}
\hypersetup{
  colorlinks=true,
  linkcolor=black,
  citecolor=black,
  filecolor=black,
  urlcolor=black,
  linktoc = all,
  unicode,
  bookmarksnumbered,
  pdftitle = {AGT/Z\string_2},
  pdfauthor = {Bruno Le Floch and Gustavo J. Turiaci}}

\numberwithin{equation}{section}

\usepackage{environ}

\newcommand{\beq}{\begin{equation}}
\newcommand{\eeq}{\end{equation}}
 
\newcommand{\bea}{\begin{eqnarray}}
\newcommand{\ea}{\end{eqnarray}}
\newcommand{\barr}{\begin{array}}
\newcommand{\earr}{\end{array}}
\newcommand{\lb}{{\langle}}
\newcommand{\rb}{{\rangle}}
\newcommand{\bra}[1]{\langle#1\rvert}
\newcommand{\ket}[1]{\lvert#1\rangle}
\newcommand{\bbra}[1]{\langle\!\langle#1\rVert}
\newcommand{\bket}[1]{\lVert#1\rangle\!\rangle}

\newcommand{\block}[4]{\mathcal{#1}_{#2}\bigl(\begin{smallmatrix}#3\end{smallmatrix};#4\bigr)}
\newcommand{\Nsusy}{{\mathcal{N}}}
\newcommand{\Qcharge}{{\mathcal{Q}}}
\newcommand{\ZZ}{\texorpdfstring{\mathbb{Z}}{Z}}
\newcommand{\RR}{\texorpdfstring{\mathbb{R}}{R}}
\newcommand{\CC}{\texorpdfstring{\mathbb{C}}{C}}
\newcommand{\PP}{\texorpdfstring{\mathbb{P}}{P}}
\newcommand{\RP}{\RR\PP}
\newcommand{\HS}{{\rm HS}}
\newcommand{\lie}{\mathfrak}
\newcommand{\Lie}{\operatorname}

\newcommand{\B}{{\cal B}}

\newcommand{\repr}{\mathcal{R}}
\newcommand{\supercharge}{{\mathcal{Q}}}
\newcommand{\Weyl}{\mathcal{W}}
\DeclareMathOperator{\rank}{rank}
\DeclareMathOperator{\Pexp}{Pexp}
\DeclareMathOperator{\id}{id}

\DeclareMathOperator{\diag}{diag}
\DeclareMathOperator{\weights}{weights}
\DeclareMathOperator{\roots}{roots}
\DeclareMathOperator{\Hom}{Hom}

\DeclareMathOperator*{\JKres}{JK-res}
\DeclareMathOperator{\re}{Re}
\DeclareMathOperator{\im}{Im}
\DeclareMathOperator{\Tr}{Tr}
\DeclarePairedDelimiter{\abs}{\lvert}{\rvert}
\newcommand{\dd}[1]{\mathop{\mathrm{d}#1}}

\newcommand{\GaiottoTheory}[2]{\mathcal{T}\ifstrempty{#2}{}{_{#2}}[#1]}
\newcommand{\holonomy}{\mathfrak{m}}

\makeatletter
\newsavebox{\uprightarrowbox}
\savebox{\uprightarrowbox}{\scalebox{.5}{\tikz[scale=.36,color=gray]{\draw[fill](0,0) rectangle (.4,.4);\draw[ultra thick,->](0.2,0.2)--(.8,.8);}}}
\let\my@current@biblabel\empty
\let\oldbibcite\bibcite
\def\bibcite#1{\gdef\my@current@biblabel{#1}\oldbibcite{#1}}
\newcommand{\hiddenlink}[1]{%
  \immediate\write\@auxout{\unexpanded{\csgdef{my@link@to@\my@current@biblabel}{#1}}}}
\newcommand{\arxivlink}[2][]{%
  \hiddenlink{https://arxiv.org/abs/#2}%
  \href{https://arxiv.org/abs/#2}{{\ttfamily arXiv:#2\ifstrempty{#1}{}{ [#1]}}}}
\renewcommand{\citenumfont}[1]{\hbox{#1}\ifcsdef{my@link@to@\@citeb}{\textsuperscript{\;\href{\csname my@link@to@\@citeb\endcsname}{\!\usebox\uprightarrowbox\!}\,}}{}}
\makeatother

\title{\boldmath AGT/$\ZZ_2$}
\author[a]{Bruno Le Floch,}
\emailAdd{blefloch@princeton.edu}
\author[b]{Gustavo J. Turiaci,}
\emailAdd{turiaci@princeton.edu}
\affiliation[a]{Princeton Center for Theoretical Science,}
\affiliation[b]{Physics Department,\\
Princeton University, Princeton NJ 08544, USA}

\abstract{%
  We relate Liouville/Toda CFT correlators on Riemann surfaces with boundaries and cross-cap states to supersymmetric observables in four-dimensional $\Nsusy=2$ gauge theories.  Our construction naturally involves four-dimensional theories with fields defined on different $\ZZ_2$~quotients of the sphere (hemisphere and projective space) but nevertheless interacting with each other.
  The six-dimensional origin is a $\ZZ_2$~quotient of the setup giving rise to the usual AGT correspondence.
  To test the correspondence, we work out the $\RP^4$ partition function of four-dimensional $\Nsusy=2$ theories by combining a 3d lens space and a 4d hemisphere partition functions.  The same technique reproduces known $\RP^2$ partition functions in a form that lets us easily check two-dimensional Seiberg-like dualities on this nonorientable space.
  As a bonus we work out boundary and cross-cap wavefunctions in Toda CFT.%
}
\arxivnumber{1708.04631}

\begin{document}

\maketitle

\section{Introduction} 

The AGT correspondence relates correlation functions in Liouville/Toda 2d CFT on a Riemann surface and 4d $\Nsusy=2$ gauge theories on~${\rm S}^4$~\cite{AGT}.  The interplay between these two completely different setups has provided new ways to obtain and motivate new results on both sides, mainly for four dimensional gauge theories. From the two dimensional perspective this correspondence has been studied on Riemann surfaces with punctures and arbitrary genus.
Nevertheless, a question originally posed in~\cite{AGT} has not been addressed in the literature:

\medskip

What does CFT on surfaces with boundaries correspond to on the gauge theory side?

\medskip

In 2d CFT, adding boundaries to the surface in which the theory lives has proven to be very fruitful towards understanding their structure. Important progress in this direction was initiated by Cardy in a seminal paper~\cite{Cardy:1989ir}. Given the importance of his construction in CFT one is led to wonder what it teaches us about four dimensional gauge theories. This is the motivation for this work, and we take a first step by answering the question raised above. A second motivation is to learn more about the 6d $\Nsusy=(2,0)$ theory that gives rise to the AGT correspondence.

The 6d theory admits no supersymmetric boundary conditions, because of chirality.  Instead, 2d boundaries (and cross-caps, as we will see) arise from a $\ZZ_2$~quotient of the usual AGT setup.\footnote{A toy example to keep in mind is that the quotient $({\rm S}^1\times\RR)/\ZZ_2$ gives rise to a half-line upon reduction on~${\rm S}^1$.  This is true for both actions $\theta\to-\theta$ and $\theta\to\theta+\pi\bmod{2\pi}$ on~${\rm S}^1$, analogous to the two actions in~\eqref{reflectionandantipodal}.}
The 2d surface is a $\ZZ_2$~quotient of a closed Riemann surface~$\widehat{\Sigma}$ and we consider the 6d $\Nsusy=(2,0)$ theory on several quotients $({\rm S}^4_b\times\widehat{\Sigma})/\ZZ_2$.
Since the AGT correspondence relates (anti)holomorphic parts of 2d CFT correlators to (anti)instantons the two poles of the ellipsoid~${\rm S}^4_b$, and since the $\ZZ_2$~reverses the orientation on~$\widehat{\Sigma}$ it must concurrently exchange the poles of~${\rm S}^4_b$.

Discrete quotients of the AGT setup considered in previous works fixed the poles of the ellipsoid, where instantons and anti-instantons are located.  For example \cite{Belavin:2011pp, Belavin:2011tb,Belavin:2011sw,Bonelli:2012ny} considered $\ZZ_p$~subgroups of $(x_1,x_2)$ and $(x_3,x_4)$ rotations; instanton contributions matching Virasoro conformal blocks are then changed to those of a different chiral algebra.  In contrast, our~$\ZZ_2$ exchanges poles of the ellipsoid, hence identifies the instanton and anti-instanton sums.  Correspondingly, the $\ZZ_2$~action identifies left-moving and right-moving modes of the CFT\@.  An interesting extension would be to combine this $\ZZ_p$~with our $\ZZ_2$~quotients to learn about boundary states of the parafermionic CFTs corresponding to ${\rm S}^4_b/\ZZ_p$, including $\Nsusy=1$ super Liouville for $p=2$, and $\Nsusy=2$ super Liouville obtained by adding a surface operator on the gauge theory side~\cite{Belavin:2012uf}.

Our main tool to probe the proposed correspondence is supersymmetric localization.  The $\ZZ_2$~action must thus leave the localizing supercharge invariant, and in particular its square.  In terms of embedding coordinates $x\in\RR^5$ the ellipsoid is defined by
\begin{equation}\label{ellipsoid}
  x_0^2+b^{-2}(x_1^2+x_2^2)+b^2(x_3^2+x_4^2)=1 ,
\end{equation}
with poles at $x_0=\pm 1$, and the supercharge squares to rotations in the $(x_1,x_2)$ and $(x_3,x_4)$ planes.  We thus restrict our attention to the reflection $x_0\to -x_0$ around the ${\rm S}^3_b$ equator of ${\rm S}^4_b$ and to the antipodal map:
\begin{equation}\label{reflectionandantipodal}
  x_0\to -x_0 \text{ (reflection)} \quad\text{and}\quad x\to -x \text{ (antipodal map)}.
\end{equation}
The quotient ${\rm S}^4_b/\ZZ_2$ is the (squashed) hemisphere $\HS^4_b$ and projective space $\RP^4_b$, respectively.

\begin{figure}[t!]\centering\capstart
\begin{tikzpicture}[scale=1,rotate=0]\scriptsize
\node[inner sep=0pt] (whitehead) at (-9,2)
 {\includegraphics[width=.15\textwidth]{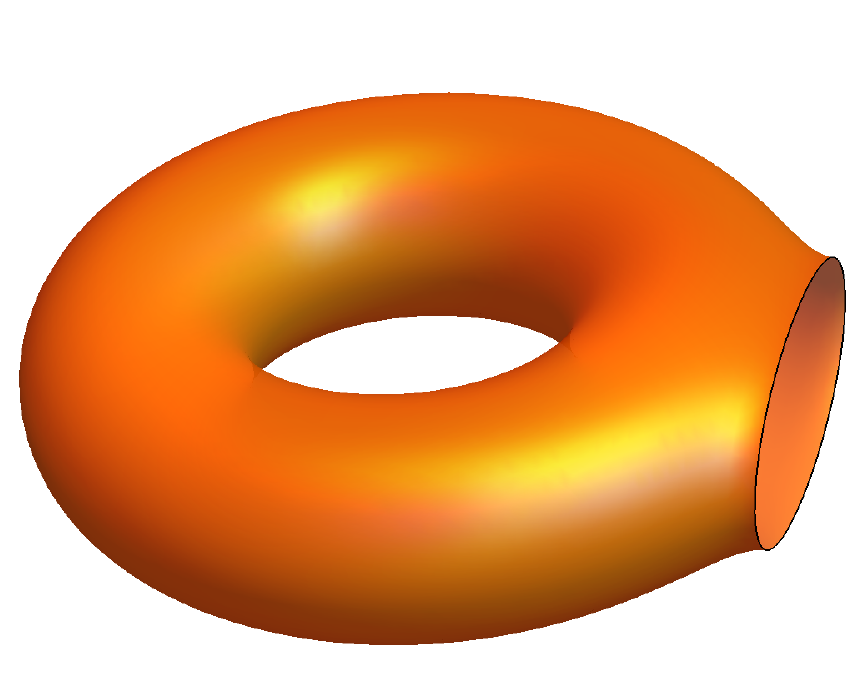}};
\draw (-6.5,2.3) node {\large $\Sigma=\widehat{\Sigma}/\ZZ_2$};
 \node[inner sep=0pt] (whitehead) at (-8,0)
    {\includegraphics[width=.3\textwidth]{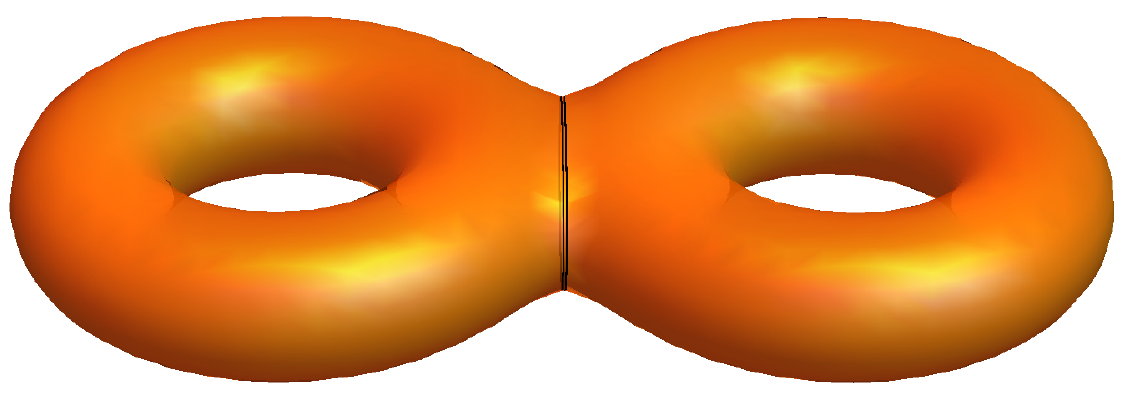}};
    \draw (-7,1.2) node {\large $\widehat{\Sigma}$};
    \draw (-8,-1.3) node {\large (a)};
\node[inner sep=0pt] (whitehead) at (0,0)
    {\includegraphics[width=.3\textwidth]{Sigma.pdf}};
    \draw (0,-1.3) node {\large (b) $({\rm S}^4_b\times\widehat{\Sigma})/\ZZ_2$};
 \draw (-2-0.6,2) arc (180:360:0.6cm and 0.3cm);
    \draw[dashed] (-2-0.6,2) arc (180:0:0.6cm and 0.3cm);
    \draw (-2,2) circle (0.6cm);
    \shade[ball color=blue!100!,opacity=0.50] (-2,2) circle (0.6cm);
    \draw[thick, dashed] (-2,2-0.6) -- (-2,0);
    \draw[fill] (-1.9-0.65,2.25) circle (0.05cm);
    \draw (-2.9,2.35) node {$P_1$};
    \draw (-0-0.6,2) arc (180:360:0.6cm and 0.3cm);
    \draw[dashed] (-0-0.6,2) arc (180:0:0.6cm and 0.3cm);
    \draw (-0,2) circle (0.6cm);
    \shade[ball color=blue!100!,opacity=0.50] (0,2) circle (0.6cm);
     \draw[thick, dashed] (0,2-0.6) -- (0,0);
          \draw[fill] (-0.4,2.45) circle (0.05cm);
    \draw (-0.5,2.75) node {$P_2$};
         \draw[fill] (-0.4,1.55) circle (0.05cm);
    \draw (-0.5,1.2) node {$P_2$};
    \draw (2-0.6,2) arc (180:360:0.6cm and 0.3cm);
    \draw[dashed] (2-0.6,2) arc (180:0:0.6cm and 0.3cm);
    \draw (2,2) circle (0.6cm);
    \shade[ball color=blue!100!,opacity=0.50] (2,2) circle (0.6cm);
     \draw[thick, dashed] (2,2-0.6) -- (2,0);
     \draw[fill] (2-0.51,1.7) circle (0.05cm);
    \draw (1.2,1.7) node {$P_1$};
\end{tikzpicture}
\vspace{-0.4cm}
\caption{\label{fig:6dorbifold}{\small (a)~An involution turns a Riemann surface without boundaries $\widehat{\Sigma}$ into one with boundaries $\Sigma$.  (b)~The lift of this construction to a 6d orbifold. We indicate how it acts on the ellipsoid~${\rm S}^4_b$. The singular locus is ${\rm S}^3_b\times {\rm S}^1$, where the first factor corresponds to the equator of the middle ellipsoid and the second factor to the boundary of the Riemann surface.}}
\end{figure}

In \autoref{fig:6dorbifold} we show a simple example of our setup. We begin with a Riemann surface~$\Sigma$ composed of a torus with a single circular boundary. In the same figure we show the Schottky double~$\widehat{\Sigma}$ which has no boundary but has genus~$2$. In the right panel we show how this construction naturally lifts to 6d, with the involution acting simultaneously on $\widehat{\Sigma}$ and~$S^4_b$.

\bigskip

The 4d theory corresponding to the surface with boundaries $\widehat{\Sigma}/\ZZ_2$ is then obtained by a $\ZZ_2$~identification of fields in the 4d $\Nsusy=2$ theory $\GaiottoTheory{\widehat{\Sigma}}{}$ associated to~$\widehat{\Sigma}$.
Recall how the 4d theory $\GaiottoTheory{\widehat{\Sigma}}{}$ is constructed for a choice of pants decomposition of~$\widehat{\Sigma}$.  One associates isolated CFTs (matter theories) to three-punctured spheres and one gauges flavor symmetries using vector multiplets associated to tubes connecting these spheres.  Given two three-punctured spheres or two tubes mapped to each other by the~$\ZZ_2$, the corresponding matter theories or vector multiplets are identified up to the space-time symmetry~\eqref{reflectionandantipodal}.  The vector multiplet corresponding to a tube invariant under~$\ZZ_2$, namely to a boundary of~$\Sigma$, for instance the middle tube in \autoref{fig:6dorbifold}, is instead identified to itself under this space-time symmetry.

The theory can be described more parsimoniously by keeping only one of each pair of fields identified by the~$\ZZ_2$.  This associates a vector multiplet to each tube in the bulk of $\Sigma=\widehat{\Sigma}/\ZZ_2$, and to each boundary of~$\Sigma$ it associates a vector multiplet living on~${\rm S}^4_b/\ZZ_2$ with the identification~\eqref{reflectionandantipodal}.  The recipe in this description is to start from the theory on~${\rm S}^4_b$ associated to the Riemann surface obtained by replacing all boundaries (and cross-caps) of~$\Sigma$ by punctures, then to gauge each corresponding flavor symmetry using a vector multiplet on~${\rm S}^4_b/\ZZ_2$.  From this point of view there is no reason for all of these vector multiplets to be subject to the same $\ZZ_2$~identification among~\eqref{reflectionandantipodal}.  This more general setup is obtained by replacing ${\rm S}^4_b\times\widehat{\Sigma}$ by a fibration over~$\widehat{\Sigma}$.

Apart from this description based on the fundamental domain ${\rm S}^4_b\times\Sigma$ of the $\ZZ_2$~action, there is another description based on the fundamental domain $\HS^4_b\times\widehat{\Sigma}$ where $\HS^4_b$ is the squashed hemisphere $x_0>0$.  Then the full field contents of $\GaiottoTheory{\widehat{\Sigma}}{}$ are kept, but restricted to~$\HS^4_b$, and with a boundary condition implementing the $\ZZ_2$~identification.  When $\ZZ_2$~acts as $x_0\to-x_0$ the vector multiplet associated to a tube joining $\Sigma$ to its image is given Neumann boundary conditions.  When $\ZZ_2$~acts as the antipodal map the boundary condition is non-local; it relates fields at opposite points of the equator so that 4d~fields effectively live on the projective space~$\RP^4_b$.

\bigskip

Two-dimensional CFTs have a rich set of boundary conditions.  In Cardy's formalism, which we review and apply to Liouville/Toda CFT in \autoref{app:wavefunctions}, boundary conditions are labeled by primary operators.  So far, in our construction, boundaries only depend on whether $\ZZ_2$~acts on the ellipsoid by the reflection or antipodal map.  The reflection case turns out to correspond to identity branes, labeled by the identity operator.  For this $\ZZ_2$~action $x_0\to-x_0$, there is a fixed-point locus\footnote{A priori it may be problematic to consider the 6d $\Nsusy=(2,0)$ theory on an orbifold, but our concrete match of 2d and 4d observables in the main text suggests otherwise.} ${\rm S}^3_b\times {\rm S}^1\subset {\rm S}^4_b\times\widehat{\Sigma}$ for each boundary of~$\Sigma$.  Inserting along this singular locus some codimension~$2$ and~$4$ operators of the 6d theory has the effect of changing the boundary condition of the CFT\@.

\begin{table}\capstart
  \centering
  \begin{tabular}{lm{.47\linewidth}l}
    \toprule
    CFT & Gauge theory & Labels \\ \midrule
    Ishibashi state & Dirichlet boundary condition (scaled) & Momentum~$\alpha$\\ \midrule
    ZZ brane (identity) & Neumann boundary condition & None \\
    ZZ brane (all) & Neumann $+$ Wilson line at equator & $(R_1,R_2)$ \\
    FZZT brane & Symmetry-breaking boundary condition \par \mbox{}$\qquad+$ FI parameters~$m$ $+$ Wilson line & $(H\subset G,m,R_1^H,R_2^H)$ \\
    Unusual brane & Antipodal boundary condition ($\vartheta=\pi$ or $0$) & None \\ \midrule
    Cross-cap state & Antipodal boundary condition ($\vartheta=0$ or $\pi$) & None \\
    Unusual cross-cap & Neumann boundary condition ($\vartheta=\pi$) & None \\\bottomrule
  \end{tabular}
  \caption{\label{tab:OpenAGTsummary}
    Correspondence between 3d $\Nsusy=2$ preserving boundary conditions for 4d $\Nsusy=2$ theories on the squashed hemisphere~$\HS^4_b$ and 2d CFT Ishibashi states, branes (Cardy states), and cross-caps.
    ZZ branes are labeled by a fully-degenerate primary hence by a pair of representations of the Lie algebra.  FZZT branes are labeled by semi-degenerate or non-degenerate primaries.
    The 6d construction also singles out an unusual boundary state, whose wavefunction is that of a cross-cap state, and an unusual cross-cap state, whose wavefunction is that of an identity brane.  On the gauge theory side these correspond to turning on a discrete theta angle.}
\end{table}

Strikingly, the basis of boundary conditions in Liouville/Toda CFT obeying the Cardy condition are in one-to-one correspondence with these operators of the 6d theory.  The explicit correspondence is listed in \autoref{tab:OpenAGTsummary}.
In analogy to branes in Liouville CFT~\cite{FZZ, ZZ} we call ZZ~branes the branes labeled by a fully degenerate primary (including the identity) and FZZT those labeled by a semi-degenerate or a non-degenerate primary.  The ZZ~brane is labeled by a pair of representations $(R_1,R_2)$ of the simply-laced Lie algebra~$\lie{g}$ that defines the 6d $\Nsusy=(2,0)$ theory.  It is obtained by inserting a pair of Wilson lines at $x_3=x_4=0$ and $x_1=x_2=0$ in representations $R_1$ and~$R_2$ of the gauge group associated to the boundary.  The FZZT~branes correspond to symmetry-breaking boundary conditions, essentially identical to symmetry-breaking walls obtained for topological defects~\cite{Drukker:2010jp}: the gauge group is broken to a subgroup $H\subset G$ at the boundary and FI~parameters for the remaining $U(1)$'s are turned on, as well as Wilson lines of~$H$ stuck to the boundary by gauge invariance.
Each of these 2d CFT boundary states is a linear combinations of Ishibashi states, and we match the latter (up to a scaling) to Dirichlet boundary conditions for the corresponding vector multiplet of the theory $\GaiottoTheory{\widehat{\Sigma}}{}$.  Our task then boils down to equating the wavefunction of the identity brane to the one-loop determinant of a vector multiplet with $x_0\to-x_0$ identification.

To circumvent the well-known issue of infinite fusion multiplicities in Toda CFT, the concrete examples of correlators that we give are for the 6d $\Nsusy=(2,0)$ theory of A-type with ``degenerate-enough'' insertions (see the main text).  They correspond in gauge theory to quivers of $SU(N)$ gauge groups.  Our results however apply to all branes in ADE Toda CFT\@.\footnote{We did not find in the literature a clear dictionary relating D-type and E-type Toda CFT to gauge theory.  The main missing ingredient seems to be finding any Toda CFT three-point functions beyond the A-type case.}

Besides having boundaries, $\Sigma$~can also be nonorientable.  Such surfaces can also be assembled by inserting cross-caps in an orientable surface, so they fit in Cardy's formalism.  We work out the ADE Toda cross-cap wavefunction in \autoref{app:wavefunctions}, namely what linear combination of cross-cap Ishibashi states is needed to describe the nonorientable surface.
The cross-cap turns out to correspond to an antipodal identification of the corresponding vector multiplet.
To identify in detail the Liouville/Toda cross-cap to a vector multiplet on~$\RP^4_b$ we worked out the partition function of such a multiplet.  We obtained it by a gluing prescription that combines ingredients from~$\HS^4_b$ and lens space~${\rm S}^3_b/\ZZ_2$ partition functions.  For completeness we extended this to any 4d $\Nsusy=2$ Lagrangian gauge theory (see \autoref{sec:rp4}, that can be read independently).

The 6d construction leaves a choice of whether to construct boundaries or cross-caps of the 2d CFT, and of whether $\ZZ_2$~acts by reflection or the antipodal map on the ellipsoid.  Before listing the four cases let us make some comments.  In 2d CFT\@, boundary and cross-cap Ishibashi states are formally related by an analytic continuation of a real cross-ratio to its opposite.  The AGT correspondence relates cross-ratios to complexified gauge couplings, and this analytic continuation corresponds to turning on a theta angle $\vartheta=\pi$.  Note that the ${\rm S}^4_b$ integral of the topological term $\Tr F\wedge F$ vanishes for vector fields that are invariant under the reflection or antipodal maps.  The theta term must thus only be integrated over a fundamental domain of the $\ZZ_2$ action.  To avoid dependence on the choice of fundamental domain, $\vartheta$ must be invariant (up to the $2\pi$ periodicity) under $\vartheta\to-\vartheta$, so $\vartheta=0$ or $\pi$.
The four cases are as follows.
\begin{enumerate}
\item Boundary and reflection: this gives rise to a vector multiplet on $\HS^4_b$~with Neumann boundary conditions, and on the CFT side to an identity brane.
\item Cross-cap and reflection: this gives the same vector multiplet, but with a discrete theta angle $\vartheta=\pi$.  On the CFT side it gives an unusual cross-cap state whose wavefunction is that of an identity brane.
\item Boundary and antipodal map: this gives an $\RP^4_b$~vector multiplet and on the CFT side an unusual boundary state whose wavefunction (given by the $\RP^4_b$ vector multiplet one-loop determinant) is equal to the cross-cap wavefunction.
\item Cross-cap and antipodal map: this also gives rise to an $\RP^4_b$~vector multiplet, but the theta angles in the two cases must differ by~$\pi$.  On the CFT side it gives a standard cross-cap.
\end{enumerate}
The 6d~quotient only has fixed points in the first case.  In other cases there is no preferred locus where codimension $2$ or $4$ operators are naturally placed; correspondingly they do not have the additional label of a primary operator.
It would be very interesting to understand whether the unusual cross-cap and boundary states coming out of the 6d construction are singled out from the 2d CFT perspective.
In the main text we only consider cases 1.\@ and 4.

All of the usual bells and whistles of the AGT correspondence can be inserted: defects of the 6d theory have the same 2d and 4d interpretations as before provided they are away from the fixed point locus.  We discuss these and a few other cases in \autoref{sec:generalizations}.
On the CFT side, besides correlation functions of bulk operators in the presence of a boundary one can also insert boundary operators. Their effect on correlation functions can be obtained by bootstrap methods and we find cases that correspond to duality walls put on the equator of the 4d geometry.  We were informed by Bawane, Benvenuti, Bonelli, Muteeb and Tanzini of their upcoming work~\cite{Bawane} in this direction.

\bigskip

The paper is organized as follows.
In \autoref{sec:bcft} we set up Liouville/Toda CFT notations and explain the construction of boundary states.
In \autoref{sec:hs4} we describe how 4d $\Nsusy=2$ SQCD on $\HS^4_b$ and 2d CFT disk two-point functions correspond for various choices of boundaries.
In \autoref{sec:rp4} we write the $\RP^4_b$ partition function~\eqref{ZRP4} of 4d $\Nsusy=2$ (Lagrangian) gauge theories as a combination of $\HS^4_b$ and ${\rm S}^3_b/\ZZ_2$ localization results, then detail the correspondence between $\RP^4_b$ and cross-caps.
In \autoref{sec:generalizations} we give generalizations: we explain how arbitrary Riemann surfaces with boundaries give 4d $\Nsusy=2$ quiver gauge theories, how to include the usual loop, surface, and domain wall operators, and discuss boundary-changing operators.
We conclude with some outlook in \autoref{sec:conclusion}.

Appendices collect interesting results we obtained in the course of our investigations.
In \autoref{app:wavefunctions} we give a streamlined derivation of Liouville/Toda boundary CFT results that exclusively uses modular bootstrap of the annulus.  From the M\"obius strip bootstrap we derive the ADE Toda cross-cap wavefunction~\eqref{psicrosscapexplicit}, which appears to be new.
In \autoref{app:rp2} we motivate the gluing procedure we use for the $\RP^4_b$~partition function by checking that the same technique reproduces the $\RP^2$~partition function of~\cite{Kim:2013ola}, and we check Seiberg dualities.
In \autoref{app:quotients} we describe quotients of ${\rm S}^4_b$ consistent with localization; this suggests possible generalizations of our work.
In \autoref{app:rp4hyper} we consider hypermultiplet determinants on~$\RP^4_b$.

\section{Liouville/Toda boundary CFT}\label{sec:bcft}

We briefly review here the construction of boundary states for ADE Toda field theory.  Branes in Liouville CFT were classified in~\cite{FZZ,ZZ}.  For Toda CFT they were classified in~\cite{Fateev:2010za} (see also~\cite{Fredenhagen:2010zh,Sarkissian:2011tr}) and found to be in one-to-one correspondence with representations of the extended algebra defining the theory, following the intuition of the Cardy construction.  We describe the modular bootstrap of these boundary states in \autoref{app:wavefunctions}.  In this subsection we cherry-pick details that are essential for the main text and describe how the results reduce to more familiar Liouville ones.

The Toda field theory is a nonrational CFT with diagonal spectrum.  It depends on a Lie algebra~$\lie{g}$ that we take to be simply-laced, and it reduces to Liouville theory when $\lie{g}=\lie{su}(2)$.  We let $\lie{t}$ denote the Cartan algebra of~$\lie{g}$ and $r=\rank\lie{g}$.  The theory has an extended W-algebra symmetry generated by currents 
\beq
W^{(s_i)},\qquad i=1,\ldots, r
\eeq
with spins $s_i=l_i+1$ determined by the set of exponents~$\{l_i\}$ of the Lie algebra.  This includes the stress tensor, of spin $s=2$.  The algebra of these currents has a central charge $c=r + 12\lb Q,Q\rb$, where we defined the background charge to be $Q=(b+1/b)\rho$, where $\rho$ is the Weyl vector of the algebra.  In the Lagrangian formulation the theory consists of a scalar field~$\phi$ in the Cartan subalgebra $\lie{t}$ of~$\lie{g}$, with exponential potentials.  In terms of the rescaled cosmological constant
\begin{equation}
  \hat{\mu}=(\pi\mu\gamma(b^2)b^{2-2b^2})^{1/b}
\end{equation}
the theory is invariant under $(b,\hat{\mu})\to(1/b,\hat{\mu})$.

Let us summarize the different W-algebra representations that we use in this work,\footnote{Yuji Tachikawa pointed out to us that the D-type and E-type 6d $\Nsusy=(2,0)$ theories have codimension~$2$ operators that fall outside this classification.  These should correspond to other W-algebra representations but the dictionary does not seem to be available in the literature.} classifying the primary operators of the CFT\@.  The W-algebra has non-degenerate, semi-degenerate, and fully degenerate representations with various amounts of null vectors.
\begin{itemize}
\item Non-degenerate representations.  These representations have no null states.  In the Lagrangian formulation they are constructed as $e^{\lb \mu, \phi \rb}$ in terms of a vector $\mu=Q+m$ with $m\in i \lie{t}$.  Their conformal weight is $\bar{\Delta}(\mu) = \Delta(\mu) = \lb 2Q-\mu, \mu\rb/2$.  Representations whose momenta~$m$ are related by the Weyl group are identified up to a reflection amplitude as~\eqref{reflectionidentification}.

\item Semi-degenerate representations.  They are labeled by a choice of a full-rank (hence simply-laced) subalgebra $\mathfrak{h}\subset \mathfrak{g}$ and two representations $R_1$, $R_2$ of~$\mathfrak{h}$ with highest weights $\lambda_1$, $\lambda_2$.  We denote by~$\mathcal{I}$ the set of simple roots of the subalgebra $\mathfrak{h}$ and indicate group theory quantities related to this algebra by an index $\mathcal{I}$. The momentum of these states is
\beq
m  = - (\rho_\mathcal{I} + \lambda_1) b - (\rho_\mathcal{I} + \lambda_2) /b + \tilde{m},
\eeq
where $\tilde{m}$ is a vector with imaginary components that is orthogonal to all roots in~$\mathcal{I}$.

\item Fully degenerate representations, namely semi-degenerate representations with $\mathfrak{h}=\mathfrak{g}$. They are labeled by two highest weights $\lambda_1$, $\lambda_2$ of representations $R_1$ and~$R_2$. The momentum is discrete,
\beq
m = - (\rho+ \lambda_1) b - (\rho + \lambda_2)/b.
\eeq
When $R_1$ and $R_2$ are the trivial representation ($\lambda_1=\lambda_2=0$) this operator is the identity.
\end{itemize}
The Virasoro algebra (corresponding to $\lie{g}=\lie{su}(2)$) only has fully degenerate and non-degenerate representations.\footnote{\label{foot:LiouvilleVsTodaNotations}It is worth noting a discrepancy between standard Liouville momenta~$\alpha_{\text{L}}$ and Toda notations: while $\alpha=\alpha_{\text{L}}e$ where $e$~is the positive root of~$\lie{su}(2)$, the parameters~$Q$ are related by $Q=Q_{\text{L}}e/2$.  Hence for instance $\Delta=\alpha_{\text{L}}(Q_{\text{L}}-\alpha_{\text{L}})=\langle\alpha,2Q-\alpha\rangle/2$ where the overall factor of~$2$ is due to roots having length-squared $\langle e,e\rangle=2$.}

To define boundary states associated to holes in the Riemann surface, we need to specify the boundary condition on the currents. We take untwisted Ishibashi states~$\bket{\alpha}$, namely formal sums of descendants of a primary~$\ket{\alpha}$ obeying\footnote{The insertion of this Ishibashi state gives a hole of unit size in the plane. A hole of arbitrary size~$e^r$ is obtained as $\bket{\alpha}_r = e^{2r L_0} \bket{\alpha}$ by applying the dilatation operator $L_0+\overline{L}_0$.  It obeys $\bigl( e^{-nr}W_n^{(s)} - (-1)^s e^{nr}\overline{W}_{-n}^{(s)}\bigr) \bket{\alpha}_r =0$.  The size of the hole simply alters cross-ratios.}\textsuperscript{,}\footnote{\label{foot:twistIshibashi}Ishibashi states can be twisted by an automorphism $\hat{\rho}$ of the W-algebra, replacing the condition~\eqref{Ishibashicondition} by $\bigl( W_n^{(s)} - (-1)^s \hat{\rho}\bigl(\overline{W}_{-n}^{(s)}\bigr)\bigr) \bket{\alpha} =0$.  In the Schottky double described in \autoref{sec:disk2pt}, the image momenta $2Q-\beta$ are replaced by $\hat{\rho}(2Q-\beta)$.}
\begin{equation}\label{Ishibashicondition}
\Bigl( W_n^{(s)} - (-1)^s \overline{W}_{-n}^{(s)}\Bigr) \bket{\alpha} =0.
\end{equation}
Branes are linear combinations of Ishibashi states with non-degenerate momenta,
\begin{equation}\label{boundarystateintermsofIshibashi}
  \ket{\B} = \frac{1}{|\Weyl|} \int\dd{(\im\alpha)} \Psi_{\B}(\bar{\alpha}) \bket{\alpha} .
\end{equation}
Here and throughout the paper, $\int\dd{\alpha}=\int\dd{(\im\alpha)}$ means an integral over all imaginary $\alpha-Q$ with standard real measure in the basis of simple roots~$e_j$: for any function~$F$,
\begin{equation}\label{intdalpha}
  \frac{1}{|\Weyl|} \int \dd{(\im\alpha)} F(\alpha)
  = \frac{1}{|\Weyl|} \int_{\RR^r}\dd{x^1}\cdots\dd{x^r} \!\sqrt{C} \, F\biggl(Q+i\sum_{j=1}^rx^je_j\biggr) ,
\end{equation}
where $C$~is the determinant of the Cartan matrix, since that matrix $C_{ij}=\langle e_i,e_j\rangle$ is the metric on the Cartan algebra in the coordinates~$x$.  While we divide all our integrals by the order $|\Weyl|$ of the Weyl group to reduce them to a Weyl chamber (the integrands are always Weyl-invariant), we keep this factor explicit in equations such as~\eqref{boundarystateintermsofIshibashi}.

Let us explain the appearance of $\bar{\alpha}=2Q-\alpha$ in~\eqref{boundarystateintermsofIshibashi}.
Vertex operators with momenta related by Weyl symmetry are proportional:
\begin{equation}\label{reflectionidentification}
  V_{\alpha} = R_w(\alpha) V_{Q+w(\alpha-Q)}
\end{equation}
for some reflection amplitude $R_w(\alpha)$ that we do not need explicitly~\cite{Fateev:2001mj}.  The two-point function on the plane is normalized as
\begin{equation}
  \langle V_{\alpha_1}(z_1)V_{\alpha_2}(z_2)\rangle
  = \frac{1}{\abs{z_1-z_2}^{4\Delta(\alpha_2)}} \sum_{w\in\Weyl} R_w(\alpha_2)\, \delta\bigl((\alpha_1-Q)+w(\alpha_2-Q)\bigr)
\end{equation}
where $\int\dd{(\im\alpha)}\delta(\alpha-\alpha_0)f(\alpha)=f(\alpha_0)$.  From~\eqref{boundarystateintermsofIshibashi} we then deduce that the one-point function on the plane with a unit-sized hole~\eqref{vertexhole} is\footnote{We use the natural normalization coming from the modular bootstrap, which is such that $\lb \mathbf{1} \rb_{\B} = \Psi_{\B}(0) \neq 1$. Another common choice that we will not use is to normalize $\lb \mathbf{1} \rb_\B =1$.}
\begin{equation}\label{vertexhole}
  \langle V_{\alpha}(z) \rangle_{\B} =\abs{1-z\bar{z}}^{-2\Delta(\alpha)} \Psi_{\B}(\alpha).
\end{equation}
This justifies the convention of using $\bar{\alpha}$ in~\eqref{boundarystateintermsofIshibashi}.  By a conformal transformation, the disk one-point function is also proportional to~$\Psi_{\B}(\alpha)$.  In particular, $\Psi_{\B}(\alpha)$~must transform like $V_{\alpha}$ \eqref{reflectionidentification} under the Weyl group.
It is straightforward to check that the wavefunctions below have this property.

\subsection{Boundary wavefunctions}

As we explain in \autoref{app:wavefunctions}, all the boundary states that come out of the Cardy construction result from adding to the identity brane a Verlinde loop operator along the boundary.  The wavefunction of the identity brane is~\eqref{psi1explicit}
\begin{equation}\label{psi1explicitmaintext}
  \Psi_1(\alpha)
  = (\hat{\mu}b^{2b-2/b})^{-\lb \rho, a\rb} \prod_{e>0}\frac{2\pi \lb e, a\rb }{\Gamma(1- b \lb e, a\rb)\Gamma(1 - b^{-1} \lb e, a\rb)}
\end{equation}
where we recall $\alpha=Q+a$.
Then explicit expressions of the S-matrix elements give all the branes wavefunctions. By adding a Verlinde loop operator corresponding to a degenerate vertex operator labeled by $(R_1, R_2)$ we obtain the degenerate brane
\beq\label{eq:todaCD}
\Psi_{(R_1, R_2)}(\alpha) = \Psi_1 (\alpha) \chi_{R_1}(e^{2 \pi i b a}) \chi_{R_2}(e^{2\pi i a/b})
\eeq
in terms of characters of~$\lie{g}$.
The semi-degenerate brane is given by 
\beq\label{eq:todaSD}
\Psi_{\mu, R_1, R_2}(\alpha) = \Psi_1(\alpha) \sum_{w\in\Weyl/\Weyl_{\mathcal{I}}} \frac{e^{2 \pi i \lb w(\tilde{m}),a \rb}  \chi_{R_1}(e^{w^{-1}(2 \pi i b a)}) \chi_{R_2}(e^{w^{-1}(2\pi i a/b)})}{\prod_{e\in \Delta^+ - w(\Delta^+_\mathcal{I})} \bigl(-4 \sin ( \pi b \lb a, e\rb) \sin (\pi \lb a, e\rb/b)\bigr)}.
\eeq
Finally, the analogue of the FZZT brane associated to a non-degenerate operator is given by 
\beq\label{eq:todaND}
\Psi_\mu(\alpha) = \Psi_1(\alpha) \sum_{w\in\Weyl} \frac{e^{2 \pi i \lb w(m),a \rb} }{\prod_{e>0} \bigl( -4 \sin ( \pi b \lb a, e\rb) \sin (\pi \lb a, e\rb/b)\bigr)}.
\eeq
On a related note, we give the cross-cap wavefunction in~\eqref{psicrosscapexplicitmaintext}.
All of the wavefunctions obey $\overline{\Psi_{\B}(\alpha)} = \Psi_{\B}(\overline{\alpha})$ and are invariant under Weyl group actions on their labels.  Like vertex operators, they are multiplied by a reflection amplitude when the Weyl group acts on~$\alpha$.  

For the Liouville theory, these branes are the ZZ branes \cite{ZZ} and FZZT branes \cite{FZZ} found long ago.  We parametrize Liouville momenta as $\alpha=\frac{1}{2}(b+1/b)+iP$ and reproduce the well-known results (with a pleasant normalization thanks to our choice of measure):
\begin{align}
\Psi_1^{\text{Liouville}}(P) & =  (\hat{\mu}b^{2b-2/b})^{-iP} \frac{2 \pi i P }{\Gamma(1-2 i P b) \Gamma(1-2 i P/b)}, \\
\Psi_{\text{ZZ}(m,n)}^{\text{Liouville}}(P) & = \Psi_1 (\tfrac{1}{2}(b+1/b)+iP) \,\frac{\sinh(2 \pi m P/b)}{\sinh(2 \pi P/b)}\frac{\sinh(2 \pi n b P)}{\sinh(2 \pi b P)}, \\
\Psi_{\text{FZZT}(s)}^{\text{Liouville}}(P) & = (\hat{\mu}b^{2b-2/b})^{-iP} \frac{\Gamma(1+2 i b P) \Gamma(1+2 i P/b) }{-2 \pi i P} \frac{1}{2} \cos(4 \pi s P).
\end{align}

\subsection{Disk two-point function}\label{sec:disk2pt}

Correlators on a surface~$\Sigma$ with boundaries can be computed using the method of images.  The surface is written as a quotient $\Sigma=\widehat{\Sigma}/\ZZ_2$ of its Schottky double, a closed Riemann surface.  Each vertex operator on~$\Sigma$ is split into two chiral operators on~$\widehat{\Sigma}$ in conjugate representations of the W-algebra, labeled by momenta $\alpha$ and $2Q-\alpha$.  The original correlator is then a chiral correlation function on~$\widehat{\Sigma}$, integrated over the internal Liouville/Toda momenta with certain weights.  For each brane~$\B$, the momentum~$\alpha$ that flows along the corresponding tube from the Riemann surface to its double is integrated with a weight~$\Psi_{\B}(2Q-\alpha)$ given in the last subsection.

In the normalization of conformal blocks (chiral correlators) we include the square root of an OPE coefficient at each trivalent vertex. Since each trivalent vertex appears twice in the Schottky double (with all momenta mapped as $\alpha\to 2Q-\alpha$), this is equivalent to the single OPE coefficient that one would obtain when computing the correlator on~$\Sigma$ using a bulk OPE. Our formulas thus only involve these normalized conformal blocks and the boundary (and later cross-cap) wavefunctions.

The simplest observable with a boundary is the one-point function~\eqref{vertexhole}.  We now focus mainly on the simplest non-trivial observable: the two point function in the region $\abs{z}\geq 1$ with a single boundary along $\abs{z}=1$.
This readily generalizes to higher genus or to more boundaries. For example a correlator on a cylinder (genus-zero Riemann surface with two boundaries) has as its Schottky double a chiral correlator on a torus.

\begin{figure}[t!]\centering\capstart
\begin{tikzpicture}[scale=.95,rotate=0]\scriptsize
\fill [draw=none, fill=blue, opacity=0.08, smooth] (-2.5,-2.5) -- (2.5,-2.5) -- (2.5,2.5) -- (-2.5,2.5)-- cycle;
\draw[thick,fill=white] (0,0) circle (1);
\draw (0.1,-0.65) node {$\ket{\B}$};
\draw[thick,fill=black] (1.6,1.6) circle (0.05);
\draw[thick,dashed] (1.6,1.6) -- (0.3125,0.3125);
\draw[thick,fill=black] (0.3125,0.3125) circle (0.05);
\draw (1.9,1.6) node {$\alpha_1$};
\draw (0.6,0.2) node {$\bar{\alpha}_1$};
\draw[thick,fill=black] (-0.5,1.5) circle (0.05);
\draw[thick,dashed] (-0.5,1.5) -- (-0.2,0.6);
\draw[thick,fill=black] (-0.2,0.6) circle (0.05);
\draw (-0.6,1.75) node {$\alpha_2$};
\draw (-0.45,0.3*1.5) node {$\bar{\alpha}_2$};
\end{tikzpicture}
\hspace{1cm}
\begin{tikzpicture}[scale=1.1,rotate=0]
\draw (0.35,-1) node {\footnotesize $\ket{\B}$};
\draw[ultra thick] (-2,0) -- (2,0);
\draw[ultra thick] (-1,0) -- (-1,1);
\draw[ultra thick] (1,0) -- (1,1);
\draw[thick,dashed] (0,-1.3) -- (0,2);
\draw (-0.3,-0.3) node {$\alpha$};
\draw (-2,-0.3) node {$\alpha_1$};
\draw (-1,1.3) node {$\alpha_2$};
\draw (2,-0.3) node {$\bar{\alpha}_1$};
\draw (1,1.3) node {$\bar{\alpha}_2$};
\draw[white] (0,-1.5) circle (0.1);
\end{tikzpicture}
\caption{\small Left: Schottky double. Right: Trivalent graph of the bulk two-point function.  In both figures we denote $\bar{\alpha}=2Q-\alpha$ for brevity.}
\label{fig2ptSD}
\end{figure}
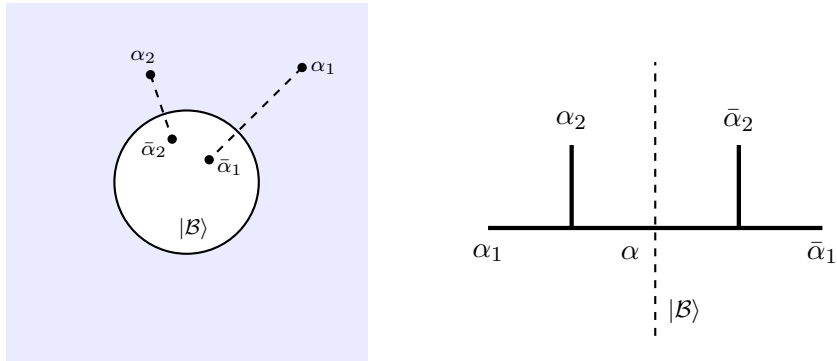

In this simple case of a single boundary, for each operator $V_\alpha(z)$, we consider a chiral correlator on the plane including the image insertion $V_{2Q-\alpha}$ at the image position $z^\star=1/\bar{z}$.  This construction is shown in the left of \autoref{fig2ptSD}.  It gives
\begin{equation}\label{eq:bulk2pt}
  \begin{aligned}
    \bigl\langle V_{\alpha_1}(z_1,\bar{z}_1) V_{\alpha_2}(z_2,\bar{z}_2)\bigr\rangle_{\B} &=   \bigl\langle V_{\alpha_1}(z_1) V_{\alpha_2}(z_2) V_{2Q-\alpha_1}(1/\bar{z}_1)V_{2Q-\alpha_2}(1/\bar{z}_2)\bigr\rangle_{\B}^{\text{chiral}}\\
    &= \frac{|1 - z_2 \bar{z}_2|^{2\Delta_1-2\Delta_2}}{|1-z_1\bar{z}_2|^{4\Delta_2}} \frac{1}{|\Weyl|}\int\dd{(\im\alpha)} \Psi_{\B}(\alpha)\,
    \block{F}{\alpha}{\alpha_2&2Q-\alpha_2\\\alpha_1&2Q-\alpha_1}{z},
  \end{aligned}
\end{equation}
where the cross-ratio~$z$ is real and $\Psi_{\B}$ is the wavefunction of the boundary state.  To avoid infinite fusion multiplicities we restrict concrete calculations to Liouville CFT or to A-type Toda CFT with one momentum $\alpha_2=\kappa\omega_1$ proportional to the first fundamental weight~$\omega_1$, hence in particular semi-degenerate.

As summarized in the right panel of \autoref{fig2ptSD}, we are computing the two-point function in the channel that involves the OPE of the two operators and the OPE of their images.  The result must be equivalent to the one computed using the OPE of an operator with its image.  This constraint is one of the sewing (crossing symmetry) constraints that are known~\cite{Lewellen:1991tb} to be sufficient to imply consistency of the CFT with boundaries.  Cardy's condition, which we use in \autoref{app:wavefunctions} to derive wavefunctions, is insufficient to imply full consistency in the most general case.

\section{Hemisphere and branes}\label{sec:hs4}

In this section we explain the dictionary between Liouville/Toda boundary states and 4d $\Nsusy=2$ gauge theories.
We first reproduce the identity brane~$\B_1$ from gauge theory.  As described in the introduction, this comes from an orbifold $({\rm S}^4_b \times \widehat{\Sigma})/\ZZ_2$ of the 6d theory, where $\ZZ_2$ acts by a $\ZZ_2$~involution of~$\widehat{\Sigma}$ such that $\widehat{\Sigma}/\ZZ_2$ is the surface with boundaries that we are interested in, and by $x_0\to-x_0$ on the ellipsoid~\eqref{ellipsoid}.
We analyze in \autoref{sec:orbid} how the latter identification affects the 4d $\Nsusy=2$ vector multiplet: this gives a vector multiplet on the squashed hemisphere~$\HS^4_b$ with Neumann boundary condition.  In \autoref{sec:Zhemi} we write the hemisphere partition function of~\cite{HS4} (generalized from $b=1$ to $b\neq 1$).

We match in \autoref{sec:matchZZ} the $\HS^4_b$~partition function with Neumann boundary conditions to the wavefunction of the identity brane.
This is the main result of this section: the identity brane~$\B_1$ corresponds to a vector multiplet with a $\ZZ_2$~identification that makes it effectively live on~$\HS^4_b$ with Neumann boundary.  All other fields live on the whole ellipsoid.
As a concrete example we match the partition function of 4d $\Nsusy=2$ $SU(N)$ SQCD with $N_f=2N$ fundamental hypermultiplets of pairwise equal masses~$m_j$ to a disk two-point function in $A_{N-1}$~Toda CFT:
\begin{equation}\label{HS4SQCDeq2ptdisk}
  Z_{\HS^4_b,\text{Neumann}}^{\text{SQCD},N_f=2N}\bigl(m_1,m_1,\dots,m_N,m_N; q)
  = \lb V_{\mu}(z_1) V_{\kappa\omega_1}(z_2)\rb_{\B_1}
\end{equation}
up to unimportant factors.  The masses~$m_j$ are encoded in the labels $\mu=Q+i\sum_jm_jh_j$ and $\kappa\omega_1=\bigl(\tfrac{1}{2}N(b+1/b)-i\sum_jm_j\bigr)\omega_1$ of primary operators on the Toda side, where $h_j$~are weights of the fundamental representation of~$SU(N)$.  The cross-ratio of their positions $z_1$, $z_2$ and their $\ZZ_2$~images $z_1^\star$ and~$z_2^\star$ is the instanton-counting parameter~$q$.

As we review in \autoref{app:wavefunctions}, all boundary states are obtained from the identity brane (also called the ZZ~brane) by inserting Verlinde loops along the boundary.  The gauge theory interpretation of these is already understood~\cite{Drukker:2010jp} and we recall it in \autoref{sec:otherbranes}.

We end with a brief discussion of S-duality of boundary conditions in \autoref{sec:sduality}.

\subsection{Orbifold identification}\label{sec:orbid}

The first step is to rewrite fields on the 4d orbifold ${\rm S}^4_b/\ZZ_2$ as fields on the squashed hemisphere~$\HS^4_b$ with appropriate boundary conditions at $x_0=0$.  We find that the correct prescription is to take Neumann boundary conditions, as defined for example in~\cite{Drukker:2010jp, Gaiotto:2008sa}.

In a 4d $\Nsusy=2$ gauge theory the vector multiplet is composed of
\beq
\text{vector~multiplet}:\quad A_\mu,~ \lambda^A,~ \bar{\lambda}_A,~ \sigma = \phi_1 + i \phi_2 ,~ \bar{\sigma} = \phi_1 - i\phi_2 ,
\eeq
where we used the notation $\phi_1$ and~$\phi_2$ to ease comparison with~\cite{HS4}.
Following~\cite{Hama:2012bg} we parametrize the ellipsoid as 
\begin{equation}
  \begin{aligned}
    x_0 &= \cos \rho, \\
    x_1+i x_2 &= b \sin \rho \cos \theta e^{i \varphi},\\
    x_3 + i x_4 &= b^{-1} \sin \rho \sin \theta e^{i \chi},
  \end{aligned}
\end{equation}
where $0\leq\varphi,\chi\leq 2\pi$ are periodic, $0\leq \rho \leq \pi$ and $0\leq\theta\leq\pi/2$.  The singular locus of the 6d orbifold is located along the equator $\rho=\pi/2$ of ${\rm S}^4_b$, and the $\ZZ_2$ identification acts as 
\beq
(\rho, \theta, \varphi, \chi) \leftrightarrow (\pi -\rho, \theta, \varphi, \chi).
\eeq
Since $\dd{\rho}\to-\dd{\rho}$ and $\dd{\theta}$, $\dd{\varphi}$, $\dd{\chi}$ are unchanged, $A_{\rho}$~and $F_{\rho\mu}$ for $\mu\in\{\theta,\varphi,\chi\}$ change sign under this~$\ZZ_2$ while other components do not.

The $\ZZ_2$ changes chirality hence must exchange gauginos $\lambda$ and~$\bar{\lambda}$ up to some invertible $2\times 2$ matrix.  Supersymmetry then constrains this matrix: the supersymmetry variation
\begin{equation}
  \delta A_\mu=i\xi^A\tau_\mu\bar{\lambda}_A-i\bar{\xi}^A\bar{\tau}_\mu\lambda_A
\end{equation}
should be $\ZZ_2$~invariant for $\mu\in\{\theta,\varphi,\chi\}$ and change sign for $\mu=\rho$.  Here $\tau_\mu$~are Pauli matrices.  This holds for
\begin{align}
  \label{lambdasym}
  i \tau^3 \lambda_A (\rho, \theta, \varphi, \chi) &= \bar{\lambda}_A (\pi -\rho, \theta, \varphi, \chi), \\
  \label{xisym}
  i \tau^3 \xi_A (\rho, \theta, \varphi, \chi) & = \bar{\xi}_A(\pi -\rho, \theta, \varphi, \chi),
\end{align}
where $\tau^3$ is $\tau^\rho$, normalized by~$\sqrt{g_{\rho\rho}}$.  The Killing spinors used in~\cite{Hama:2012bg,HS4} do indeed obey~\eqref{xisym}.\footnote{In the 6d setup~\cite{Cordova:2016cmu} with partial topological twist along the Riemann surface~$\widehat{\Sigma}$, supercharges are constant scalars on~$\widehat{\Sigma}$ hence are $\ZZ_2$-invariant regardless of the $\ZZ_2$~action on~$\widehat{\Sigma}$.}  The supersymmetry variations $\delta\sigma=\bar{\xi}^A\bar{\lambda}_A\leftrightarrow-\xi^A\lambda_A=\delta\bar{\sigma}$ so the $\ZZ_2$~exchanges $\sigma\leftrightarrow\bar{\sigma}$.  We get the following $\ZZ_2$~identification together with~\eqref{lambdasym}
\begin{equation}\label{hs4identificationofvectormultiplet}
  \begin{aligned}
    A^\mu (\rho, \theta, \varphi, \chi) &=
    \begin{cases}
      A^\mu (\pi -\rho, \theta, \varphi, \chi), & \text{if } \mu= \theta, \varphi, \chi,\\
      -A^\mu (\pi -\rho, \theta, \varphi, \chi), & \text{if } \mu=\rho,
    \end{cases}\\
    \sigma(\rho, \theta, \varphi, \chi) &= \bar{\sigma} (\pi-\rho, \theta, \varphi, \chi),
  \end{aligned}
\end{equation}
up to gauge transformations, of course.  This identification can be twisted by an outer automorphism~$\hat{\rho}$ of the gauge algebra, changing for instance the last equation above to $\sigma(\rho, \theta, \varphi, \chi) = \hat{\rho}\bigl(\bar{\sigma}(\pi-\rho, \theta, \varphi, \chi)\bigr)$.  We only consider the untwisted case: the twisted case should correspond to twisted boundary states discussed in \autoref{foot:twistIshibashi}.

The (untwisted) identification implies Neumann boundary conditions along the equator,
\begin{equation}\label{Neumannboundarycondition}
  F_{\rho j}\big|_{\rho=\pi/2} =0,\quad
  \phi_2\big|_{\rho=\pi/2} = 0,\quad
  \partial_\rho \phi_1\big|_{\rho=\pi/2} = 0,\quad
  \bigl( i \tau^3 \lambda_A - \bar{\lambda}_A \bigr)\big|_{\rho=\pi/2} =0 ,
\end{equation}
where $j\in\{\theta, \varphi, \chi\}$, together with infinitely many conditions on higher derivatives.  These match with the Neumann boundary condition (5.9) of~\cite[v2]{HS4} when their~$a$ is set to zero and the relation between their $\Lambda$ and~$\phi_1$ is taken into account.
The way Neumann boundary conditions are conventionally defined \cite{Drukker:2010jp, Gaiotto:2008sa} is $(4\pi/g^2) F_{\rho i} + (\vartheta/(2\pi)) \frac{1}{2} \epsilon_{ijk} F^{jk} =0$ where $i$, $j$, $k$ are transverse to the equator and $\vartheta$~is the theta angle.  We take $\vartheta=0$, consistent with the fact that for CFT with boundaries the cross-ratio is real and of a definite sign.
In fact, under $\ZZ_2$ the theta angle changes sign, so $\vartheta=0$ and $\vartheta=\pi$ are both sensible choices (integrated over a hemisphere); on the CFT side this changes the sign of cross-ratios, interchanging cross-cap and boundary states.

On the hemisphere $0\leq \rho \leq \pi/2$ the variation of the action, both the original and the localization term, vanishes up to boundary terms.  As shown in~\cite{HS4}, the boundary terms vanish upon imposing the Neumann boundary conditions~\eqref{Neumannboundarycondition}, but also upon imposing Dirichlet boundary conditions, namely the supersymmetric completion of $\bigl( i \tau^3 \lambda_A + \bar{\lambda}_A \bigr)\big|_{\rho=\pi/2}=0$.  The sign change compared to~\eqref{Neumannboundarycondition} leads to $A_j|_{\rho=\pi/2}=0$ in some gauge and
\begin{equation}\label{Dirichletboundarycondition}
  F_{ij}\big|_{\rho=\pi/2} =0,\quad
  \phi_1\big|_{\rho=\pi/2} = \text{constant},\quad
  \partial_\rho \phi_2\big|_{\rho=\pi/2} = 0,\quad
  \bigl( i \tau^3 \lambda_A + \bar{\lambda}_A \bigr)\big|_{\rho=\pi/2} =0 ,
\end{equation}
where $i,j\in\{\theta,\varphi,\chi\}$.
One may wonder whether the Dirichlet boundary conditions can be obtained from a $\ZZ_2$~orbifold of the whole~${\rm S}^4_b$.  To obtain $A_j|_{\rho=\pi/2}=0$, the $A_j$~components should change sign under~$\ZZ_2$ and $A_\rho$~not.  Then the first two terms in $F_{\rho j} \sim \partial_\rho A_j - \partial_j A_\rho + [A_\rho ,A_j]$ are invariant under $\rho\to\pi-\rho$ (since $\partial_\rho\to-\partial_\rho$) but the commutator changes sign.  Therefore we cannot get a consistent identification of the field strength unless the gauge group is abelian.  (In the abelian case the Dirichlet boundary arises from the orbifold~\eqref{hs4identificationofvectormultiplet} twisted by charge conjugation.)

From the CFT side this impossibility may be related to Cardy's condition. We check below that a hemisphere with Neumann boundary conditions corresponds to the identity brane, and Dirichlet boundary conditions to an Ishibashi state (up to normalization). As we review in Appendix \ref{app:wavefunctions} Ishibashi states do not satisfy Cardy's condition and therefore do not correspond to physical boundary conditions. It would be very interesting to understand Cardy's condition from the 4d~point of view in a more general setting.

Before moving on we mention what changes for the $\ZZ_2$~action that creates an unoriented surface~$\RP^4_b$, used in \autoref{sec:rp4}. In this case the points that are to be identified are 
\beq
(\rho, \theta, \varphi, \chi) \leftrightarrow (\pi -\rho,\theta, \pi+ \varphi,\pi+ \chi).
\eeq
The identification is consistent with supersymmetry if the fields transform as\footnote{Here we use~\cite{HS4}'s conventions for the Killing spinors; the $\lambda$~identification would have opposite sign with conventions in~\cite{Hama:2012bg}.}
\begin{equation}\label{rp4identificationofvectormultiplet}
  \begin{aligned}
    A^\mu (\rho, \theta, \varphi, \chi) &=
    \begin{cases}
      A^\mu (\pi -\rho, \theta, \pi+\varphi, \pi +\chi), &\text{if } \mu= \theta, \varphi, \chi\\
      -A^\mu (\pi -\rho, \theta, \pi+\varphi, \pi+\chi), &\text{if } \mu=\rho
    \end{cases}\\
    i \tau^3 \lambda_A (\rho, \theta, \varphi, \chi) &= \bar{\lambda}_A (\pi -\rho, \theta, \pi+\varphi, \pi+\chi), \\
    \sigma(\rho, \theta, \varphi, \chi) &= \bar{\sigma} (\pi-\rho, \theta,\pi+ \varphi, \pi+\chi).
  \end{aligned}
\end{equation}
As they should, the Killing spinors $\xi$ and~$\bar{\xi}$ obey the same constraint as $\lambda$ and~$\bar{\lambda}$.

\subsection{Partition function on hemisphere}\label{sec:Zhemi}

The partition function of a 4d $\Nsusy=2$ gauge theory on a squashed hemisphere differs from the one on the full ellipsoid in two ways.

On the one hand, while on an ellipsoid the instanton partition function appears in pairs as $Z_{\text{inst}} (q)\bar{Z}_{\text{inst}} (\bar{q})$ due to the presence of instantons localized at the north pole and anti-instantons localized at the south pole, for the squashed hemisphere a single instanton contribution $Z_{\text{inst}}(q)$ appears, localized at a single pole.\footnote{While one may select a given instanton sector (power of~$q$) by imposing the value of the Chern--Simons term integrated along the boundary, this is unnatural since it is a nonlocal boundary condition.} At the same time, the classical action is half of that on the ellipsoid.\footnote{If one localized on the region $\rho<\rho_0$ instead of $\rho<\pi/2$, one-loop determinants would be unchanged and the classical action would be proportional to the volume of the region.} Both are expected if we want this partition function to match with a CFT correlator on a disk, as that is given by a chiral correlator on the Schottky double and therefore a single conformal block appears for the channel associated to the boundary state.

On the other hand, the one-loop determinants are different. The one-loop determinants for Dirichlet and Neumann boundary conditions are respectively
\begin{align}
  \label{oneloopHS4D}
  Z_{\HS^4_b,\text{Dirichlet}}^{\text{vector}} & = \prod_e\frac{1}{\Gamma_b(b+1/b-\lb e,i\sigma\rb)}= \prod_{e>0}\biggl[\frac{\prod_{\pm}\Gamma(1+b^{\pm 1}\lb e,i\sigma\rb)}{2\pi\lb e,i\sigma\rb b^{(b-1/b)\lb e,i\sigma\rb}} \Upsilon_{b}(\lb e, i\sigma\rb)\biggr],\\
  \label{oneloopHS4N}
  Z_{\HS^4_b,\text{Neumann}}^{\text{vector}} &= \prod_e\frac{1}{\Gamma_b(\lb e,i\sigma\rb)} = \prod_{e>0}\biggl[\frac{-2\pi\lb e,i\sigma\rb b^{-(b-1/b)\lb e,i\sigma\rb}}{ \prod_{\pm} \Gamma(1-b^{\pm 1} \lb e,i\sigma\rb)} \Upsilon_{b}(\lb e,i\sigma\rb)\biggr],
\end{align}
where products range over all roots~$e$ of~$G$ or only positive roots $e>0$.  In the Neumann case the Coulomb branch parameter~$\sigma$ is integrated over the Cartan algebra~$\lie{t}$ with the measure~\eqref{intdalpha} normalized to match with CFT\@.  The usual Jacobian factor (Vandermonde determinant) converting the integral from $\lie{g}$ to $\lie{t}$ is included in the one-loop determinant displayed here. The properties of the functions appearing in these one-loop determinants are as follows.
\begin{itemize}
\item $\Gamma_b$~is the Barnes double gamma function; it has poles at $b\ZZ_{\leq 0}+b^{-1}\ZZ_{\leq 0}$, is invariant under $b\to 1/b$ and obeys $\Gamma_b(x+b)=\Gamma_b(x)\sqrt{2\pi}b^{xb-1/2}/\Gamma(bx)$ and $\Gamma_b((b+1/b)/2)=1$;
\item $\Upsilon_b(x)=1/(\Gamma_b(x)\Gamma_b(b+1/b-x))$ has zeros at $x\in(b\ZZ_{\leq 0}+b^{-1}\ZZ_{\leq 0})\cup(b\ZZ_{\geq 1}+b^{-1}\ZZ_{\geq 1})$;
\item $S_b(x)=\Gamma_b(x)/\Gamma_b(b+1/b-x)$ used later has poles at $x\in b\ZZ_{\leq 0}+b^{-1}\ZZ_{\leq 0}$ and zeros at $x\in b\ZZ_{\geq 1}+b^{-1}\ZZ_{\geq 1}$.
\end{itemize}
For $b=1$ these one-loop determinants were computed in~\cite{HS4}: these authors used spherical harmonics to find eigenvalues of the operator associated to quadratic fluctuations of the localizing term in the action, and restricted to modes that are consistent with the boundary condition or equivalently with the $\ZZ_2$~identification (in the Neumann case).  The precise forms of the localizing term and quadratic fluctuations are given there.

To find the generalization to $b\neq 1$ we kept track of zeros and imposed $b\to b^{-1}$ and $\sigma\to-\sigma$ invariance.  The $b=1$ result from~\cite{HS4} in the Dirichlet case has zeros of order $n-1$ at $\pm\lb e,i\sigma\rb=n$ for integers $n\geq 2$ and positive roots~$e$.  Given the ellipsoid answer~\cite{Hama:2012bg} these multiple zeros should split into two lattices $\pm\lb e,i\sigma\rb \in t(b)+b\ZZ_{\geq 0}+b^{-1}\ZZ_{\geq 0}$ for some function~$t(b)$.  It is natural to expect points in the lattice and its opposite to differ by multiples of $b$ and $b^{-1}$, so $2t(b)=kb+\ell b^{-1}$ for some integers $k$ and~$\ell$.  Using $t(b)=t(b^{-1})$ and $t(1)=2$ fixes $k=\ell=2$, telling us the set of zeros of~\eqref{oneloopHS4D}.

The normalization is not fixed by these arguments; we choose it to have two properties (besides being nonnegative, reducing to the known $b=1$ case and being $\sigma\to-\sigma$ invariant).  First, the determinant for the Neumann case is equal to the Dirichlet one times an ${\rm S}^3_b$~vector multiplet determinant $\prod_e 1/S_b(\lb e,i\sigma\rb)$.  Second, this 3d determinant times the square of the Dirichlet one gives the vector multiplet determinant on the whole ellipsoid~${\rm S}^4_b$.  As discussed more near~\eqref{S4bgluingvector}, these two properties are justified by gauging along ${\rm S}^3_b$ the global symmetry of a vector multiplet on one or two squashed hemispheres with Dirichlet boundary conditions.

Let us quote the result for the one-loop determinant for matter fields, even though we do not use it in our work. Supersymmetry leads to mixed Dirichlet/Neumann boundary conditions. Then the one-loop determinant of a hypermultiplet in the representation~$\repr$ of~$G$ is
\begin{equation}
  Z_{\HS^4_b}^{\text{hyper}} = \prod_{w\in\weights(\repr)} \frac{1}{\Upsilon_{b}\big(\frac{1}{2}(b+1/b)+\lb w,i\sigma\rb \big)^{1/2}} .
\end{equation}

\subsection{Identity brane}\label{sec:matchZZ}

We now match the gauge theory results to an identity (ZZ) brane.

The momentum integral translates through $\alpha=Q+i\sigma$ to the Coulomb branch integral of the gauge theory partition function with Neumann boundary conditions.  In the vector multiplet determinant~\eqref{oneloopHS4N} we recognize the wavefunction $\Psi_1(\alpha)$ of the identity brane (ZZ brane):
\begin{equation}\label{vectormatchesZZ}
  Z_{\HS^4_b,\text{Neumann}}^{\text{vector}} = \Psi_1(Q+i\sigma) \, \hat{\mu}^{\lb\rho,i\sigma\rb} \prod_{e>0}\Bigl(-\Upsilon_{b}(\lb e,i\sigma\rb)\Bigr) .
\end{equation}
The factor could be absorbed into a normalization of vertex operators that makes them Weyl invariant.  In correlators it cancels with parts of the OPE coefficients that depend separately on each momentum, specifically the numerator of the DOZZ formula or its Toda CFT analogue.

For concreteness consider the disk two-point function~\eqref{HS4SQCDeq2ptdisk} of $A_{N-1}$~Toda CFT, with a non-degenerate momentum~$\alpha_1$ and a semi-degenerate momentum $\kappa\omega_1$.  The result of the boundary state formalism translates to gauge theory as~\eqref{disk2pt}.  Besides an instanton partition function and classical contribution, the conformal block contains a normalization since we included in it a single OPE coefficient~$C^\alpha_{\alpha_1,\kappa\omega_1}$.  From the AGT correspondence on~${\rm S}^4_b$ we know that this OPE coefficient is equal to hypermultiplet determinants on~${\rm S}^4_b$, up to the factor discussed below~\eqref{vectormatchesZZ}.  More precisely, this is the one-loop determinant on ${\rm S}^4_b$ of $N^2$~hypermultiplets that transform in the bifundamental representation of the gauge group $SU(N)$ and a flavor symmetry group $U(N)$.  They can alternatively be described as pairs of hypermultiplets on~${\rm S}^4_b$ pairwise-identified at the boundary.  The real cross-ratio~$q$ translates to an instanton counting parameter.  Altogether,
\begin{equation}\label{disk2pt}
  \begin{aligned}
    \lb V_{\alpha_1}(z_1) V_{\kappa\omega_1}(z_2)\rb_{\B_1} & = \frac{1}{|\Weyl|} \int \dd{(\im\alpha)} \Psi_1(\alpha) \block{F}{\alpha}{\kappa\omega_1&2Q-\kappa\omega_1\\\alpha_1&2Q-\alpha_1}{q}
    \\
    & = \frac{1}{|\Weyl|} \int\dd{\sigma}Z_{\HS^4_b,\text{Neu.}}^{\text{vector}}(\sigma) Z_{{\rm S}^4_b}^{\text{hyper}}(\sigma) Z_{\text{instanton}}(\sigma,q) ,
  \end{aligned}
\end{equation}
where we omitted some powers of $\abs{z_2-z_2^\star}$ and $\abs{z_1-z_2^\star}$ given in~\eqref{eq:bulk2pt}.

We get the partition function of a theory constructed as follows: start with $N^2$ hypermultiplets living on~${\rm S}^4_b$ (this is the theory associated to a three-punctured sphere with momenta $\alpha_1$, $\kappa\omega_1$, $\alpha$), then gauge an $SU(N)$ symmetry using a vector multiplet that has the $\ZZ_2$~identification~\eqref{hs4identificationofvectormultiplet}.  In other words, gauge it using a vector multiplet living on~$\HS^4_b$ with Neumann boundary conditions.  Alternatively the whole theory can be considered on~$\HS^4_b$ by describing the hypermultiplets as a pair of hypermultiplets on the squashed hemisphere.

Finally, consider the gauge theory we just described on the squashed hemisphere, namely a pair of identical-mass hypermultiplets coupled to a vector multiplet, but change the vector multiplet boundary conditions to Dirichlet.  Up to a normalization, the partition function is a disk two-point function with Ishibashi boundary state:
\beq
Z_{{\rm HS}^4_b} = \block{F}{\alpha}{\kappa\omega_1&2Q-\kappa\omega_1\\\alpha_1&2Q-\alpha_1}{q}
= \lb V_{\alpha_1}(z_1) V_{\kappa\omega_1}(z_2)\rb_{\text{Ishibashi}(\alpha)} .
\eeq
This is a special case of the CFT interpretation of Dirichlet boundary conditions~\eqref{Dirichletboundarycondition} as selecting internal momenta of conformal blocks~\cite{Drukker:2010jp}.

\subsection{Other branes}\label{sec:otherbranes}

As we review in \autoref{app:wavefunctions} the Cardy construction of boundary CFT implies that a general boundary state labeled by~$\alpha$ is obtained from the identity brane by acting with a Verlinde loop operator along the boundary.  The gauge theory counterpart of Verlinde loops was worked out in \cite{Drukker:2010jp} and here we present their results for completeness.
From the 6d orbifold perspective, Verlinde loops correspond to the insertion of defects on the singular locus ${\rm S}^3_b\times {\rm S}^1$.  For $\alpha$ fully-degenerate the defects are codimension~$4$ defects of the 6d theory and span a great circle of~${\rm S}^3_b$, times the~${\rm S}^1$ boundary; they reduce to Wilson loops in~4d.  Otherwise they are codimension~$2$ and span the whole singular locus; they reduce to symmetry-breaking boundary conditions in~4d.

Writing the squashed sphere in coordinates
\beq
x_0^2 + \frac{x_1^2+x_2^2}{b^2} +  b^2(x_3^2+x_4^2)=1,
\eeq
the supercharge used for localization squares to
\beq
Q^2 = R_{12}+R_{34},
\eeq 
where $R_{ij}$ corresponds to a rotation in the $(x_i,x_j)$ plane.  These rotations admit two invariant circles, and we put there two supersymmetric Wilson lines in representation $R_1$~and $R_2$ of~$\lie{g}$:
\begin{equation}\label{greatcircles}
  \begin{aligned}
    & R_1 \quad\text{along}\quad x_1^2 + x_2^2 = b^2,~x_3 = x_4=0, \\
    & R_2 \quad\text{along}\quad x_1=x_2=0,~ x_3^2+x_4^2 = b^{-2}.
  \end{aligned}
\end{equation}
From \cite{Pestun:2007rz} we know the effect of these insertions is to include the character of these representations inside the partition function 
\beq
\chi_{R_1}(e^{2 \pi i b a}) \chi_{R_2}(e^{\frac{2\pi i}{b} a}).
\eeq
Since we know the $\HS^4_b$ result with Neumann boundary conditions reproduces the identity brane, these insertions add the factors necessary to turn the boundary state into a completely degenerate one, as in equation~\eqref{eq:todaCD}.

Another type of defect we can insert along the equator is a symmetry-breaking wall. This was defined for $\Nsusy=4$ SYM in  \cite{Gaiotto:2008sa} and extended to $\Nsusy=2$ in \cite{Drukker:2010jp}. They have the property of breaking the group $G$ along the equator to a subgroup $H$ by requiring the vector multiplet to take values in~$\lie{h}$ rather than~$\lie{g}$ at the wall. This can be done in a way that preserves the 3d $\Nsusy=2$ supersymmetry at the equator. In \cite{Drukker:2010jp} it is argued that the action of this symmetry-breaking wall is to introduce the following factor in the partition function (when $H$~has full rank)
\beq
\sum_{w\in\Weyl/\Weyl_H} \frac{e^{2 \pi i \lb w(\tilde{m}),a \rb}}{\prod_{e\in \Delta^+ - w(\Delta^+_H)} \bigl(-4 \sin ( \pi b \lb a, e\rb) \sin (\pi \lb a, e\rb/b)\bigr)}.
\eeq
From the gauge theory point of view the parameter~$\tilde{m}$ consists of Fayet-Iliopoulos terms for the $U(1)$ factors localized at the wall.  Semi-degenerate momenta are parametrized by $H$ and $\tilde{m}$ but also representations $R_1$ and~$R_2$ of~$H$; correspondingly, to reproduce the action of a Verlinde loop operator one needs to insert along the symmetry-breaking wall a pair of Wilson lines in representations $R_1$ and~$R_2$ of the residual symmetry~$H$.  Combining these with the identity brane wavefunction yields the semi-degenerate wavefunction~\eqref{eq:todaSD}.

\subsection{S-duality}\label{sec:sduality}
To conclude this section, we indicate how to use CFT results to probe how boundaries transform under S-duality for some $\Nsusy=2$ gauge theories, as thoroughly studied in~\cite{Gaiotto:2008ak} for $\Nsusy=4$.  We limit ourselves to $SU(2)$ for this discussion, but keep notations adapted to Toda CFT (see \autoref{foot:LiouvilleVsTodaNotations}).  In this case, we have shown above that
\beq\label{HSNS}
Z_{\HS^4_b}^{\text{Neumann}} = \int_{-\infty}^{\infty} \frac{\dd{(\im\alpha)}}{\sqrt{2}} ~\Psi_{\text{ZZ}}(\alpha) ~ \block{F}{\alpha}{\alpha_2&\alpha_2\\\alpha_1&\alpha_1}{q}.
\eeq
The 4d $\Nsusy=2$ theory on~$\HS^4_b$ is $SU(2)$ $N_f=4$ SQCD where hypermultiplets have pairwise equal masses. Then we can use an identity found in~\cite{ZZ} between Virasoro conformal blocks (where $\alpha=0$ denotes the identity)\footnote{These authors also have a similar identity for the degenerate brane $(2,1)$ instead of the identity brane, which indicates that Wilson lines map to 't~Hooft lines, as expected under S-duality.}
\begin{equation}\label{ZZsduality}
  \begin{aligned}
    Z_{\HS^4_b}^{\text{Neumann}}(q)&=\int\frac{\dd{(\im\alpha)}}{\sqrt{2}} \Psi_{\text{ZZ}}(\alpha)\block{F}{\alpha}{\alpha_2&2Q-\alpha_2\\\alpha_1&2Q-\alpha_1}{q} \\
    &=\block{F}{\alpha=0}{2Q-\alpha_1&2Q-\alpha_2\\\alpha_1&\alpha_2}{1-q}= Z_{\HS^4_b}^{\text{Dirichlet}}(1-q).
  \end{aligned}
\end{equation}
In terms of the theories living on a squashed hemisphere this identity relates by S-duality $q\to1-q$ a gauge theory with Neumann boundary conditions with the same gauge theory but with Dirichlet boundary conditions. See \autoref{figdisk2ptchannels} on \autopageref{figdisk2ptchannels} for a CFT depiction of these two channels.

The vacuum block is known explicitly in the semiclassical limit~\cite{Fitzpatrick:2015zha}, and we can deduce from the CFT result~\eqref{ZZsduality} an explicit expression of the partition function on $\HS^4_b$ with Neumann boundary condition in that limit. In gauge theory that limit is the Nekrasov-Shatashvili limit in which the Omega deformation parameter $\epsilon_2\to 0$ with $\epsilon_1$~fixed since $c\sim 6 \epsilon_1/\epsilon_2\to\infty$ in the CFT\@.  The analysis in~\cite{Fitzpatrick:2015zha} apply when $\alpha_1$~is heavy, meaning that $\Delta_1/c$ is finite, and $\alpha_2$ light, meaning that $\Delta_2\sim \mathcal{O}(1)$ in the $c\to\infty$ limit.  This corresponds to taking all hypermultiplet masses to infinity, keeping their differences finite.  It would be interesting to understand purely from gauge theory why the simplification coming from combining~\eqref{ZZsduality} with the results of~\cite{Fitzpatrick:2015zha} occurs in this limit.\footnote{Some aspects of Liouville theory has been recently related to quantum chaos in 2d \cite{Jackson:2014nla, Turiaci:2016cvo,Yamazaki:2016aam, Mertens:2017mtv}. One might wonder whether the behavior of the out-of-time-ordered four-point function has a gauge theory interpretation.}

An exhaustive classification of $\Nsusy=2$ boundary conditions and how S-duality acts on them is not the purpose of this paper and therefore we leave it for future work. Nevertheless we see that the relation with Liouville theory with boundaries might be helpful as a guiding principle.

\section{Projective space and cross-caps}\label{sec:rp4}

We outline in \autoref{sec:rp4calc} a method to localize 4d $\Nsusy=2$ gauge theories on~$\RP^4_b$ by splitting that space into a hemisphere~$\HS^4_b$ and a lens space ${\rm S}^3_b/\ZZ_2$.  We confirm our procedure in \autoref{app:rp2} by reproducing known $\RP^2$~localization results from an analogous splitting.  For completeness we hazard a conjecture in \autoref{app:rp4hyper} for the hypermultiplet one-loop determinant.  These parts can be read independently from the rest of the paper.
In \autoref{sec:rp4agt} we discuss the AGT correspondence for a CFT two-point function with a cross-cap; this confirms $\RP^4_b$~results other than the hypermultiplet one-loop determinant.

All of the spaces are of unit size and squashed in a $U(1)\times U(1)$ invariant way: ${\rm S}^3_b/\ZZ_2 = \{x\in\RP^4_b\mid x_0=0\}$ where
\begin{equation}
  \RP^4_b = \bigl\{x_0^2+b^{-2}(x_1^2+x_2^2)+b^2(x_3^2+x_4^2)=1\bigr\}/(x\sim-x) .
\end{equation}
Of course, ${\rm S}^3_b/\ZZ_2$ is neither a boundary nor a singularity of~$\RP^4_b$ and we are simply cutting along ${\rm S}^3_b/\ZZ_2$ to reuse known localization results.  In this section we denote by $\mu,\nu,\dots$ the 4d indices and $i,j,\dots$ the 3d indices.

The 4d $\Nsusy=2$ theory we consider has gauge group~$G$, and hypermultiplets in a representation~$\repr$ (more generally, half-hypermultiplets).  Unsurprisingly its partition function reads
\begin{equation}\label{ZRP4}
  Z_{\RP^4_b}(q) = \frac{1}{\abs{\Weyl}} \sum_{y\in T,y^2=1} \int_{\lie{t}} \dd{\sigma}
  Z_{\RP^4_b,\eqref{RP4-vector}}^{\text{vector}}
  Z_{\RP^4_b,\eqref{RP4-hyper}}^{\text{hypermultiplet}}
  q^{\frac{1}{2}\Tr \sigma^2} Z_{\text{instanton}}(q) ,
\end{equation}
with a sum over the holonomy~$y$ along the nontrivial element of $\pi_1(\RP^4_b)=\ZZ_2$, with $y$~in the Cartan torus~$T$, and an integral over the Cartan algebra~$\lie{t}$ whose measure is defined in~\eqref{intdalpha}.  The classical action is half of the one on the sphere, and there is only a single instanton partition function, normalized as $Z_{\text{inst}}(q) = 1 + ({\dots})q+\dots$\footnote{Note that this differs from the normalization we chose for the conformal blocks in previous sections.}

Contrarily to ${\rm S}^4_b$, to $\HS^4_b$, or to flat space, $\int\Tr(F\wedge F)$ vanishes on~$\RP^4_b$ hence there can be no continuous theta term; the instanton counting parameter $q=\exp(-8\pi^2/g^2)$ for each simple group factor is thus real.\footnote{The discrete theta angle $\vartheta=\pi$ flips the sign of~$q$, which on the CFT side exchanges boundary and cross-cap Ishibashi states.  As briefly explained in the introduction this corresponds to different quotients of the 6d setup.}  This is a manifestation of the fact that only parity-invariant theories can be put on nonorientable surfaces, unless one allows nonconstant couplings or domain walls.  One can also turn on an FI parameter for each $U(1)$ factor in~$G$ but they play no role in our paper so we do not include such terms.

Our approach is not sensitive to phases that only depend on the holonomy~$y$, and we add such phases ``by hand'' when matching with Toda CFT\@.  In that context we also need to couple the $\RP^4_b$~fields to hypermultiplets on~${\rm S}^4_b$.  These can also be described as a pair of hypermultiplets on~$\RP^4_b$ with a non-trivial flavor holonomy $y_F=(\begin{smallmatrix}0&1\\1&0\end{smallmatrix})$, namely a periodicity that exchanges the two hypermultiplets when going along an antipodal loop.

\subsection{Partition function on projective space}\label{sec:rp4calc}

The locus ${\rm S}^3_b/\ZZ_2$ is preserved by a 3d $\Nsusy=2$ subalgebra of the 4d $\Nsusy=2$ supersymmetry algebra so the restrictions to ${\rm S}^3_b/\ZZ_2$ of 4d fields and (all of) their normal derivatives decompose into (an infinite set of) multiplets of this subalgebra.  Most importantly the 4d $\Nsusy=2$ vector multiplet with bosonic components $(A_\mu,\sigma+i\upsilon)$ includes (see for instance~\cite{Erdmenger:2002ex,Drukker:2010jp,Gaiotto:2014gha}) a 3d $\Nsusy=2$ vector multiplet whose bosonic components are $A_i$ and a real (i.e.\@ Hermitian) scalar $\sigma$; a 3d $\Nsusy=2$ adjoint chiral multiplet whose bosonic components are the real scalars $A_4$ and~$\upsilon$; infinitely more multiplets involving normal derivatives of 4d fields.  This decomposition differs from that of a 3d $\Nsusy=4$ vector multiplet into vector and chiral multiplets in the way fermions are arranged.

Let $\supercharge$ be the $\RP^4_b$~projection of the supercharge used on~${\rm S}^4_b$ by Pestun~\cite{Pestun:2007rz}.  Its restriction to ${\rm S}^3_b/\ZZ_2$ is the supercharge used in localization calculations~\cite{Gang:2009wy,Benini:2011nc,Alday:2012au,Imamura:2012rq,Imamura:2013qxa,Nieri:2015yia}.  Add to the 4d action the ($\supercharge$-exact and $\supercharge$-closed) localization term used in the latter works, supported on~${\rm S}^3_b/\ZZ_2$.  It localizes the 3d $\Nsusy=2$ vector multiplet to
\begin{equation}
  F_{ij}=0, \qquad D_i\sigma=0
\end{equation}
and fixes the auxiliary scalar of the 3d $\Nsusy=2$ vector multiplet.  It also localizes all 3d $\Nsusy=2$ chiral multiplets to zero.

Flat connections $F_{ij}=0$ are classified by $\Hom\bigl(\pi_1({\rm S}^3_b/\ZZ_2),G\bigr)=\{y\in G\mid y^2=1\}$ modulo conjugation by~$G$.  Explicitly, $y$~is (the conjugacy class of) the Wilson loop $y=\Pexp\int iA$ along the $\ZZ_2$~projection of a path joining antipodes on~${\rm S}^3_b$.  Then $D_i\sigma=0$ fixes $\sigma$ in terms of its value at one point~$x$ by parallel transport, and periodicity on~$\RP^4_b$ implies $y\sigma(x)y^{-1}=\sigma(x)$.  Thus $y$ can be taken in the Cartan torus~$T$, and the gauge field as well as $\sigma(x)$ in the Cartan Lie algebra~$\lie{t}$.  Then $\sigma$~is constant throughout ${\rm S}^3_b/\ZZ_2$.  Of course there remains an ambiguity by simultaneous action of the Weyl group~$\Weyl$ on $\sigma$ and~$y$.

We are then left with a path integral over the 4d $\Nsusy=2$ fields in the bulk~$\HS^4_b$, with boundary conditions given by the 3d BPS configurations.  Since from the $\HS^4_b$ point of view only the Wilson line $y^2=1$ is gauge-invariant, the holonomy is gauge-equivalent to a trivial holonomy and does not affect the $\HS^4_b$~path integral.  We recognize the Dirichlet boundary conditions given in equation (5.7) of \cite[v2]{HS4}: $F_{ij}=0$ in directions parallel to the boundary, and constant $\sigma$ in the Cartan algebra~$\lie{t}$.  The bulk path integral then localizes to constant $\sigma$ and to $F_{\mu\nu}=0$ except for point-like instantons at the pole $x=(1,0,0,0,0)$.

We reached the form
\begin{equation}\label{ZRP4prelim}
  Z_{\RP^4_b}(q) = \frac{1}{\abs{\Weyl}} \int_{\lie{t}} \dd{\sigma} \Biggl( \sum_{y\in T,y^2=1} Z_{{\rm S}^3_b/\ZZ_2}^{\text{one-loop}}(\sigma,y) \Biggr) Z_{\HS^4_b}^{\text{Dirichlet}}(\sigma,q)
\end{equation}
in terms of one-loop determinants on ${\rm S}^3_b/\ZZ_2$ and of the $\HS^4_b$~partition function, where we suppress mass parameters in the notation.  Amusingly, by treating the 4d path integral as a function of its boundary fields we can think of it as a complicated observable in the 3d theory; we are then simply localizing the 3d theory in the presence of a $\supercharge$-invariant observable.
We have not worked out from first principles which one-loop determinants to include: instead we rely on consistency checks with~${\rm S}^4_b$ and on an analogy with the $\RP^2$~case treated in \autoref{app:rp2}.

Consider first the vector multiplet.  We notice that one-loop determinants obey
\begin{equation}\label{S4bgluingvector}
  Z_{{\rm S}^4_b}^{\text{vector}}(\sigma)
  = \prod_{e>0} \Bigl(\Upsilon(\lb e,\sigma\rb) \Upsilon(-\lb e,\sigma\rb)\Bigr)
  = \Bigl( Z_{\HS^4_b,\text{Dir}}^{\text{vector}}(\sigma) \Bigr)^2 Z_{{\rm S}^3_b}^{\text{vector}}(\sigma)
\end{equation}
with $Z_{{\rm S}^3_b}^{\text{vector}}(\sigma) = \prod_e 1/S_b(\lb e,\sigma\rb) = \prod_{e>0} -4\sin(\pi b\lb e,\sigma\rb)\sin(\pi\lb e,\sigma\rb/b)$ and determinants on $\HS^4_b$ with Dirichlet boundary are given in~\eqref{oneloopHS4D}.  The $\RP^4_b$ one-loop determinant of a 4d $\Nsusy=2$ vector multiplet is thus its $\HS^4_b$~determinant times a 3d $\Nsusy=2$ vector multiplet determinant on~${\rm S}^3_b/\ZZ_2$.  The latter is easiest to write in the important case of a (compact) connected gauge group: the exponential map is then surjective so up to conjugation the holonomy is $y=\exp(\pi i\holonomy)$ for some $\holonomy\in\Lambda/(2\Lambda)$, where $\Lambda=\{\mathfrak{w}\in\lie{t}\mid\exp(2\pi i\mathfrak{w})=1\}$ is the coweight lattice of~$G$.  Then
\begin{equation}\label{RP4-vector}
  Z_{\RP^4_b}^{\text{vector}} = \frac{\prod_{e>0} \prod_{\pm} 2\sinh\bigl((\pi/2)\langle e,b^{\pm 1}\sigma\pm i\holonomy\rangle\bigr)}{\prod_{e\in\roots(G)} \Gamma_b(b+1/b-i\langle e,\sigma\rangle)} .
\end{equation}
Expanding the $\sinh$ in terms of exponentials and expanding the product over $\pm$ gives a formula in terms of $y^e=\exp(\pi i\langle e,\holonomy\rangle)$, valid also for disconnected~$G$.
Changing $\holonomy\mapsto-\holonomy$ does not affect $y^e$ hence leaves the result invariant.
The result is also manifestly invariant under $(\sigma,\holonomy)\mapsto(-\sigma,-\holonomy)$.

An interesting possibility is to turn on a nontrivial flavor symmetry holonomy~$y_F$, namely a holonomy for a background vector multiplet.  This replaces $y$ by~$yy_F$ in all formulas.
In the description of $\RP^4_b$~as a quotient ${\rm S}^4_b/\ZZ_2$, these holonomies are equivalent to a nontrivial identification of hypermultiplets at antipodal points, namely instead of $\Phi(-x)=\Phi(x)$ one imposes $\Phi(-x)=y\cdot y_F\cdot\Phi(x)$.  (Of course, if $y_F^2\neq 1$ this requires summing over gauge holonomies obeying $y^2y_F^2=1$ instead of $y^2=1$.)

For example, a hypermultiplet on~${\rm S}^4_b$ is equivalent to a pair of hypermultiplets with a flavor holonomy $y_F=(\begin{smallmatrix}0&1\\1&0\end{smallmatrix})$.  This holonomy is itself conjugate to $\diag(1,-1)$, so we learn that the product of hypermultiplet one-loop determinants on $\RP^4_b$~with opposite holonomies must be equal to the known determinant on~${\rm S}^4_b$:
\begin{equation}\label{S4-as-RP4}
  \prod_{y=\pm} Z_{\RP^4_b}^{\text{hypermultiplet}}(\sigma,y) = \prod_{w\in\weights} \frac{1}{\Upsilon_b(\tfrac{1}{2}(b+1/b)+\lb w,i\sigma\rb)}
\end{equation}
where $w$~are weights under the gauge and flavor symmetry groups, and $a$~combines the vector multiplet scalar and background vector multiplet scalars (masses).  The conjecture that we make for completeness in \autoref{app:rp4hyper} obeys this property.  Our main results do not rely on this conjecture since they use directly the ellipsoid result for the CFT comparison.

\subsection{Cross-cap two-point function}\label{sec:rp4agt}

We now write explicitly the AGT relation between cross-cap states and $\RP^4_b$~vector multiplets that stems from the quotient $({\rm S}^4_b\times\Sigma)/\ZZ_2$ where the $\ZZ_2$ acts on ${\rm S}^4_b$ by the antipodal map and on the Riemann surface by an involution without fixed point.

Even though a cross-cap does not introduce a boundary, it is useful to study it using boundary CFT techniques. For the case of a single cross-cap it is equivalent to putting the CFT on $\RP^2$, for two cross-caps the Klein bottle, etc. The main difference is how the corresponding Ishibashi state is defined, which now is
\beq
( W_n^{(s)} - (-1)^{s+n} \bar{W}_{-n}^{(s)}) \bket{\alpha}_{\otimes} = 0
\eeq
for $\alpha=Q+a$ non-degenerate.  Besides this $(-1)^n$ difference, the cross-cap state can be expanded in a similar way to boundary states.  We worked out the generalization of Liouville results for ADE Toda CFT in \autoref{app:wavefunctions}: the wavefunction we obtain for the cross-cap state is~\eqref{psicrosscapexplicit}
\begin{equation}\label{psicrosscapexplicitmaintext}
  \begin{aligned}
    \Psi_{\otimes}(\alpha) &= (\hat{\mu}b^{2b-2/b})^{-\lb \rho, a\rb} \prod_{e>0} \frac{\Gamma(1+b\lb e,a\rb)\Gamma(1+b^{-1}\lb e,a\rb)}{ -2\pi \lb e,a\rb} \\
    &\quad \times \sum_{w,w'\in \Weyl} (-1)^{\lb\rho,\rho\rb-\lb w(\rho),w'(\rho)\rb}\epsilon(w) e^{ \pi i b \lb w(\rho), a\rb} \epsilon(w')e^{\pi i \lb w'(\rho), a\rb/b}.
  \end{aligned}
\end{equation}
The two-point function on~$\RP^2$ is given by~\eqref{eq:bulk2pt} with the wavefunction replaced by~\eqref{psicrosscapexplicitmaintext}.

Just as in \autoref{sec:matchZZ}, we focus on an $A_{N-1}$~Toda CFT two-point function with momenta $\alpha_1$ and $\kappa\omega_1$, but on~$\RP^2$ instead of the plane with a hole. The boundary state formalism gives
\beq\label{crosscap2pt}
\lb V_{\alpha_1}(z_1) V_{\kappa\omega_1}(z_2)\rb_{\RP^2} = \frac{1}{|\Weyl|} \int \dd{(\im\alpha)} \Psi_{\otimes}(\alpha) \mathcal{F}_\alpha (q), 
\eeq
where $\Psi_{\otimes}(\alpha)$ is~\eqref{psicrosscapexplicitmaintext}, while $\mathcal{F}_a(q)$ is the conformal block normalized in a way that includes a single OPE coefficient, and $q$ is the real cross-ratio corresponding to the cross-cap identification.

Each piece in~\eqref{crosscap2pt} translates readily to gauge theory.  Integrals coincide through $\alpha=Q+i\sigma$.  The conformal block is an instanton partition function (plus a classical piece) due to instantons at the single pole of~$\RP^4_b$.  The instanton counting parameter~$q$ is real.  Thanks to the usual AGT correspondence we know that the OPE coefficient normalizing the conformal block is equal to the one-loop determinant~\eqref{S4-as-RP4} of a (bifundamental) hypermultiplet on~${\rm S}^4_b$.  Such a hypermultiplet can also be described by a pair of hypermultiplets on~$\RP^4_b$ with opposite flavor holonomy if one wants a purely~$\RP^4_b$ theory.

The only nontrivial step is to reproduce the cross-cap wavefunction by summing the $\RP^4$~vector multiplet determinant over holonomies: we need to introduce phases depending on the holonomies.  For simplicity we do the case of $G=U(N)$ rather than $SU(N)$.  Holonomies are labeled by $\holonomy\in\{0,1\}^N$ through $y=\exp(\pi i\diag(\holonomy))$.
The numerator of the vector multiplet determinant~\eqref{RP4-vector} reads (using the $\sigma\to-\sigma$ symmetry)
\begin{equation}
  \prod_{e>0} \prod_{\pm} 2\sinh\bigl((\pi/2)\langle e,-b^{\pm 1}\sigma\pm i\holonomy\rangle\bigr)
  =
  \prod_{\pm} \sum_{w\in\Weyl} \epsilon(w) e^{\pi\langle w(\rho),(iab^{\pm 1}\pm i\holonomy)\rangle}
\end{equation}
where we used the Weyl character formula.  We sum it with extra phases $\frac{\pi}{2} \lb \holonomy,\holonomy\rb$ using
\begin{equation}
  \begin{aligned}
    \sum_{\holonomy} e^{\frac{i \pi}{2}\lb \holonomy,\holonomy\rb}e^{i \pi \lb \holonomy,w(\rho) - w'(\rho)\rb } & = (1+i)^N e^{\frac{i \pi}{2} \lb w(\rho)-w'(\rho), w(\rho)-w'(\rho)\rb} \\
    & = (1+i)^N (-1)^{\lb\rho,\rho\rb-\lb w(\rho),w'(\rho)\rb} .
  \end{aligned}
\end{equation}
In the $SU(N)$ case it would be natural to also impose the value of $\sum_i\holonomy_i\mod 2$; the above expression is then replaced by its real or imaginary part, which differ by a constant factor that may vanish.
In any case, we get precisely the signs in the cross-cap wavefunction~\eqref{psicrosscapexplicitmaintext}.  The rest of the wavefunction, multiplied by $\hat{\mu}^{\langle\rho,i\sigma\rangle}\prod_{e>0}\bigl(-\Upsilon_b(\langle e,i\sigma\rangle)\bigr)$ as in~\eqref{vectormatchesZZ}, gives the rest of the one-loop determinant.

Combining these ingredients, we get the cross-cap analogue of~\eqref{HS4SQCDeq2ptdisk}: the two-point function on~$\RP^2$ is the partition function of 4d $\Nsusy=2$ $SU(N)$ SQCD with $2N$ hypermultiplets with pairwise-equal masses:
\begin{equation}
  Z_{\RP^4_b}^{\text{SQCD},N_f=2N}\bigl(m_1,m_1,\dots,m_N,m_N; q)
  = \lb V_{\mu}(z_1) V_{\kappa\omega_1}(z_2)\rb_{\RP^2} .
\end{equation}

\section{Generalizations}\label{sec:generalizations}

We describe here several generalizations, discussing Riemann surfaces~$\Sigma$ with an arbitrary collection of boundaries and cross-caps, and the addition of various observables.

As in the AGT correspondence without boundary the 4d $\Nsusy=2$ theory $\GaiottoTheory{\Sigma}{G}$ corresponding to~$\Sigma$ admits one description for each cusp in the moduli space of Riemann surfaces, in other words for each pants decomposition\footnote{A pants decomposition is a collection of curves in~$\Sigma$ whose complement consists of spheres with three punctures.} of~$\Sigma$.  The various descriptions are S-dual.  Recall the standard dictionary:
\begin{equation}\label{picturedictionary}
  \vcenter{\hbox{\begin{tikzpicture}[scale=.6]
        \draw (1,.3) arc (270:210:1.43205);
        \draw[rotate=120] (1,.3) arc (270:210:1.43205);
        \draw[rotate=-120] (1,.3) arc (270:210:1.43205);
        \draw (1,0) circle (.15 and .3);
        \draw[rotate=120] (1,0) circle (.15 and .3);
        \draw[rotate=-120] (1,0) circle (.15 and .3);
      \end{tikzpicture}}}
  = \text{matter} , \qquad
  \vcenter{\hbox{\begin{tikzpicture}[scale=.6]
        \draw (1,0) circle (.15 and .3);
        \draw (-1,0) circle (.15 and .3);
        \draw (-1,.3) -- (1,.3);
        \draw (-1,-.3) -- (1,-.3);
      \end{tikzpicture}}}
  = \text{vector multiplet} .
\end{equation}
Namely, each three-punctured sphere (also called pair of pants) corresponds to a given matter theory (an isolated CFT, for instance free hypermultiplets) where each puncture describes a flavor symmetry group, and connecting two punctures together via a tube corresponds to gauging the two flavor symmetries diagonally using a vector multiplet.  Internal momenta in the Toda CFT give Coulomb branch parameters of these vector multiplets, while external momenta give mass parameters, obtained by turning on non-zero background vector multiplet scalars for the remaining flavor symmetries.

\begin{figure}\centering\capstart
\begin{tikzpicture}[scale=1.1,rotate=0]
\draw (0.35,-1.2) node {\footnotesize $\ket{\B}$};
\draw[ultra thick] (-2,0) -- (2,0);
\draw[ultra thick] (-1,0) -- (-1,1);
\draw[ultra thick] (1,0) -- (1,1);
\draw[thick,dashed] (0,-1.3) -- (0,2);
\draw (-0.3,-0.3) node {$\alpha_s$};
\draw (-2,-0.3) node {$\alpha_1$};
\draw (-1,1.3) node {$\alpha_2$};
\draw (2,-0.3) node {$\bar{\alpha}_1$};
\draw (1,1.3) node {$\bar{\alpha}_2$};
\draw[white] (0,-1.5) circle (0.1);
\end{tikzpicture}
\hspace{1cm}
\begin{tikzpicture}[scale=1.1,rotate=0]
\draw (0.35,-1.2) node {\footnotesize $\ket{\B}$};
\draw[ultra thick] (-1,-0.5) -- (1,-0.5);
\draw[ultra thick] (0,-0.5) -- (0,1);
\draw[ultra thick] (-1,1) -- (1,1);
\draw[thick,dashed] (0,-1.3) -- (0,2);
\draw (-0.3,0.2) node {$\alpha_t$};
\draw (-1.2,-0.75) node {$\alpha_1$};
\draw (-1,1.3) node {$\alpha_2$};
\draw (1.2,-0.75) node {$\bar{\alpha}_1$};
\draw (1,1.3) node {$\bar{\alpha}_2$};
\draw[white] (0,-1.5) circle (0.1);
\end{tikzpicture}
  \caption{Disk two-point function in two channels.  Left: s-channel, with a cut along the disk's boundary.  Right: t-channel, with a cut intersecting the disk's boundary.}
  \label{figdisk2ptchannels}
\end{figure}
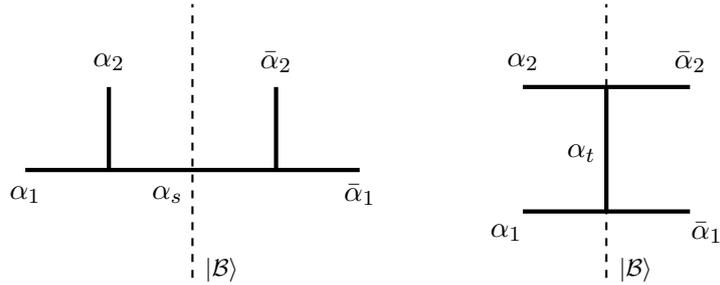

Pants decompositions of~$\Sigma$ are simply $\ZZ_2$-invariant decompositions of its Schottky double~$\widehat{\Sigma}$.  In \autoref{sec:quivers} we consider decompositions with cuts along all of the boundaries of~$\Sigma$.  In \autoref{sec:boundaryoperators} we discuss other decompositions, which involve bulk-to-boundary OPEs (see \autoref{figdisk2ptchannels}) hence boundary-changing operators.

\subsection{Quivers}\label{sec:quivers}

When there are only ZZ branes and cross-caps, the easiest way to obtain the theory corresponding to~$\Sigma$ is to perform a $\ZZ_2$ identification of fields in the theory $\GaiottoTheory{\widehat{\Sigma}}{G}$.  Three-punctured spheres in the decomposition of~$\widehat{\Sigma}$ are paired up by the $\ZZ_2$~identification since none can be mapped to itself; the corresponding matter theories in $\GaiottoTheory{\widehat{\Sigma}}{G}$ are thus identified to obtain $\GaiottoTheory{\Sigma}{G}$.  Similarly, cuts lying entirely in~$\Sigma$ are paired up with those lying entirely outside~$\Sigma$, and the corresponding vector multiplets are identified to get $\GaiottoTheory{\Sigma}{G}$.  It is then natural that the vector multiplet corresponding to a given cross-cap or ZZ brane boundary of~$\Sigma$ is identified with itself (at a different point) under~$\ZZ_2$.

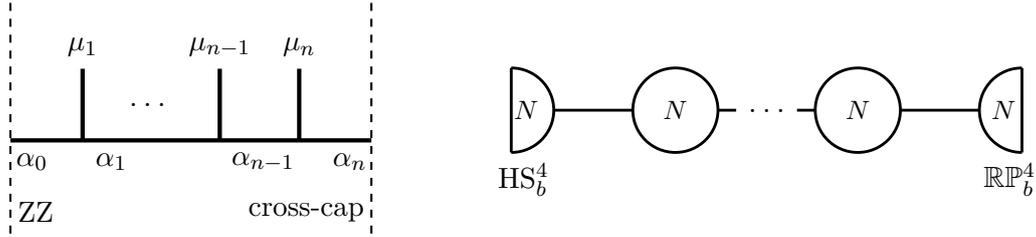
\begin{figure}\centering\capstart
  \begin{tikzpicture}
    \begin{scope}[scale=.95]
      \draw[thick,dashed] (-5,-1.3) -- (-5,2);
      \draw[ultra thick] (-5,0) -- (0,0);
      \draw[ultra thick] (-1,0) -- (-1,1);
      \draw[ultra thick] (-2.1,0) -- (-2.1,1);
      \draw[ultra thick] (-4,0) -- (-4,1);
      \draw (-3.1,0.5) node {\large $\ldots$};
      \draw[thick,dashed] (0,-1.3) -- (0,2);
      \draw (-4.65,-1) node {ZZ};
      \draw (-4.7,-0.3) node {$\alpha_0$};
      \draw (-3.6,-0.3) node {$\alpha_1$};
      \draw (-1.5,-0.3) node {$\alpha_{n-1}$};
      \draw (-0.3,-0.3) node {$\alpha_n$};
      \draw (-.9,-1) node {cross-cap};
      \draw (-1,1.3) node {$\mu_n$};
      \draw (-2.1,1.3) node {$\mu_{n-1}$};
      \draw (-4,1.3) node {$\mu_1$};
    \end{scope}
    \begin{scope}[shift={(4,.4)},scale=0.4]
      \draw (-5,-2.3) node {$\HS^4_b$};
      \draw[line width=1.07pt] (-5.4,-1.4) arc (-90:90:1.4);
      \draw[line width=1.07pt] (-5.4,-1.4) -- (-5.4,1.4);
      \draw (-4.9,0) node {\small $N$};
      \draw[line width=1.07pt] (-1.4,0) -- (-4,0); 
      \draw[line width=1.07pt] (0,0) circle (1.4);
      \draw (0,0) node {\small $N$};
      \draw[line width=1.07pt] (1.4,0) -- (2,0);
      \draw (3,0) node {\large $\ldots$};
      \draw[line width=1.07pt] (4,0) -- (4.6,0);
      \draw[line width=1.07pt] (6,0) circle (1.4);
      \draw (6,0) node {\small $N$};
      \draw[line width=1.07pt] (7.4,0) -- (10,0);
      \draw[line width=1.07pt] (11.4,1.4) arc (90:270:1.4);
      \draw[line width=1.07pt] (11.4,-1.4) -- (11.4,1.4); 
      \draw (10.8,0) node {\small $N$};
      \draw (11,-2.3) node {$\RP^4_b$};
    \end{scope}
  \end{tikzpicture}
  \caption{\small Left: Conformal block of the $\lie{g}=\lie{su}(N)$ W-algebra for the $n$-point function on a M\"obius strip, with semi-degenerate external momenta $\mu_i=\kappa_i\omega_1$ (this keeps fusion multiplicities finite).  Right: Corresponding quiver, built from $SU(N)$ vector multiplets and bifundamental hypermultiplets defined on~${\rm S}^4_b$.  The leftmost and rightmost vector multiplets are subject to $\ZZ_2$~identifications turning them effectively into fields on $\HS^4_b$ (with Neumann boundary conditions) and $\RP^4_b$ respectively.  The momenta $\alpha_i$ are Coulomb branch parameters of the corresponding gauge groups, and the external momenta~$\mu_i$ encode mass parameters for the matter theories.}
  \label{fignptmobius}
\end{figure}

Altogether, the dictionary for cuts and three-punctured spheres in the bulk of~$\Sigma$ is the usual one~\eqref{picturedictionary}. Each ZZ brane boundary of~$\Sigma$ corresponds to a vector multiplet on~${\rm S}^4_b$ constrained by a $\ZZ_2$~identification~\eqref{hs4identificationofvectormultiplet} at opposite points around the equator, in other words a vector multiplet on~$\HS^4_b$ gauging a flavor symmetry of a matter theory on~${\rm S}^4_b$. A cross-cap corresponds to a vector multiplet on~${\rm S}^4_b$ constrained by a $\ZZ_2$~identification~\eqref{rp4identificationofvectormultiplet} at antipodal points, which makes it equivalent to a vector multiplet on~$\RP^4_b$.  We use this description in \autoref{fignptmobius} when drawing the quiver corresponding to an $n$-point function on a M\"obius strip, a case where $\Sigma$~has both a ZZ brane and a cross-cap.

Any field on ${\rm S}^4_b$ can be described as a pair of fields on $\HS^4_b$ identified to each other at the equator.  In this way, the theory $\GaiottoTheory{\Sigma}{G}$ associated to a Riemann surface with boundaries but no cross-cap can be described purely as the theory $\GaiottoTheory{\widehat{\Sigma}}{G}$ on~$\HS^4_b$ with an appropriate boundary condition.  When there are cross-caps, it is also possible to describe the $\RP^4_b$~vector multiplets in terms of vector multiplets on $\HS^4_b$ with a boundary condition relating the vector multiplet at opposite points of the equator, but that is nonlocal.

The 6d description is simplest when there are only ZZ branes.  Let $x$~be the embedding coordinates of ${\rm S}^4_b\subset\RR^5$ with poles at $x_0=\pm 1$, and let $\omega:\widehat{\Sigma}\to\widehat{\Sigma}$ be the involution that defines $\Sigma=\widehat{\Sigma}/\ZZ_2$.  The $\ZZ_2$~involution $(z,x_0,x_1,\ldots,x_4)\to(\omega(z),-x_0,x_1,\ldots,x_4)$ of $\widehat{\Sigma}\times {\rm S}^4_b$ preserves orientation and the required supersymmetry (we discuss other quotients in \autoref{app:quotients}).  The three descriptions we gave above of the 4d theory correspond respectively to
\begin{itemize}
\item reducing the 6d theory on $\widehat{\Sigma}\times {\rm S}^4_b$ then performing the $\ZZ_2$ identification;
\item writing $(\widehat{\Sigma}\times {\rm S}^4_b)/\ZZ_2$ as $(\Sigma\times {\rm S}^4_b)/\sim$ with a certain identification ``$\sim$'' at the boundary;
\item writing $(\widehat{\Sigma}\times {\rm S}^4_b)/\ZZ_2$ as $(\widehat{\Sigma}\times \HS^4_b)/\sim$ with a certain identification ``$\sim$'' at the boundary.
\end{itemize}
When there are cross-caps the appropriate $\ZZ_2$~action to consider is locally $(z,x)\to(-1/\bar{z},-x)$; when there are both cross-caps and boundaries the $\ZZ_2$ action can only be defined globally if the 6d manifold is modified to be an ${\rm S}^4_b$ fibered over $\widehat{\Sigma}$.

Recall that the 6d $\Nsusy=(2,0)$ theory has codimension~$2$ defects (M5~brane intersections) and codimension~$4$ defects (M2~brane endings).  The orbifold locus in $({\rm S}^4_b\times\widehat{\Sigma})/\ZZ_2$ is ${\rm S}^3_b\times{\rm S}^1$ for each boundary of~$\Sigma$; codimension~$2$ defects can be added there, as well as codimension~$4$ defects along the $(x^1,x^2)$ and $(x^3,x^4)$ circles.  On the CFT side this acts with a Verlinde loop on the ZZ brane, constructing all branes obeying Cardy's conditions (see \autoref{sec:otherbranes}).  On the gauge theory side, codimension~$2$ defects insert symmetry-breaking walls and codimension~$4$ defects insert Wilson lines~\cite{Drukker:2010jp}.

\subsection{Verlinde loops}

In the AGT correspondence, a Verlinde loop or web\footnote{As shown in~\cite{Drukker:2010jp} Verlinde loops are equivalent to topological defects.  Both are labeled by a representation of the chiral algebra of the CFT\@.  Verlinde webs are collections of lines connected by trivalent junctions, where each line is labeled by a representation and each junction by an invariant in the tensor product of the three representations associated to lines ending there.} only acts on the holomorphic or on the antiholomorphic conformal block.  Likewise, it is natural in our setting to consider a single Verlinde loop or web on the Schottky double $\widehat{\Sigma}$ of the Riemann surface $\Sigma$ with boundaries.

As discussed in \autoref{sec:otherbranes}, Verlinde loops along boundaries of~$\Sigma$ change ZZ branes to other branes.\footnote{More generally, acting with a Verlinde loop on a brane produces a superposition of branes according to the OPE of the corresponding vertex operators.}  On the other hand if the loop lies in the bulk of~$\Sigma$, its gauge theory interpretation as a Wilson--'t~Hooft operator or as a symmetry-breaking wall is identical to the setting without boundary already treated in~\cite{Drukker:2010jp}.

The interesting case is when the Verlinde loop intersects the boundary of~$\Sigma$.  Consider for simplicity the 4d $\Nsusy=2$ $SU(N)$ gauge theory with $N_f=2N$ hypermultiplets of pairwise equal masses.  We reproduced its Neumann $\HS^4_b$~partition function in~\eqref{disk2pt} as a disk two-point function.  Inserting in the (Schottky double) chiral correlator a Verlinde loop with degenerate momentum is known~\cite{Gomis:2010kv,Gomis:2011pf} to correspond in gauge theory to inserting a pair of supersymmetric 't~Hooft loops along great circles~\eqref{greatcircles} of the equator.  Each of the 't~Hooft loops is labeled by a representation of the gauge group ($G$~is simply-laced).  Such operators create quantized magnetic flux by imposing the behaviour
\begin{equation}
  F = \frac{B}{4} \frac{\epsilon_{ijk} x^i\dd{x}^j\wedge\dd{x}^k}{\abs{x}^3} + \cdots
\end{equation}
in directions $i,j,k$ transverse to the support.  This singular behaviour near a loop on~${\rm S}^4_b$ is compatible with the $\ZZ_2$~identification~\eqref{hs4identificationofvectormultiplet}.  From the squashed hemisphere point of view, the loop alters the Neumann boundary condition.

Note that one can also define a supersymmetric 't~Hooft loop operator supported at fixed latitude inside the bulk of~$\HS^4_b$, and in $\supercharge$-cohomology it is independent of the latitude.  However, the limit of such a bulk loop is not the boundary loop discussed here.  This is most easily seen on ${\rm S}^4_b$ before $\ZZ_2$~identification: then one has two 't~Hooft loops at opposite latitudes and bringing them to the equator requires an OPE.  The boundary 't~Hooft loop is very analogous to a fractional brane stuck at an orbifold point.

It would be good to clarify the interplay between boundary-changing operators of \autoref{sec:boundaryoperators} and Verlinde loops straddling $\Sigma$ and its double.  Another operator we did not investigate is a Verlinde line along the non-trivial cycle on~$\RP^2$, or more generally connecting a point in~$\Sigma$ to its $\ZZ_2$~image by passing through a cross-cap.  For a degenerate momentum this could plausibly correspond to Wilson--'t~Hooft lines joining antipodal points in~$\RP^4_b$, and for a non-degenerate one it could be a domain wall supported on a lens space ${\rm S}^3_b/\ZZ_2\subset\RP^4_b$.

\subsection{Surface operators}

As in the usual AGT correspondence our story can be enriched by adding half-BPS surface operators in the 4d $\Nsusy=2$ gauge theory~\cite{Alday:2009fs} (see the review~\cite{Gukov:2014gja} and references therein).  They are inserted along the two squashed~${\rm S}^2$ located at $x^3=x^4=0$ and at $x^1=x^2=0$.  The two spheres intersect at poles of~${\rm S}^4_b$.  There are two types of surface operators.

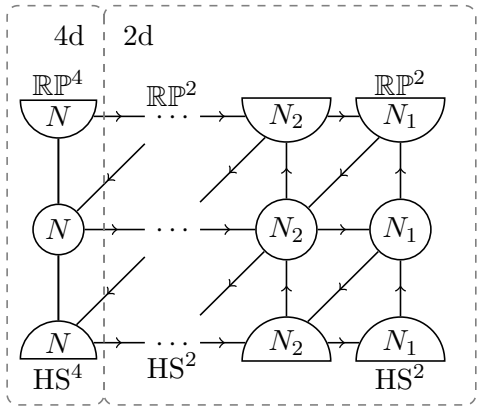
\begin{figure}
  \centering
  \begin{tikzpicture}[scale=1.5,semithick,
    color-group/.style  = {shape = circle , minimum size = 2.5ex, inner sep = .5ex, draw},
    flavor-group/.style = {shape = rectangle, minimum size = 2.5ex, inner sep = .5ex, draw},
    cf-group/.style     = {shape = rounded rectangle, rounded rectangle right arc = none, draw},
    fc-group/.style     = {shape = rounded rectangle, rounded rectangle left arc = none, draw},
    ->-/.style = {decoration = {markings, mark = at position #1 with {\arrow{>}}}, postaction = {decorate}}]
    \node (A) at (0,0)   [color-group,semicircle,shape border rotate=180]                    {$N$};
    \node (B) at (0,-1)  [color-group]         {$N$};
    \node (C) at (0,-2)  [color-group,semicircle]         {$N$};
    \node (Adots) at (1,0)               {${\cdots}$\strut};
    \node (A2) at (2,0) [color-group,semicircle,shape border rotate=180]     {$N_2$};
    \node (A1) at (3,0) [color-group,semicircle,shape border rotate=180]        {$N_1$};
    \node (Bdots) at (1,-1)               {${\cdots}$\strut};
    \node (B2) at (2,-1) [color-group]     {$N_2$};
    \node (B1) at (3,-1) [color-group]        {$N_1$};
    \node (Cdots) at (1,-2)               {${\cdots}$\strut};
    \node (C2) at (2,-2) [color-group,semicircle]     {$N_2$};
    \node (C1) at (3,-2) [color-group,semicircle]        {$N_1$};
    \draw (A) -- (B) -- (C);
    \draw[->-=.55] (A) -- (Adots);
    \draw[->-=.55] (Adots) -- (A2);
    \draw[->-=.55] (A2) -- (A1);
    \draw[->-=.55] (Adots) -- (B);
    \draw[->-=.55] (A2) -- (Bdots);
    \draw[->-=.55] (A1) -- (B2);
    \draw (B) -- (A);
    \draw[->-=.55] (B2) -- (A2);
    \draw[->-=.55] (B1) -- (A1);
    \draw[->-=.55] (B) -- (Bdots);
    \draw[->-=.55] (Bdots) -- (B2);
    \draw[->-=.55] (B2) -- (B1);
    \draw[->-=.55] (Bdots) -- (C);
    \draw[->-=.55] (B2) -- (Cdots);
    \draw[->-=.55] (B1) -- (C2);
    \draw (C) -- (B);
    \draw[->-=.55] (C2) -- (B2);
    \draw[->-=.55] (C1) -- (B1);
    \draw[->-=.55] (C) -- (Cdots);
    \draw[->-=.55] (Cdots) -- (C2);
    \draw[->-=.55] (C2) -- (C1);
    \node at (0,.3) {$\RP^4$};
    \node at (1,.2) {$\RP^2$};
    \node at (3,.3) {$\RP^2$};
    \node at (0,-2.3) {$\HS^4$};
    \node at (1,-2.2) {$\HS^2$};
    \node at (3,-2.3) {$\HS^2$};
    \node (top-corner) at ($(A.north east)+(.25,5ex)$) {};
    \node (bottom-corner) at ($(C.south west)+(-0.3,-0.4)$) {};
    \node (4d)  [below left = -1pt of top-corner] {4d};
    \node (2d)  [below right = -1pt of top-corner] {2d};
    \draw [color=gray, dashed, rounded corners]
          (top-corner) rectangle (bottom-corner);
    \draw [color=gray, dashed, rounded corners]
          (top-corner) rectangle ($(bottom-corner -| A1.east)+(0.3,0)$);
  \end{tikzpicture}
  \caption{\label{fig:M5surfacedefect}M5 brane defect described as a 2d $\Nsusy=(2,2)$ quiver gauge theory (nodes are $U(N_i)$ gauge groups, arrows are bifundamental chiral multiplets) coupled to the 4d $\Nsusy=2$ theory (nodes are $SU(N)$ gauge groups, edges bifundamental hypermultiplets).  The defect breaks $SU(N)$ to $S[U(N_1)\times U(N_2-N_1)\times\cdots\times U(N-N_{M-1})]$ on its support.  In the example depicted here, one of the 4d vector multiplets is $\ZZ_2$ identified to live on~$\RP^4$, and the 2d fields that implement its breaking (the first row of nodes and arrows) likewise live on~$\RP^2$; another vector multiplet lives on~$\HS^4$ and the last row of 2d fields on~$\HS^2$.  The quiver description itself will be studied in~\cite{LeFloch1}.}
\end{figure}

The first type descends from a codimension~$2$ (``M5~brane'') defect wrapping~$\Sigma$ of the 6d $\Nsusy=(2,0)$ theory.  It is described in the 4d $\Nsusy=2$ gauge theory by imposing for all gauge groups a singular behaviour $A\sim\alpha\dd{\chi}$ where $\chi$ is the angular coordinate around the support and $\alpha\in\lie{g}$.  At the location of this surface operator all gauge groups of the theory are broken in the same way.  The one-form $\dd{\chi}$ (and $\dd{\varphi}$ for an operator on the other sphere) is invariant under the $\ZZ_2$~involutions \eqref{hs4identificationofvectormultiplet} and~\eqref{rp4identificationofvectormultiplet} hence the surface operator is well-defined in our setup.  Let us mention that the defects have a description as 2d $\Nsusy=(2,2)$ theories supported on the defect and coupled to the 4d $\Nsusy=2$ theory (see \autoref{fig:M5surfacedefect}).  On the CFT side, the effect of this type of defect is only known at the level of conformal blocks~\cite{Alday:2010vg,Kanno:2011fw}: the defect changes the W-algebra to a different Drinfeld--Sokolov reduction of the affine Lie algebra $\widehat{\lie{su}}(N)$, that depends on discrete data associated to the defect.  In the absence of defect the S-matrix elements~$S_{1\alpha}$ \eqref{Sidnondeg} are equal to the one-loop determinant of a 3d vector multiplet on the equator~${\rm S}^3_b$.  The defect is a vortex loop in the 3d theory, and its expectation value should give~$S_{1\alpha}$ for the altered W-algebras.  Our analysis in \autoref{app:wavefunctions} of boundary conditions in terms of the S-matrix should go through unchanged for each of these algebras.  The relation between correlators on surfaces with boundaries and gauge theories on ${\rm S}^4_b$ with a $\ZZ_2$-identification should also not change.

The second type descends from a codimension~$4$ (``M2~brane'') defect at a point on~$\Sigma$.  On the CFT side it corresponds to the insertion of a degenerate vertex operator labeled by a representation $R_1$ (and $R_2=1$) at that point.  It is described in the 4d $\Nsusy=2$ gauge theory by adding on the support a 2d $\Nsusy=(2,2)$ gauge theory whose field content depends on~$R_1$.  The 2d theory is coupled by identifying its flavor symmetries to 4d flavor and gauge symmetries through a cubic superpotential coupling involving a bifundamental hypermultiplet of the 4d theory (different descriptions are related by the node-hopping duality).  Degenerate vertex operators labeled by a pair of representations correspond to intersecting defects (see~\cite{Gomis:2016ljm} and references therein) described by a 2d $\Nsusy=(2,2)$ labeled by~$R_1$ on the sphere at $x^3=x^4=0$ and one labeled by~$R_2$ on the other sphere, with additional degrees of freedom on the intersection.  Boundaries and cross-caps on~$\Sigma$ change nothing substantial to this discussion: the 4d vector multiplets to which the 2d theory couples may simply be subject to some $\ZZ_2$~identification.  That identification does not affect 2d fields.  When writing the gauge theory as a theory on a squashed hemisphere instead, the 2d fields get doubled like the 4d ones.  It is then natural to ask whether one can consider 2d fields that live only on a (squashed) hemisphere.  We come back to this at the end of the next subsection.

\subsection{Boundary operators}\label{sec:boundaryoperators}

As depicted in \autoref{figdisk2ptchannels} there are $\ZZ_2$-invariant pants decompositions of the Schottky double~$\widehat{\Sigma}$ that include cuts intersecting with boundaries of~$\Sigma$.  In such a channel, which we used in \autoref{sec:sduality} to discuss S-duality, correlators are expressed using some bulk-to-boundary OPEs rather than only bulk OPE.  This leads us to the question of how boundary operators are encoded in gauge theory.  Before addressing it we review boundary operators in CFT\@.

Boundary operators~$\Phi^{(ab)}_\beta$ interface between boundary conditions $a$ and~$b$ and carry an additional label~$\beta$, where $a$, $b$ and~$\beta$ label primary states of the CFT\@.  Using bulk or bulk-boundary OPE the most general correlation function can be computed from the two building blocks $\lb V_\alpha (z,\bar{z})\Phi^{(aa)}_\beta (x) \rb_{\B_a}$ and $\lb \Phi^{(ab)}_{2Q-\beta_3} (x_1)\Phi^{(bc)}_{\beta_2}(x_2) \Phi^{(ca)}_{\beta_1}(x_3)\rb$, where now $x$ labels a point on the boundary. The space-dependence of these functions is completely fixed by conformal symmetry so we can focus on overall factors: the bulk-boundary OPE coefficient $B_{\alpha \beta}^a$ and the boundary OPE coefficient $C_{\beta_1 \beta_2 (2Q-\beta_3)}^{a b c}$. This information is encoded in the OPE 
\begin{align}
  V_\alpha(z,\bar{z}) & \sim \sum_k B_{\alpha\beta}^a \Phi^{(aa)}_\beta(x)\abs{z-x}^{\Delta_\beta- 2 \Delta_\alpha},\\
  \Phi^{(ab)}_{\beta_1}(x)\Phi^{(bc)}_{\beta_2}(y) &\sim \sum_k  C_{\beta_1 \beta_2 (2Q-\beta_3)}^{a b c} \Phi^{(ac)}_{\beta_3} (y) \abs{x-y}^{\Delta_{\beta_3} - \Delta_{\beta_1} -\Delta_{\beta_2}}.
\end{align}
In these expressions all labels $(a,b,c,\alpha,\beta,\beta_1,\beta_2,\beta_3)$ run over primary fields. We differentiate between Greek/Roman letters only to distinguish insertions labels from boundary conditions. 

For the A-series minimal models the Cardy-Lewellen sewing constraints were solved \cite[(21),(31)]{Runkel:1998he} explicitly in terms of the bulk OPE coefficient~$C_{\alpha \beta \gamma}$, the S-matrix $S_a^b(c)$ of the torus one-point function with an insertion labeled by~$c$, and the usual fusion matrix, defined schematically by $\mathcal{F}_\alpha^{\text{(s)}}(z) = \int d\beta F_{\alpha\beta} \mathcal{F}_\beta^{\text{(t)}}(z)$.  The results are
\begin{equation}
  \label{eq:boundaryOPEcoeff}
  B_{\alpha \beta}^a = e^{(i \pi/2) \Delta_\beta} C_{\alpha \beta a} S_a^\alpha(\beta)/S_a^1 , \qquad
  C_{\beta_1 \beta_2 (2Q-\beta_3)}^{a b c} =  F_{b\beta_3} \left[\begin{smallmatrix}a&c\\\beta_1&\beta_2\end{smallmatrix}\right] .
\end{equation}
It was noticed long ago \cite{Ponsot:2003ss,Teschner:2000md, Ponsot:2001ng} that the bulk-boundary and boundary OPE coefficients of Liouville CFT satisfy these relations, despite Liouville not being rational.\footnote{A quick way to rederive this is to consider the modular bootstrap of the boundary one-point function of $\Phi^{(\alpha\alpha)}_\beta$ in the annulus with the other brane being labeled by~$\gamma$.  In the closed string channel, the boundary state $\ket{\B_\gamma}=\int\dd{\mu}\Psi_\gamma(\mu)\bket{\mu}$ is a sum over descendants of~$\ket{\mu}$; inserting this expansion in the correlator gives the disk correlator of $\Phi^{(\alpha\alpha)}_\beta$ with the bulk operator~$V_{\mu}$, multiplied by a character of the once-punctured torus.  In the open string channel it is a trace over the open string spectrum, and using known commutators of the Virasoro algebra with the primary~$\Phi$, contributions of descendants are repackaged into the once-punctured torus character at the conjugate modular parameter $\tau\to-1/\tau$.  The two are related by the S-matrix.}

Recall that branes labeled by a momentum~$a$ are obtained from the identity brane by acting with a topological defect operator~\eqref{wavefunctiongeneralformula}.  Boundary operator insertions can in fact be realized similarly in the framework of~\cite{Gaiotto:2014lma}.  Let~$\Sigma$ be a punctured Riemann surface and $M$~be a three-manifold whose boundary is~$\Sigma$.  Gaiotto introduces a method to associate a conformal block on~$\Sigma$ to any trivalent graph embedded in~$M$ with external legs attached at punctures of~$\Sigma$.  (For WZW models this is the standard construction~\cite{Witten:1988hf} of conformal blocks as states in Chern--Simons theory prepared using a network of Wilson lines.)  In this perspective, Verlinde loops arise as disconnected circle components of the trivalent graph, that run near the boundary of~$M$, away from punctures.

A correlator on a Riemann surface with boundaries is obtained from the Schottky double chiral correlator with identity branes by acting with a Verlinde loop, which is realized in Gaiotto's construction as a circle near the boundary of~$M$.
We propose that a boundary operator~$\Phi^{(ab)}_{\beta}$ is realized by adding a segment connecting the puncture~$\beta$ to the circle describing the brane.  The labels on both sides of the resulting trivalent vertex are $a$ and~$b$.  We check this explicitly for the bulk-boundary propagator.
\begin{equation}
  \begin{aligned}
    & \mathtikz{
      \draw (0,0) -- (.95,0) node [pos=0.15,below] {$\alpha$};
      \draw (1.05,0) -- (1.6,0);
      \draw (0.5,-0.05) arc (-174:174:0.25 and 0.5) node [pos=.85,above] {$s$};
      \draw (0.9,0.4) -- (1.1,0.8) node [pos=1, below] {$\beta$};
    }
    =
    \mathtikz{
      \draw (0,0) -- (.95,0) node [pos=0.15,below] {$\alpha$};
      \draw (1.05,0) -- (1.6,0);
      \draw (0.5,-0.05) arc (-174:174:0.25 and 0.5) node [pos=.85,above] {$s$};
      \draw (0.9,0.4) -- (1.1,0.8) node [pos=1, below] {$\beta$};
      \draw[dashed] (0.75,0) -- (0.75,-0.5) node [midway,left=-.4em] {\small $0$};
    }
    =
    \sum_\gamma F_{0\gamma}\!\begin{bmatrix}s&\alpha\\s&\alpha\end{bmatrix}
    \mathtikz{
      \draw (0,0) -- (1.45,0) node [pos=0.15,below] {$\alpha$} node [pos=0.7,above=-1ex] {$\gamma$};
      \draw (1.55,0) -- (2,0) node [pos=0.8,below] {$\alpha$};
      \draw (1.2,0) arc (-180:0:0.15) node [pos=.5,below=-.3ex] {$s$} arc (0:174:0.5) node [pos=.8,above] {$s$};
      \draw (.8,0) arc (0:-160:0.15) node [pos=.5,below=-.3ex] {$s$};
      \draw (1.3,0.4) -- (1.6,0.8) node [pos=1, below] {$\beta$};
    }
    \\
    & =
    \sum_\gamma F_{0\gamma}\!\begin{bmatrix}s&\alpha\\s&\alpha\end{bmatrix}
    e^{i \pi (\Delta_{\gamma}-\Delta_{\alpha}-\Delta_{s})}
    \mathtikz{
      \draw (0,0) -- (2,0) node [pos=0.1,below] {$\alpha$} node [pos=0.5,below] {$\gamma$} node [pos=0.9,below] {$\alpha$};
      \draw (1.5,0) arc (0:180:0.5) node [pos=.8,above] {$s$};
      \draw (1.3,0.4) -- (1.6,0.8) node [pos=1, below] {$\beta$};
    }
    \\
    & =
    \sum_{\gamma,\beta'} F_{0\gamma}\!\begin{bmatrix}s&\alpha\\s&\alpha\end{bmatrix}
    e^{i \pi (\Delta_{\gamma}-\Delta_{\alpha}-\Delta_{s})}
    F_{\gamma\beta'}\!\begin{bmatrix}\alpha&\alpha\\s&s\end{bmatrix}
    \mathtikz{
      \draw (0,0) -- (2,0) node [pos=0.1,below] {$\alpha$} node [pos=0.9,below] {$\alpha$};
      \draw (1.1,.6) circle [radius=.3] node {$s$};
      \draw (.6,0) -- (.8,.6) node [pos=0.5,left=-.4em] {$\beta'$};
      \draw (1.4,.6) -- (1.7,.6) node [pos=.8, right] {$\beta$};
    }
  \end{aligned}
\end{equation}
Equation (52) of~\cite{Runkel:1998he} gives the expected S-matrix element, up to some normalization we did not track down.  As a sanity check, notice that for $\beta=0$ (identity insertion), the construction we give reduces to the Verlinde loop along the boundary.

We now give an explicit formula in the case of Liouville theory. If we fix the normalization of boundary operators such that $B_{\alpha_1 0}^{\alpha_3} = D_{\alpha_1 \alpha_3}$ (the Verlinde loop operator) then the bulk-boundary OPE coefficient is~\cite{Hosomichi:2010vh}
\beq\label{bulkboundaryOPE}
B_{\alpha_1 \alpha_2}^{\alpha_3} =\frac{2^{3/2}}{S_b(\tfrac{1}{2}(b+1/b)-iP_2)} \int\dd{r}
e^{2 \pi i r P_3} \prod_{\pm\pm} S_b\Bigl(\frac{b+1/b}{4} - \frac{iP_2}{2} \pm iP_1 \pm ir\Bigr).
\eeq
The boundary OPE coefficient is likewise given by a fusion matrix, derived in \cite{Teschner:2000md,Ponsot:2001ng,Ponsot:2003ss}.

Both of these OPE coefficients take the form of the ${\rm S}^3_b$~partition function of some 3d $\Nsusy=2$ theory, which is easy to read off from explicit formulas.  This was done for~\eqref{bulkboundaryOPE} in~\cite{Hosomichi:2010vh} and for the boundary OPE coefficient in~\cite{Teschner:2012em}.  For instance the former is obtained by coupling a mass deformation of the 3d $T[SU(2)]$ theory at the equator to the 4d theory.  This simplicity is misleading however: when there are more than three boundary operators, the cross-ratio of their position is likely an instanton-counting parameter, which requires 4d~fields on the gauge theory side.  It would be interesting to clarify the general story for non-degenerate boundary operators, in particular how they arise from the 6d interpretation.

For degenerate boundary operators there is a well-motivated guess, based on the fact that degenerate operators in the bulk correspond to surface operators supported on spheres~${\rm S}^2_b$.  Recall that a boundary operator $\Phi_{\beta}^{ab}$ interpolating between boundaries labeled by momenta~$a$ and~$b$ only exists provided $a$, $b$, and~$\beta$ are compatible with fusion.  In particular there cannot be any boundary operator on an identity brane.  A given degenerate brane only allows finitely many different degenerate boundary operators.  Such branes are realized in gauge theory as Wilson loops at the equator, and it is plausible that degenerate boundary operators correspond to surface operators supported on $\HS^2_b$ that end on Wilson lines.  One would need to elucidate the interplay between possible field contents of the 2d theory (especially the rank of 2d gauge groups) and the charge of the Wilson line.

\section{Conclusion}\label{sec:conclusion}

In this paper we matched correlation functions in Liouville/Toda in the presence of ZZ brane boundaries to partition functions of 4d $\Nsusy=2$ gauge theories, where the channel associated to the boundary corresponds to a vector multiplet living on the orbifold ${\rm S}^4/\ZZ_2=\HS^4$.  We have a 6d interpretation as putting the theory on an orbifold $({\rm S}^4 \times\widehat{\Sigma})/\ZZ_2$ depicted in \autoref{fig:6dorbifold}.  This orbifold has fixed points ${\rm S}^3\times {\rm S}^1$ for each boundary of $\widehat{\Sigma}/\ZZ_2$. As opposed to previous constructions, this acts on the 4d space by exchanging the poles of S$_b^4$. We obtained correlators with other CFT boundary states by adding defects inside this singular locus, as was done for topological defects in~\cite{Drukker:2010jp}. We also matched cross-cap states to gauge theories with an antipodal identification on~${\rm S}^4$.  In order to do this we proposed a formula for localization on~$\RP^4$.  Let us conclude by mentioning some applications and open questions.

In our 6d motivation, the $\ZZ_2$~actions on 4d space and 2d space (in both cases, reflection or antipodal map) do not seem constrained to be the same.  Nevertheless, it seems they need to be in order to give branes and cross-caps in the 2d CFT.  Given that theta angles measures Dehn twists of the Riemann surface and that cross-caps and branes are related by a half twist, the $\ZZ_2$~quotients of 6d that we did not consider should give rise to gauge theories with discrete theta terms.  On the CFT side however one would end up with a boundary state with a wavefunction given by the cross-cap wavefunction, or viceversa.  The resulting states seem to violate the Cardy condition.

This is especially puzzling given that we got all Liouville/Toda CFT branes (solutions of Cardy's condition) from the 6d setups we considered so M-theory brane intersections are somehow sensitive to this condition.  It would be very interesting to give a 4d or 6d perspective on the Cardy condition.  One approach may be to derive the results presented in this work along the lines of~\cite{Cordova:2016cmu}, see also \cite{Nekrasov:2010ka}. The procedure involves starting from the 6d geometry, writing ${\rm S}^4$ as a fibration of ${\rm S}^3$ on an interval, and compactifying on this ${\rm S}^3$, obtaining 3d complex Chern-Simons that can be further reduced to Liouville or Toda. Orbifolding this setup should produce Chern-Simons on a 3d orbifold.  This was studied originally by Horava~\cite{Horava:1990ee} in relation to boundary CFT and later in~\cite{Felder:1999cv,Felder:1999mq}. We may learn in this way what is the higher dimensional counterpart of Cardy's condition selecting physical boundary conditions. Unfortunately, the 6d theory is indirectly defined as a low energy limit of string theory on a stack of M5-branes, and we do not know what becomes of it in the presence of singular geometries.\footnote{We thank Herman Verlinde for pointing this out to us.}

We studied the standard form of the AGT correspondence that can be though of coming from 6d $\Nsusy=(2,0)$ on ${\rm S}^4 \times \Sigma$.  One can also study what happens when the 6d theory lives on ${\rm S}^3\times {\rm S}^1 \times \Sigma$.  In this case it was found that the partition function of a gauge theory living in ${\rm S}^3 \times {\rm S}^1$ (an index) corresponds to 2d Yang-Mills theory living on~$\Sigma$~\cite{Gadde:2011ik}.  Does this scenario admit an interesting $\ZZ_2$~orbifold, perhaps with a ${\rm S}^3 \times {\rm S}^1/\ZZ_2$ partition function?  Similarly, what can be learned from orbifolds of the 3d-3d correspondence?

We proposed a technique to build up the partition function of gauge theories on (squashed)~$\RP^4$ by gluing hemisphere and lens space results, and confirmed it using the AGT correspondence.  We leave for future work the detailed proof of these statements along the lines of~\cite{HS4}. Such a gluing prescription applies more broadly than the $\RP^2$ and~$\RP^4$ cases we considered; for instance instead of gluing a single lens space, we can consider the region between two constant latitude spheres in ${\rm S}^4$ and have an antipodal identification on each of these two boundaries. A detailed study of the gluing procedure in 3d and other dimensions is being carried out in~\cite{Dedushenko1,Dedushenko2,Dedushenko3}.

An interesting application of the $\RP^4$ results would be to study Wilson loops or other observables of $SU(N)$ $\Nsusy=4$ gauge theories with large $N$ living on projective space.  This should be compared with the AdS$_5$ orbifold that is dual to it.  An interesting AdS$_3$/CFT$_2$ version of this problem was studied in \cite{Maloney:2016gsg}. Another approach could be to use cross-caps in the 2d CFT to represent local bulk fields~\cite{Miyaji:2015fia,Nakayama:2015mva, Verlinde:2015qfa, Lewkowycz:2016ukf} and apply the 4d perspective to this problem.

We commented on the action of S-duality on a boundary condition in one example of 4d $\Nsusy=2$ gauge theory.  It would be useful to generalize this construction along the lines of~\cite{Gaiotto:2008ak}.  To our knowledge an exhaustive classification of $\Nsusy=2$ invariant boundary conditions has not been developed.

On a related note, our $\RP^4$ results could be confirmed by checking that the partition function is S-duality invariant.  The match with Liouville/Toda CFT gives how instanton partition function transform and one needs to prove an integral identity on double sine functions coming from one-loop factors.  It would also be interesting to derive them from first principles.

Another generalization would be to study irregular boundary state obtained from colliding a primary with its Schottky image in the Gaiotto--Teschner scaling limit \cite{Gaiotto:2012sf}.

\acknowledgments

We thank Mykola Dedushenko, Jaume Gomis, Kazuo Hosomichi, Sungjay Lee, Thomas Mertens, M\'ark Mezei, Muteeb Nouman, Sylvain Ribault, Mauricio Romo, Yuji Tachikawa, Herman Verlinde and Yifan Wang for useful discussions and correspondence.

\appendix

\section{ADE Toda boundary states}\label{app:wavefunctions}

We review ADE Toda CFT in~\autoref{sec:bcft}.  In \autoref{sec:boundarystates} we use modular invariance of the annulus partition function to completely determine the wavefunctions of boundary states in terms of S-matrix elements and two-point functions, including a phase that is traditionally fixed using disk correlators with degenerate insertions.  In \autoref{sec:Walgebracharacters} and \autoref{sec:modularsmatrix} we write down the characters and S-matrix components; the wavefunctions of all branes are deduced from them.  After deriving the wavefunctions we check in \autoref{sec:CardyCondition} that they obey Cardy's condition when the open string spectrum is discrete.  Cross-cap states relevant for nonorientable surfaces are discussed briefly in \autoref{sec:crosscapstate}.

\subsection{Boundary CFT}\label{sec:boundarystates}

In this section we construct boundary states associated to branes in Liouville and ADE Toda CFT\@.  We work in a rather general setting and restrict to Liouville/Toda only when needed.
We follow ideas from the Cardy construction of CFT boundary states for RCFTs~\cite{Cardy:1989ir, Behrend:1999bn}.  When the CFT is placed on a Riemann surface with boundaries, modular invariance of the annulus (cylinder) partition function constrains the possible boundary conditions.  We do not check other sewing constraints, namely crossing symmetry of the disk two point function~\cite{Lewellen:1991tb}.

We assume that the theory on a closed Riemann surface has a chiral symmetry $\mathcal{A}\otimes \bar{\mathcal{A}}$ that includes the left-moving and right-moving Virasoro algebras and is such that the Hilbert space can be decomposed in irreducible representations $\mathcal{V}_{\alpha}\otimes\mathcal{V}_{\alpha^\star}$, where $\alpha^\star$ labels the representation conjugate to~$\alpha$.  Furthermore, we assume that the theory is diagonal, meaning that the Hilbert space on a circle decomposes as $\mathcal{H}_{\text{closed}}= \bigoplus_{\alpha\in\mathbb{S}} \mathcal{V}_\alpha \otimes \mathcal{V}_{\alpha^\star}$ for some spectrum~$\mathbb{S}$.  For RCFTs, $\mathcal{A}$~has finitely many representations and the direct sum is finite, while for Liouville and ADE Toda cases the label~$\alpha$ is continuous and we shall not specify in this paper what topology is used to complete.

Start with a CFT on an interval with boundary conditions $\B_1$ and~$\B_2$ at each end. The partition function of this configuration with periodic time can be written in the following way 
\beq\label{eq:Zopen}
Z_{\B_1 \B_2}(\tau) = \Tr_{\mathcal{H}_{\B_1 \B_2}} e^{2 \pi i \tau (L_0 - \frac{c}{24})} = \sum_\alpha N_\alpha^{\B_1 \B_2} \chi_\alpha (\tau),
\eeq 
where the trace is taken over the open-string states propagating between the boundaries. When boundary conditions preserve a diagonal subalgebra of $\mathcal{A}\otimes\bar{\mathcal{A}}$ this Hilbert space is a direct sum $\mathcal{H}_{\B_1 \B_2}= \bigoplus_\alpha N_\alpha^{\B_1 \B_2}\mathcal{V}_{\alpha}$.  Each term contributes the corresponding character~$\chi_\alpha(\tau)$ with a multiplicity~$N_\alpha^{\B_1 \B_2}$.  When the spectrum is discrete (in particular in RCFTs) these multiplicities must be positive integers.  When the spectrum is continuous, the sum is replaced by an integral and $N$~becomes a density of states.

Alternatively one can quantize the theory along the closed channel, and rewrite $Z_{\B_1 \B_2}$ as a cylinder amplitude between two boundary states $\ket{\B_1}, \ket{\B_2}$ in $\mathcal{H}_{\text{closed}}$. This gives
\beq\label{eq:cardycond}
\bra{\B_1} e^{- i \pi /\tau (L_0 + \bar{L}_0 - \frac{c}{12})} \ket{\B_2} = Z_{\B_1 \B_2}(\tau) =\sum_\alpha N_\alpha^{\B_1 \B_2} \chi_\alpha (\tau).
\eeq
The goal of Cardy's construction is to classify states~$\ket{\B}$ such that multiplicities of the spectrum between two branes are integers.

The first step consists of finding a proper basis to expand the boundary states, such that they are invariant under conformal transformations that preserve the boundary. If the boundary is placed on the real axis then this condition means classically that 
\beq
W(z) = \hat{\rho}\bigl(\overline{W}(\bar{z})\bigr), \quad z=\bar{z},
\eeq
where $W$, $\overline{W}$ are the currents generating the algebra $\mathcal{A}\otimes\mathcal{A}$.  Here, $\hat{\rho}$ denotes an automorphism of the algebra that leaves the stress-tensor invariant. In terms of modes in the closed channel this condition translates to
\beq
\Big(W_n^{(s)} - (-1)^{h_s} \hat{\rho}\Big(\overline{W}_{-n}^{(s)}\Big)\Big)\ket{\B} =0, 
\eeq
where $h_s$ is the conformal dimension of $W$ and $s$ labels the generators of $\mathcal{A}$. For $\hat{\rho}=\id$ a basis of solutions were found by Ishibashi~\cite{Ishibashi:1988kg}, labeled by representations~$\mathcal{V}_\alpha$ of~$\mathcal{A}$.  They are given by formal sums over an orthonormalized basis of descendants in~$\mathcal{V}_\alpha$,
\beq
\bket{\alpha} \equiv \sum_{\{ k \} } \ket{\alpha; k} \otimes \overline{\ket{\alpha^\star ; k}} .
\eeq
The contribution of this state to the partition function~\eqref{eq:cardycond} is proportional to the character of the corresponding representation~$\mathcal{V}_\alpha$.  Namely, it is
\beq
\bbra{\beta} e^{2\pi i(-1/\tau)H_{\text{closed}}} \bket{\alpha} = \lb\beta|\alpha\rb \chi_\alpha(-1/\tau),
\eeq
where $H_{\text{closed}}$ is the closed-channel Hamiltonian.  The two-point function is proportional to Kronecker/Dirac delta as appropriate: $\lb\beta|\alpha\rb=\langle V_{\beta^\star}V_{\alpha}\rangle=\mathcal{R}(\alpha)\delta_{\beta\alpha}$.  We deduce $\mathcal{R}(\alpha)=\mathcal{R}(\alpha^\star)$.

Boundary states can then be expanded in Ishibashi states as\footnote{In ADE Toda CFT labeled by~$\lie{g}$, the spectrum is a Weyl chamber $\mathbb{S}=\lie{t}/\Weyl$ of the Cartan algebra, and we use the measure~\eqref{intdalpha} on that Weyl chamber rather than integrating throughout~$\lie{t}$ with a factor of $1/|\Weyl|$.  The two-point function $\mathcal{R}(\alpha)$ given in~\eqref{TodaReflection} is called a maximal reflection amplitude.  In the main text we use transformation properties under the Weyl group to recast $\Psi_{\B}(\alpha^\star)/\mathcal{R}(\alpha^\star)=\Psi_{\B}(2Q-\alpha)$, which for non-degenerate momenta is $2Q-\alpha=\bar{\alpha}$.}
\beq
\ket{\B} = \int_{\alpha\in\mathbb{S}} \dd{\alpha} \frac{\Psi_{\B}(\alpha^\star)}{\mathcal{R}(\alpha^\star)} \bket{\alpha},
\quad \text{that is}, \quad
\bra{\B} = \int_{\alpha\in\mathbb{S}} \dd{\alpha} \frac{\Psi_{\B}(\alpha)}{\mathcal{R}(\alpha)} \bbra{\alpha},
\eeq
where the boundary state wavefunction $\Psi_{\B}$ characterizes the boundary condition in the closed channel.  The normalization by $\mathcal{R}(\alpha)=\mathcal{R}(\alpha^\star)$ ensures that the disk one-point function and the one-point function with a hole are equal to the wavefunction: $\langle V_\alpha \ket{\B}=\bra{\B}V_\alpha\rangle=\Psi_{\B}(\alpha)$ up to cross-ratios.
Then Cardy's condition~\eqref{eq:cardycond} reads
\beq
\int_{\alpha\in\mathbb{S}}\dd{\alpha} \frac{\Psi_{\B_1}(\alpha) \Psi_{\B_2}(\alpha^\star)}{\mathcal{R}(\alpha)} \chi_\alpha(-1/\tau)
= \sum_\gamma N_\gamma^{\B_1 \B_2} \chi_\gamma(\tau) .
\eeq
When $\mathcal{A}$~is the Virasoro algebra, $\chi_\gamma(\tau)=\int_{\alpha\in\mathbb{S}}\dd{\alpha} S_{\gamma\alpha} \chi_\alpha(-1/\tau)$ so by linear independence
\beq\label{eq:cardycond2}
\frac{\Psi_{\B_1}(\alpha^\star) \Psi_{\B_2}(\alpha)}{\mathcal{R}(\alpha)}
=\sum_\gamma N_\gamma^{\B_1 \B_2} S_{\gamma\alpha}.
\eeq
For the W-algebra (for Toda CFT) the S-matrix is far from being fixed by how it acts on characters since those are not linearly independent.  However, there is a concrete conjecture~\cite{Drukker:2010jp} and we assume that~\eqref{eq:cardycond2} holds.  Their formula is supported by a relation with gauge theory but it would be valuable to check it for characters refined by the insertion of modes~$W^{(s)}_0$.

We assume that there exists a unique brane such that the open string spectrum between two such brane consists only of the identity.  By uniqueness it is invariant under~${}^\star$.  Cardy's condition implies
\beq\label{psi1asSR}
\Psi_1(\alpha) = \sqrt{S_{1\alpha}\mathcal{R}(\alpha)}
\eeq
Consider next a brane~$\ket{B}$ such that the spectrum of an open string between that brane and the identity brane is discrete.  Cardy's condition in the form~\eqref{eq:cardycond2} implies that the wavefunction is an integer linear combination of basic branes
\beq\label{wavefunctiongeneralformula}
\Psi_{\B}(\alpha)
=\sum_\gamma N_\gamma^{1\B} \Psi_{\gamma}(\alpha) ,
\qquad
\Psi_\gamma(\alpha)
= \frac{S_{\gamma\alpha}\mathcal{R}(\alpha)}{\Psi_1(\alpha)}
= \frac{S_{\gamma\alpha}\sqrt{\mathcal{R}(\alpha)}}{\sqrt{S_{1\alpha}}}
= \Psi_1(\alpha) \frac{S_{\gamma\alpha}}{S_{1\alpha}} .
\eeq
These basic branes are labeled by representations $\mathcal{V}_\gamma$ of~$\mathcal{A}$.  The second expression for~$\Psi_\gamma(\alpha)$ is the well-known Cardy ansatz ($\mathcal{R}(\alpha)$ is a phase).  The third shows that the brane can also be obtained from the identity brane by inserting a Verlinde loop~\cite{Verlinde:1988sn} labeled by~$\gamma$ along the boundary.  Indeed, the ratio of S-matrix elements $D_{\gamma\alpha}=S_{\gamma\alpha}/S_{1\alpha}$ is precisely the action of a topological defect or Verlinde loop~\cite{Drukker:2010jp}.  This observation makes it easier to compare boundary states in Toda to 4d gauge theory calculations later on.

Cardy's solution also specifies multiplicities in the Hilbert space of an open string stretching between two branes.  For the Toda CFT we check in \autoref{sec:CardyCondition} that they are integers when the spectrum is discrete, in other words when one of the branes is labeled by a fully degenerate representation (see definition in \autoref{sec:bcft}).  Using the explicit solution~\eqref{wavefunctiongeneralformula}, equation~\eqref{eq:cardycond2} takes the form of the Verlinde formula~\cite{Verlinde:1988sn}
\beq
\frac{S_{\mu_1^\star\alpha} S_{\mu_2\alpha}}{S_{1\alpha}} = \sum_\beta S_{\alpha\beta} \mathcal{N}_{\mu_1^\star\mu_2}^\beta,
\eeq
and therefore the degeneracy of the representation~$\alpha$ between branes~$\B_{\mu_1}$ and~$\B_{\mu_2}$ is equal to the fusion rule coefficient, $N^{\B_{\mu_1}\B_{\mu_2}}_\alpha=\mathcal{N}_{\mu_1^\star\mu_2}^\alpha$, which is automatically a nonnegative integer in cases where Verlinde's formula holds.

\subsection{Representations and characters}\label{sec:Walgebracharacters}

First we give a brief summary of highest-weight representations of W-algebras, which classify the primary operators of ADE Toda CFT\@.  We follow the notations of~\cite{Drukker:2010jp}.  The theory is labeled by a simply-laced Lie algebra~$\lie{g}$ (for Liouville CFT, $\lie{sl}_2$).  It has central charge $c=\rank\lie{g}+12\langle Q,Q\rangle$, where $Q=(b+1/b)\rho$ is a multiple of~$\rho$, the half-sum of positive roots.  Rather than the cosmological constant~$\mu$ we use $\hat{\mu} = \bigl(\pi \mu \gamma(b^2)b^{2-2b^2}\bigr)^{1/b}$.  The theory has a symmetry $(b,\hat{\mu})\to(1/b,\hat{\mu})$.

First, we consider the generic (non-degenerate) representation which consists of the full Verma module of the W-algebra and has no null states.  It is labeled by a vector
\beq
\mu = Q + m,
\eeq
where the momentum $m$ is an imaginary vector in the Cartan algebra of $\lie{g}$.  The corresponding primary operator of ADE Toda theory has dimension $\Delta(\mu)=\langle\mu,2Q-\mu\rangle/2=\langle Q,Q\rangle/2-\langle m,m\rangle/2$.  The character of this representation of the W-algebra is (in terms of $q=\exp 2\pi i\tau$)
\beq\label{characternondegenerate}
\chi_\mu (\tau) \equiv \Tr(q^{L_0-c/24}) = \frac{q^{-\lb m,m\rb/2}}{\eta(\tau)^{\rank\lie{g}}}.
\eeq

The W-algebra also has completely degenerate representations denoted $\Omega=(R_1,R_2)$, labeled by two representations $R_1$ and $R_2$ of $\lie{g}$ with highest weights respectively $\lambda_1$ and $\lambda_2$.  They have one null vector for each positive root of~$\lie{g}$ and their character is 
\begin{equation}
  \begin{aligned}
    \chi_\Omega (\tau)
    & = \sum_{w\in\Weyl} \epsilon(w) \chi_{Q-bw(\rho+\lambda_1)-(\rho+\lambda_2)/b}(\tau) \\
    & = \frac{q^{\Delta(\mu_\Omega)-(c-\rank\lie{g})/24}}{\eta(\tau)^{\rank\lie{g}}}\sum_{w\in \Weyl} \epsilon(w) q^{\lb\rho+\lambda_1-w(\rho+\lambda_1),\rho + \lambda_2\rb}
  \end{aligned}
\end{equation}
where the Toda momentum is $\mu_{\Omega} - Q = - b(\rho+\lambda_1) -(\rho + \lambda_2)/b$.

Finally, there are also semi-degenerate representations for which the null states are associated to a subset $\mathcal{I}\in\Delta^+$ of the positive roots ($\mathcal{I}=\Delta^+$ for completely degenerate representations and $\mathcal{I}=\emptyset$ for non-degenerate ones).  Following~\cite{Drukker:2010jp} we only consider the case where $\mathcal{I}$ is the set of positive roots of a (simply-laced) Lie subalgebra $\lie{g}_{\mathcal{I}}\subset\lie{g}$ with full rank.  These representations are labeled by~$\lie{g}_{\mathcal{I}}$, two representations $R_1$ and $R_2$ of~$\lie{g}_{\mathcal{I}}$ with highest weights $\lambda_1$ and~$\lambda_2$, and an imaginary vector $\tilde{m}$ orthogonal to all roots in~$\mathcal{I}$.  In terms of their Toda momentum
\beq
\mu - Q = m = \tilde{m} - b(\rho_{\mathcal{I}}+\lambda_1)-(\rho_{\mathcal{I}} + \lambda_2)/b
\eeq
where $\rho_{\mathcal{I}}$ is the half-sum of the roots in~$\mathcal{I}$, the character of these representations is
\begin{equation}
  \begin{aligned}
    \chi_{\tilde{m},\Omega,\mathcal{I}}(\tau)
    & = \sum_{\mathclap{w\in\Weyl_{\mathcal{I}}}} \epsilon(w) \chi_{Q+\tilde{m}-bw(\rho_{\mathcal{I}}+\lambda_1)-(\rho_{\mathcal{I}}+\lambda_2)/b}(\tau) \\
    & = \frac{q^{-\lb m,m\rb/2}}{\eta(\tau)^{\rank\lie{g}}}\sum_{w\in \Weyl_{\mathcal{I}}} \epsilon(w) q^{\lb \rho_{\mathcal{I}} + \lambda_1 - w(\rho_{\mathcal{I}} + \lambda_1), \rho_{\mathcal{I}}+ \lambda_2\rb} .
  \end{aligned}
\end{equation}

\subsection{Modular S-matrix}\label{sec:modularsmatrix}

In this subsection we list the elements of the S-matrix found in~\cite{Drukker:2010jp} giving the transformation of the W-algebra characters under a modular transformation, meaning
\beq\label{todasmatrix}
\chi_\mu(\tau) = \int_{\lie{t}/\Weyl}\dd{\alpha} \, S_{\mu\alpha} \chi_\alpha(-1/\tau)
\eeq
with the real measure~\eqref{intdalpha}.  This choice removes from S-matrix elements all powers of the Cartan matrix determinant and of the Weyl group order compared to~\cite{Drukker:2010jp}, as well as the factor $i^{\rank\lie{g}}$ \big(which was caused by a sign $-i\tau\to i\tau$ in their (3.30)\big).  These matrix elements are building blocks of ADE Toda boundary states (see \autoref{sec:CardyCondition}).  The results can be checked using modularity of~$\eta$ and a Gaussian integral.  Denoting by $|\Weyl|$~the order of the Weyl group and $C$ the determinant of the Cartan matrix~$C_{ij}$, they are as follows.
\begin{itemize}
\item Between non-degenerate representations with momenta $m=\mu-Q$ and $a=\alpha-Q$: 
\beq\label{Snondegnondeg}
S_{\mu\alpha} = \sum_{w\in \Weyl} e^{2 \pi i \lb w(m), a\rb}.
\eeq

\item Between the identity (denoted~$1$) and non-degenerate:
\beq\label{Sidnondeg}
S_{1\alpha} = \prod_{e>0} \Bigl( -4 \sin(\pi b\lb a,e\rb) \sin(\pi b^{-1} \lb a, e\rb) \Bigr).
\eeq 

\item Between fully degenerate $\Omega = (R_1,R_2)$ and non-degenerate:
\beq
S_{\Omega \alpha} = S_{1\alpha} \, \chi_{R_1}\bigl(e^{2 \pi i b a}\bigr) \, \chi_{R_2}\bigl(e^{2 \pi i a/b}\bigr).
\eeq
Here, $\chi_{R}(e^x)$ is the character of the representation~$R$ of~$\lie{g}$.  It is defined as a sum over weights of the representation~$R$ or, using the Weyl character formula, as a sum over Weyl group:
\beq
\chi_R(e^x) = \sum_{\mu\in\weights(R)} e^{\lb \mu , x\rb}
= \frac{\sum_{w\in\Weyl} \epsilon(w) e^{\lb w(\rho+\lambda),x\rb}}{\sum_{w\in\Weyl} \epsilon(w) e^{\lb w(\rho),x\rb}}.
\eeq

\item Between semi-degenerate $(\mathcal{I},\tilde{m},\Omega)$ and non-degenerate (note that $\chi_{R_i}$ are characters of~$\lie{g}_{\mathcal{I}}$):
\begin{equation}\label{Sseminon}
  \begin{multlined}
    S_{(\mathcal{I},\tilde{m},\Omega)\alpha}= \sum_{w\in \Weyl/\Weyl_{\mathcal{I}}} e^{2 \pi i \lb w(\tilde{m}),a\rb}\chi_{R_1}\bigl(e^{w^{-1}(2\pi i b a)}\bigr)\chi_{R_2}\bigl(e^{w^{-1}(2\pi i a/b)}\bigr)\\
    \times\prod_{e \in w(\Delta_{\mathcal{I}}^+)} \Bigl( -4 \sin\bigl(\pi b \lb a, e\rb\bigr) \sin\bigl(\pi b^{-1} \lb a, e\rb\bigr) \Bigr) .
  \end{multlined}
\end{equation}
\end{itemize}

\subsection{Cardy condition for other branes}\label{sec:CardyCondition}

From the general formula~\eqref{psi1asSR} in terms of S-matrix elements~\eqref{Sidnondeg} and the reflection amplitude
\begin{equation}\label{TodaReflection}
  \mathcal{R}(\alpha)=-(\hat{\mu}b^{2b-2/b})^{-2\langle\rho,a\rangle} \prod_{e>0} \frac{\Gamma(1+\langle a,e\rangle b) \Gamma(1+\langle a,e\rangle/b)}{\Gamma(1-\langle a,e\rangle b) \Gamma(1-\langle a,e\rangle/b)}
\end{equation}
where $a=\alpha-Q$, one works out the wavefunction
\begin{equation}\label{psi1explicit}
  \Psi_1(\alpha)
  = (\hat{\mu}b^{2b-2/b})^{-\lb \rho, a\rb} \prod_{e>0}\frac{2\pi \lb e, a\rb }{\Gamma(1- b \lb e, a\rb)\Gamma(1 - b^{-1} \lb e, a\rb)}
\end{equation}
for the identity brane, and other branes given below~\eqref{psi1explicitmaintext}.  This generalizes the $\mathfrak{su}(2)$ (Liouville) \cite{FZZ, ZZ} and $\mathfrak{su}(N)$ \cite{Fateev:2010za,Sarkissian:2011tr} results.  While these wavefunctions are in principle only defined in the fundamental Weyl chamber $\lie{t}/\Weyl$, their analytic continuation obeys a reflection formula
\begin{equation}\label{psireflection}
  \Psi_{\B}(\alpha) = R_w(\alpha) \Psi_{\B}\bigl(Q+w(\alpha-Q)\bigr)
\end{equation}
for the same reflection amplitude $R_w(\alpha)$ as~\eqref{vertexhole}, independent on the brane.  That independence is not surprising since the S-matrix elements themselves, analytically continued, are Weyl invariant.

We now check Cardy's condition~\eqref{eq:cardycond2} when neither brane is the identity.  To get a discrete spectrum we consider the case where one brane is labeled by a degenerate primary operator.  Cardy's condition is that the multiplicities in this spectrum are integers.  Using the explicit solution~\eqref{wavefunctiongeneralformula}, the condition reads
\begin{equation}
  \frac{S_{\mu_1^\star\alpha}S_{\mu_2\alpha}}{S_{1\alpha}} = \sum_\beta N_\beta^{\mu_1 \mu_2} S_{\beta\alpha}
\end{equation}
where we used $S_{\gamma\alpha^\star}=S_{\gamma^\star\alpha}$.  The following relations are useful:
\begin{equation}\label{conjugationofS}
  S_{\Omega\alpha^\star} = S_{\Omega^\star\alpha} , \quad
  S_{(\mathcal{I},\tilde{m},\Omega)\alpha^\star} = S_{(\mathcal{I}^\star,\tilde{m}^\star,\Omega^\star)\alpha} , \quad
  S_{\mu\alpha^\star} = S_{\mu^\star\alpha} .
\end{equation}

Consider two degenerate branes with labels $\Omega_{12}=(R_1,R_2)$ and $\Omega_{34}=(R_3,R_4)$.  We first work out the Verlinde formula in the form of~\cite{Jego:2006ta}:
\begin{equation}
\frac{S_{\Omega_{12}\alpha}S_{\Omega_{34}\alpha}}{S_{1\alpha}} =
  S_{1\alpha}\chi_{R_1} (e^{2 \pi i b a}) \chi_{R_2}(e^{2 \pi i a/b})  \chi_{R_3} (e^{2 \pi i b a}) \chi_{R_4}(e^{2 \pi i a/b}) 
  = \sum_{k,\ell} N_{13}^k N_{24}^\ell S_{\Omega_{k\ell}\alpha}.
\end{equation}
Here we used $S_{\Omega_{12}\alpha}/S_{1\alpha}=\chi_{R_1}(e^{2\pi i b a})\chi_{R_2}(e^{2\pi i a/b})$ then rewrote the product of the $\lie{g}$ characters as a sum over single characters
\beq
\chi_{R_1} (e^{2 \pi i b a})\chi_{R_3} (e^{2 \pi i b a})= \sum_k  N_{13}^k\chi_{R_i}(e^{2 \pi i b a}).
\eeq
This formula comes from the tensor product rules $R_1\otimes R_3 = \sum_k N_{13}^k R_k$ for representations of the Lie algebra~$\lie{g}$, hence $N_{13}^k$ are nonnegative integers that can be worked out explicitly for any Lie group.  This version of the Verlinde formula, with $(R_1,R_2)\to(R_1^\star,R_2^\star)$ as per~\eqref{conjugationofS}, expresses the degeneracies for the open string spectrum between branes $\Omega_{12}$ and~$\Omega_{34}$, namely $\mathcal{N}_{\Omega_{12}^\star\Omega_{34}}{}^{\Omega_{k\ell}} = N_{1^\star\,3}{}^k N_{2^\star\,4}{}^\ell$.

Now we repeat this procedure with the partition function in an annulus with boundary conditions associated with a degenerate brane $\Omega=(R_1,R_2)$ and a non-degenerate brane labeled by its momentum~$m$.  The relevant Verlinde formula is now
\begin{equation}
  \begin{aligned}
    \frac{S_{\Omega\alpha}S_{(Q+m)\alpha}}{S_{1\alpha}} &= \sum_{w\in \Weyl} e^{2 \pi i \lb w(m),a\rb} \chi_{R_1}(e^{2 \pi i b a})\chi_{R_2}(e^{2 \pi i a/b}),\\
    &= \sum_{w\in \Weyl} e^{2 \pi i \lb w(m),a\rb} \sum_{\mu_1,\mu_2} N^{R_1}_{\mu_1} N^{R_2}_{\mu_2} e^{2 \pi i b\lb \mu_1, a\rb + 2 \pi i\lb \mu_2, a\rb/b},\\
    &=\sum_{\mu_1,\mu_2} N^{R_1}_{\mu_1} N^{R_2}_{\mu_2}  \sum_{w\in \Weyl} e^{2 \pi i \lb w(m+b\mu_1+b^{-1} \mu_2),a\rb},\\
    &=\sum_{\mu_1,\mu_2} N^{R_1}_{\mu_1} N^{R_2}_{\mu_2} S_{m+b\mu_1 +b^{-1}\mu_2,\alpha}.
  \end{aligned}
\end{equation}
In the second line we used the definition of the $\mathfrak{g}$ characters as a sum over weights of the representation.  The Cardy condition is obeyed because multiplicities~$N^R_\mu$ are integers.  Incidentally, when the degenerate brane is the identity, the annulus partition function reduces to a single character.  A very similar story can be done for semi-degenerate representations and we will not present it: the procedure should now be clear.

\subsection{Cross-cap state}\label{sec:crosscapstate}
We end this section by using the modular bootstrap to find the wavefunction of ADE Toda describing the boundary state that reproduces the CFT on~$\RP^2$.  The analogue of Cardy's ansatz in this case is 
\beq
\Psi_{\otimes}(\alpha) = e^{i\delta(\alpha)} \frac{P_{1\alpha}}{\sqrt{S_{1\alpha}}},
\eeq
where $P_{\alpha\beta}$ is a matrix that generates the S modular transformation on the M\"obius strip character 
\beq
\hat{\chi}_{\alpha}(\tau) = e^{- i \pi (\Delta(\alpha)-c/24)} \chi_{\alpha}((1+\tau)/2)
= e^{i\pi\rank\lie{g}/24} \frac{e^{-\pi i\tau\lb a,a\rb/2}}{\eta((1+\tau)/2)^{\rank\lie{g}}}
\eeq
where the second equality uses~\eqref{characternondegenerate} hence only holds for a non-degenerate~$\alpha$.

Consistency on the M\"obius strip requires that the overall phase~$e^{i\delta(\alpha)}$ has to be the same as the one for boundary states. Therefore we can write the cross-cap state as 
\beq\label{crosscappieces}
\Psi_{\otimes}(\alpha) = \Psi_1(\alpha) \frac{P_{1\alpha}}{S_{1\alpha}}.
\eeq
Finding the cross-cap state is thus equivalent to finding this element of the P-matrix. 

Using the following Gaussian integral (for $p$ and $s$ real)
\beq
\frac{e^{\pi i p^2(-1/\tau)/2}}{\eta((1-1/\tau)/2)} = \int\dd{s} e^{-\pi i s p} \frac{e^{\pi i \tau s^2/2}}{\eta((1+\tau)/2)}
\eeq
and its $\rank\lie{g}$ analogue, we get the P-matrix between non-degenerate states, with normalization as in~\eqref{todasmatrix}:
\beq
P_{\mu\alpha} = \sum_{w\in \Weyl} e^{ \pi i \lb w(m), a\rb}.
\eeq
We need to find~$P_{1\alpha}$.  Just like what happens in Liouville, the character of a degenerate state is a sum of non-degenerate ones.  For the identity character,
\beq
\chi_1(\tau) = \sum_{w\in \Weyl} \epsilon(w) \chi_{Q+b w(\rho) + \rho/b} (\tau).
\eeq
When we write the M\"obius strip character, besides $\tau\to(1+\tau)/2$ there is an overall phase which depends on $w\in\Weyl$ through the dimension $\Delta(Q+bw(\rho)+\rho/b)=\lb\rho-w(\rho),\rho\rb$.  Explicitly,
\beq
\hat{\chi}_1(\tau) = \sum_{w\in \Weyl} (-1)^{\lb\rho-w(\rho),\rho\rb} \epsilon(w) \hat{\chi}_{Q+b w(\rho) + \rho/b} (\tau) .
\eeq
Now we can use the knowledge of the P-matrix between non-degenerate states to deduce
\beq
P_{1 \alpha} = \sum_{w,w'\in \Weyl} (-1)^{\lb\rho,\rho\rb-\lb w(\rho),w'(\rho)\rb}\epsilon(w) e^{ \pi i b \lb w(\rho), a\rb} \epsilon(w')e^{\pi i \lb w'(\rho), a\rb/b}
\eeq
Finally, \eqref{crosscappieces}~tells us to multiply $P_{1\alpha}$ by $\Psi_1(\alpha)/S_{1\alpha}=1/\Psi_1(2Q-\alpha)$ read from~\eqref{psi1explicit}:
\begin{equation}\label{psicrosscapexplicit}
  \begin{aligned}
    \Psi_{\otimes}(\alpha) &= (\hat{\mu}b^{2b-2/b})^{-\lb \rho, a\rb} \prod_{e>0} \frac{\Gamma(1+b\lb e,a\rb)\Gamma(1+b^{-1}\lb e,a\rb)}{ -2\pi \lb e,a\rb} \\
    &\quad \times \sum_{w,w'\in \Weyl} (-1)^{\lb\rho,\rho\rb-\lb w(\rho),w'(\rho)\rb}\epsilon(w) e^{ \pi i b \lb w(\rho), a\rb} \epsilon(w')e^{\pi i \lb w'(\rho), a\rb/b}.
  \end{aligned}
\end{equation}

\section{\texorpdfstring{$\RP^2$}{\detokenize{RP^2}} partition function}\label{app:rp2}

As a consistency check of the gluing prescription, we apply it to 2d $\Nsusy=(2,2)$ gauged linear sigma models on $\RP^2$.  We naturally obtain a ``Higgs branch'' expression of the $\RP^2$~partition function (with a sum over vortices), which must be equal to the ``Coulomb branch'' integral of~\cite{Kim:2013ola}.  We show the equality for a class of theories by summing up the residues in the Coulomb branch integral.  The same two types of expressions have been found in various dimensions: 2d~\cite{Benini:2012ui,Doroud:2012xw}, 3d~\cite{Fujitsuka:2013fga,Benini:2013yva}, 4d~\cite{Peelaers:2014ima,Chen:2015fta,Pan:2015hza}, and 5d~\cite{Pan:2014bwa}.

The 2d $\Nsusy=(2,2)$ gauge theories we consider have a connected gauge group~$G$ and are constructed from vector multiplets and from chiral multiplets in a representation~$\repr$ of~$G$.  Each $U(1)$ gauge factor admits an FI parameter~$\zeta_a$.  The vortex counting parameters $z_a=\exp(-2\pi\zeta_a)$ are real, as there can be no continuous theta angle on $\RP^2$.  Indeed, $\int\Tr F=0$ on~$\RP^2$ as the topological term is parity odd.  We ignore discrete theta terms, which affect how different topological sectors are summed up, because we neglect sign factors between topological sector.  For simplicity we consider theories where FI parameters do not run, namely where the sum of charges under each $U(1)$ vanishes.

We make an important technical genericity assumption on the FI parameters, which excludes for instance $SU(n)$ gauge theories for which beautiful dualities are known.  Let $\lie{g}$~be the gauge algebra, $\lie{t}$~its Cartan subalgebra, $\lie{a}$~the abelian subalgebra of~$\lie{g}$, and $\zeta\in\lie{a}^\vee\subset\lie{t}^\vee$ be the collection of FI parameters.  We assume that $\zeta$ cannot be written as a positive linear combination of less than $\dim\lie{t}$ weights of~$\repr$ and roots of~$G$.  Another technical assumption is that the trial central charge is positive
\begin{equation}\label{trialcentralcharge}
  \frac{c}{3} = \Tr_{\repr}(1-R) - \dim G > 0 ,
\end{equation}
where $R$ denotes the R-charge.  In other words we assume that the theory has enough matter with low R-charges.  This holds for instance when the theory has a phase in which it reduces to a non-linear sigma model with some nontrivial target space~$X$; indeed $c/3=\dim_{\CC}X>0$.

When applying zeta-function regularization, we omit powers of the cutoff energy scale~$\mathbf{\Lambda}$ (in units of the inverse radius of~$\RP^2$).  They can be reinstated by the replacement
\begin{equation}
  \Gamma(u) \to \mathbf{\Lambda}^{-u}\Gamma(u) , \quad \sin\pi u\to\mathbf{\Lambda}\sin\pi u, \quad \sin\frac{\pi u}{2}\to\mathbf{\Lambda}^{1/2}\sin\frac{\pi u}{2} ,
\end{equation}
similarly for $\cos$, and multiplying factors linear in~$u$ by $\mathbf{\Lambda}^{-1}$.

\subsection{Localization by gluing}

We cut $\RP^2=\{x_0^2+x_1^2+x_2^2=1\}/\ZZ_2$ along the circle at $x_0=0$, which we denote ${\rm S}^1_{1/2}=\{x_1^2+x_2^2=1\}/\ZZ_2$.  The decomposition preserves 1d $\Nsusy=2$ supersymmetry and the 1d multiplets we encounter are vector, chiral, and Fermi multiplets.
We first localize using a $\supercharge$-exact and $\supercharge$-closed term supported on~${\rm S}^1_{1/2}$, then use the resulting constant 1d fields as a boundary condition for the $\HS^2$~partition function.

Supersymmetry equations set the 1d matter multiplets to zero.  The gauge field has a holonomy~$y$ around ${\rm S}^1_{1/2}$.  Contrarily to higher dimensions, $y$ needs not square to~$1$ since $\pi_1({\rm S}^1_{1/2})=\ZZ$ rather than~$\ZZ_2$.  To simplify notations we assume $G$ is connected (and compact), so $y=\exp(\pi i\holonomy)$ with $\tfrac{1}{2}\holonomy\in\lie{t}/\Lambda$ (after gauge transformation), where $\Lambda$ is the coweight lattice.  The normalization is due to ${\rm S}^1_{1/2}$ having circumference~$\pi$.  Supersymmetric configurations for the ${\rm S}^1_{1/2}$ vector multiplet are parametrized by an additional constant real scalar~$\sigma$ that commutes with~$\holonomy$ hence can be diagonalized.  Altogether this is
\begin{equation}
  u=\holonomy+i\sigma\in\lie{t}_{\CC}/(2\Lambda) .
\end{equation}
The 1d path integral restricts to an integral over this space with real dimension $2\dim\lie{t}$.  More precisely one has to omit neighborhoods of singular points $u=u_*$ at which (enough) chiral multiplets become massless.  In the purely 1d partition function~\cite{Hwang:2014uwa,Cordova:2014oxa,Hori:2014tda,Ohta:2014ria}, the integrand turns out to be an antiholomorphic total derivative $\partial/\partial\bar{u}$, which by Stokes theorem reduces to an integral over a middle-dimensional cycle surrounding the singular points; the latter integral computes a Jeffrey--Kirwan residue ($\JKres$) of the integrand at these~$u_*$.  As explicited below we must assume that $\zeta$ is a valid JK parameter.  In 1d there is also a contribution from the region $\lvert\sigma\rvert\to\infty$.  The $\HS^2$ partition function will turn out to depend holomorphically on~$u$ and to decay at infinity, see~\eqref{HS2decay}, so the steps above all apply and the $\RP^2$ partition function reduces to a sum over the singular points.

Localization on $\HS^2$ was performed with various combinations of Neumann and Dirichlet boundary conditions for the vector and chiral multiplets in~\cite{Sugishita:2013jca,Honda:2013uca,Hori:2013ika}.  However, none of these papers gives exactly the data we need.

The restriction of one of the 2d vector multiplet scalars ($\eta$~in the notations of the review~\cite{Benini:2016qnm}) to the equator is part of a 1d adjoint chiral multiplet hence vanishes.  The $\HS^2$ BPS equations \cite[v2, (3.40)]{Sugishita:2013jca} set $F_{12}=\eta=0$, which is incompatible with the nontrivial holonomy~$\holonomy$ thus would eliminate contributions from most $u=u_*$.  However, these BPS equations were derived under a specific reality condition on the auxiliary field~$D$ and one should relax that assumption to make the contour pass through saddle-points.  BPS equations on ${\rm S}^2$ with no reality condition on~$D$ are given in \cite[v4, (2.9)--(2.10)]{Benini:2016qnm}; the only difference for $\HS^2$ is the range of the latitude $0\leq\theta\leq\pi/2$.  There it is found (we set the radius to~$1$) that $\sigma$ is a constant, $\eta=\eta(\theta)$ and that
\begin{equation}
  A=(\eta(0)-\eta(\theta)\cos\theta)\dd{\varphi} .
\end{equation}
The gauge field $A(\pi/2)=\eta(0)\dd{\varphi}$ at the equator must have holonomy~$y^2$ hence
\begin{equation}
  \eta(0)=\holonomy+k
\end{equation}
for $k\in\Lambda$.  Note that the Wilson line joining antipodal points along the equator is not gauge-invariant on the hemisphere hence needs not give~$y$ in the gauge we chose.
Reducing the domain of integration in \cite[v4, (2.35)]{Benini:2016qnm} to the hemisphere and using that $\eta=0$ on the equator, one finds that the FI term is
\begin{equation}
  \frac{i}{2\pi} \int_{\HS^2} (D+\sigma) \dd{^2x} = i \sigma + \eta(0) = u+k.
\end{equation}
Since other terms in the action are $\Qcharge$-exact, the classical contribution is $\exp(-2\pi\zeta(u+k))$.

In principle, one should then compute one-loop determinants in the non-constant background we just found.  On manifolds without boundaries, one-loop determinants can be computed by localizing to fixed points of~$\Qcharge^2$.  On the other hand, boundary conditions typically affect one-loop determinants by selecting some modes of the Laplace/Dirac operators without altering the eigenvalues for these modes.  It is thus plausible that our non-constant vector multiplet only affects chiral multiplet one-loop determinants through its value near the poles and the gauge holonomy along the equator (this is confirmed by the fact that we reproduce $\RP^2$~results~\cite{Kim:2013ola}).  Both coincide with the constant $\eta=\eta(0)$ background studied in~\cite{Sugishita:2013jca}.  After zeta-function regularization, the one-loop determinant of chiral multiplets with Dirichlet boundary conditions is
\begin{equation}
  Z_{\HS^2,\text{Dir}}^{\text{chiral},\text{1-loop}} = \prod_{w\in\weights(\repr)}\frac{\sqrt{2\pi}}{\Gamma(1+w(-R/2+u_F+u+k))} ,
\end{equation}
Here, $w$~denotes a combined weight under all symmetry groups (vector $U(1)$ R-symmetry, flavor, gauge), $R$~is the vector R-charge of each chiral multiplet, and $u_F$~combines background holonomies and twisted masses (background scalars~$\sigma$).

The contribution from off-diagonal components of the vector multiplet must be the inverse of that of an adjoint chiral multiplet of R-charge~$0$ because in a theory containing both, the super-Higgs mechanism could make both sets of components massive.  Thus
\begin{equation}\label{ZHS2vec}
  Z_{\HS^2,\text{Dir}}^{\text{vector},\text{1-loop}} = \prod_{\alpha\in\roots(G)} \frac{1}{\sqrt{2\pi}} \Gamma(1+\alpha(u+k))
  = \prod_{\alpha>0} \frac{\alpha(u+k)}{2\sin\pi\alpha(u+k)} .
\end{equation}
This includes the Vandermonde factor that relates integration over the gauge Lie algebra and its Cartan subalgebra.  A consistency check is that the theory with Dirichlet boundary condition admits a 1d flavor symmetry corresponding to constant gauge transformations at the boundary; gauging it should give Neumann boundary conditions.  We find indeed that our result differs from~\cite{Honda:2013uca,Hori:2013ika} precisely by the one-loop determinant $\prod_\alpha\sin\pi\alpha(u)$ of a vector multiplet on the full boundary ${\rm S}^1$ (of circumference $2\pi$).  Our result also differ from~\cite{Sugishita:2013jca} since their vector multiplet and adjoint chiral multiplet determinants do not cancel, most likely because the super-Higgs mechanism is incompatible with their boundary conditions.  In any case, the one-loop determinant~\eqref{ZHS2vec} is appropriate for the gluing procedure: for instance the vector multiplet one-loop determinant on~${\rm S}^2$ is the one on~${\rm S}^1$ times the square of~\eqref{ZHS2vec}.

\subsection{Higgs and Coulomb branch results}

Combining all ingredients, the $\RP^2$ partition function reads\footnote{As our 1d contribution we included the one-loop determinant of a 1d vector multiplet and a 1d chiral multiplet, namely the 1d $\Nsusy=2$ multiplets that contain the 1d restriction of bottom components of the 2d $\Nsusy=(2,2)$ multiplets.  These are the ones needed to reproduce the ${\rm S}^2$~partition function from two hemispheres.  Since this paper is mostly about the AGT correspondence we did not track down why 1d $\Nsusy=2$ multiplets containing other components of the 2d $\Nsusy=(2,2)$ multiplets do not contribute.}
\begin{equation}\label{ZRP2Higgs}
  \begin{aligned}
    Z_{\RP^2}(\zeta) & = \frac{1}{\abs{\Weyl}}\sum_{u_*}\JKres_{u=u_*}{}_{\zeta} \Biggl(
    \frac{\prod_{\alpha} 2\sin\bigl(\frac{\pi}{2}\alpha(u)\bigr)}{\prod_w 2\sin\bigl(\frac{\pi}{2}w(-R/2+u_F+u)\bigr)}
    \\
    & \quad \times \sum_{k\in\Lambda} \frac{\prod_{\alpha>0}\bigl[\frac{1}{2}\alpha(u+k)/\sin\bigl(\pi\alpha(u+k)\bigr)\bigr]} {\prod_w\Gamma\bigl(1+w(-R/2+u_F+u+k)\bigr)/\sqrt{2\pi}}
    e^{-2\pi\zeta(u+k)} \dd{^{\dim\lie{t}}u} \Biggr)
  \end{aligned}
\end{equation}
where $\abs{\Weyl}$ is the order of the Weyl group of~$G$.

What singular points~$u_*$ does the sum run over?  Recall that this sum (and the JK prescription) comes from cutting out neighborhood of singularities in a higher-dimensional version of $\int\dd{^2u}\partial_{\bar{u}}(\dots)$.  Both the singularities due to 1d chiral multiplets (namely denominator sine factors outside the sum over~$k$) and those due to $\HS^2$~vector multiplets (namely $1/\sin$ in the last numerator) must be cut out.  Each factor has poles along hyperplanes of the form $\{u\mid w(u)=\dots\}$ where $w$ is a weight of~$\repr$ or a root of~$G$.  The $u_*\in\lie{t}_{\CC}/(2\Lambda)$ to sum over are intersections of $\dim\lie{t}$ such hyperplanes and the JK prescription selects intersections such that $\zeta\in\lie{t}^\vee$ is a positive linear combination\footnote{To be a valid JK parameter, $\zeta$ must be generic in the sense that it is not a positive linear combination of fewer than $\dim\lie{t}$ weights of~$\repr$ or roots of~$G$.  We made precisely this genericity assumption.} of the corresponding~$w$.

The sum over $k\in\Lambda$ is in fact truncated. Consider one singular~$u_*$ defined by weights~$w_i$.  By construction $w_i(R/2-u_F-u)$ are integers.  The zeros of $1/\Gamma(1+w_i(-R/2+u_F+u+k))$ truncate the sum over~$k$ to $w_i(k)\geq w_i(R/2-u_F-u)$.  Since $\zeta$ is a positive linear combination of the~$w_i$, the truncation makes the classical action $2\pi\zeta(u+k)$ bounded below, which is essential for convergence.  It is heartening to see considerations of 1d localization (the positivity required by the JK residue) play such an important role in making the 2d sum over vortices converge.

We now convert our results to the form~\eqref{ZRP2Coul} found in~\cite{Kim:2013ola}.

If the sum ran over $k\in 2\Lambda$, it could be combined with the sum over singular points $u_*\in\lie{t}_{\CC}/(2\Lambda)$ into a single sum over all singular points in~$\lie{t}_{\CC}$.  As it stands, the sum over~$k$ has many more terms, accounted for by also summing over the class of $k$ in $\Lambda/2\Lambda$.  To write explicit expressions, we pick an arbitrary lift $l\colon\Lambda/2\Lambda\to\Lambda$.  We also change variables to $v=u+k$ to get
\begin{equation}
  \begin{multlined}
    Z_{\RP^2}(\zeta) = \frac{1}{\abs{\Weyl}} \sum_{k\in l(\Lambda/2\Lambda)} \sum_{v_*\in\lie{t}_{\CC}}\JKres_{v=v_*}{}_{\zeta} \Biggl(
    e^{-2\pi\zeta(v)}
    \prod_{\alpha>0} Z_{\RP^2}^{\text{vector}}\bigl(\alpha(v),\alpha(k)\bigr) \\
    \times
    \prod_{w} Z_{\RP^2}^{\text{chiral}}\bigl(w(v+u_F-R/2),w(k)\bigr)
    \frac{\dd{^{\dim\lie{t}}v}}{\mathbf{\Lambda}^{\dim\lie{t}/2}} \Biggr)
  \end{multlined}
\end{equation}
with
\begin{equation}\label{ZRP21loop}
  \begin{aligned}
    Z_{\RP^2}^{\text{vector}}(\mathsf{v},\mathsf{k})
    & = - (-1)^{\mathsf{k}} \mathsf{v}\tan\Bigl(\frac{\pi}{2}(\mathsf{v}-\mathsf{k})\Bigr) \mathbf{\Lambda}^{-1} ,
    \\
    Z_{\RP^2}^{\text{chiral}}(\mathsf{v},\mathsf{k})
    & = \frac{\sqrt{\pi/2}}{\sin\bigl(\frac{\pi}{2}(\mathsf{v}-\mathsf{k})\bigr) \Gamma\bigl(1+\mathsf{v}\bigr)} \mathbf{\Lambda}^{1/2+\mathsf{v}} ,
  \end{aligned}
\end{equation}
where we reinstated factors of the cutoff $\mathbf{\Lambda}$ for completeness ($r\Lambda$ in the notations of~\cite{Kim:2013ola}).

Two minor sign differences are worth noting.  Our chiral multiplet determinant for $\mathsf{k}\in\{0,1\}$ is equal to (minus) what they get for even/odd holonomy, but ours changes sign under $\mathsf{k}\to\mathsf{k}+2$.  In the models we consider these signs cancel thanks to $\sum_w w=0$: shifting $k\to k+2\lambda$ with $\lambda\in\Lambda$ gives a sign $\prod_w (-1)^{w(\lambda)}=1$.  Through the relation $Z_{\RP^2}^{\text{vector}}(\mathsf{v},\mathsf{k}) \allowbreak Z_{\RP^2}^{\text{chiral}}(\mathsf{v},\mathsf{k}) \allowbreak Z_{\RP^2}^{\text{chiral}}(-\mathsf{v},-\mathsf{k}) = 1$ this first sign difference implies a sign difference in the vector multiplet determinant.  Besides an unimportant $(-1)^{(\dim\lie{g}-\dim\lie{t})/2}$ we have an additional sign $\prod_{\alpha>0} (-1)^{\alpha(k)}$.  This is equal to the sign that was originally missed in ${\rm S}^2$~localization calculations~\cite{Benini:2012ui,Doroud:2012xw,Gomis:2012wy}, pointed out by Hori and collaborators~\cite{Hori:2013ika,Hori:2013ewa}, and described in detail in~\cite{Benini:2016qnm}.  It would be good to clarify the correct sign on $\RP^2$.

Interestingly, at least for $U(K)$ quiver gauge theories, contributions from poles of the $\HS^2$ vector multiplet determinant seem to cancel out, hence one can simply sum over the set of $v_*$~used in the purely 1d localization calculation.\footnote{It would be valuable to find a 1d localization term that directly localizes the index to the discrete points~$v_*$ rather than an integral.  The calculation of the 2d elliptic genus~\cite{Gadde:2013ftv,Benini:2013nda,Benini:2013xpa} has a similar structure, giving an integral over~$v$ that is reduced afterwards to a sum over singular points and it would be interesting to get the result directly from a Higgs branch localization.}  Consider an intersection $v_*$ of $\dim\lie{t}$ hyperplanes defined by some weights $w_1,\dots,w_m$ and $n\geq 1$ roots $\alpha_1,\dots,\alpha_n$, such that $\zeta = a^i w_i + b^j \alpha_j$ (with implicit summation) is a positive linear combination of these weights and roots.  Since only $U(1)$ gauge group factors have FI parameters, $\zeta$ is orthogonal to all $\alpha_j$ and in particular to $b^j\alpha_j$, thus $a^i b^j \langle w_i,\alpha_j\rangle = - \langle b^j\alpha_j,b^{j'}\alpha_{j'}\rangle < 0$.  We deduce that $\langle w_I,\alpha_J\rangle<0$ for some $I$ and~$J$, thus $w_I+\alpha_J$ is a weight of~$\repr$.\footnote{Kenny Wong's answer to the related \url{https://math.stackexchange.com/questions/2356883/} can be adapted to show this.  Consider the $\lie{sl}_2$ subalgebra $\langle H_\alpha,E_\alpha,F_\alpha\rangle$ associated to~$\alpha_J$.   If $w_I+\alpha_J$ were not a weight of~$\repr$ then the raising operator~$E_\alpha$ would act trivially on the $w_I$~weight space, in other words this weight space would consist of highest-weight vectors of $\lie{sl}_2$.  We would then have $\langle w_I,\alpha_J\rangle\geq 0$.}  Now we use the structure of the integrand: at~$v_*$, each $\alpha_j(v_*-k)$ is an odd integer and each $w_i(v_*-k+u_F-R/2)$ is an even integer so $(w_I+\alpha_J)(v_*-k+u_F-R/2)$ is an odd integer.  Let $\delta k\in\Lambda$ be such that $w_I(\delta k)$ is odd and all other $w_i(\delta k)$ and $\alpha_j(\delta k)$ are even (this is possible for quivers).  Then shifting $k\to k+\delta k$, parities of $w_I(v_*-k+u_F-R/2)$ and $(w_I+\alpha_J)(v_*-k+u_F-R/2)$ are exchanged.  This describes another contribution to the partition function and explicit calculations in examples (see next subsection) indicate that the contribution exactly cancels the unshifted one thanks to a sign in the JK prescription.

For even~$\mathsf{k}$ the chiral multiplet determinant has poles at $\mathsf{v}=0,2,4,\dots$ while for odd~$\mathsf{k}$ it has poles at $\mathsf{v}=1,3,5,\dots$  In both cases the poles are in the half-space $\re w(v)>0$, as long as R-charges are in the standard interval $(0,2)$ and there are no flavor holonomies ($\re u_F=0$).\footnote{The contour integral~\eqref{ZRP2Coul} is analytically continued to R-charges outside $(0,2)$ and to nonzero flavor holonomies (nonimaginary~$u_F$) by deforming the contour so that no pole crosses the contour.}  The sum over singular points is then exactly the sum of residues obtained by closing the following contour integral
\begin{equation}\label{ZRP2Coul}
  \begin{aligned}
    Z_{\RP^2} & = \frac{1}{\abs{\Weyl}} \sum_{k\in l(\Lambda/2\Lambda)} \int_{i\lie{t}} \frac{\dd{^{\dim\lie{t}}v}}{(2\pi i)^{\dim\lie{t}}}\, e^{-2\pi\zeta(v)}
    \prod_{\alpha>0} Z^{\text{vector}}_{\RP^2}\bigl(\alpha(v),\alpha(k)\bigr)
    \\
    & \qquad \times
    \prod_w Z^{\text{chiral}}_{\RP^2}\Bigl(w(v+u_F-R/2),w(k)\Bigr) .
  \end{aligned}
\end{equation}
The Coulomb branch integral~\eqref{ZRP2Coul} is precisely the answer that can be assembled from ingredients given in~\cite{Kim:2013ola}.  From this point of view, the sum over $k\in l(\Lambda/2\Lambda)$ is a sum over holonomies describing flat connections on~$\RP^2$.  It would be interesting to directly obtain this formula in our approach, presumably by a different choice of 1d localizing term such as the one used in~\cite{Ohta:2015fpe} to get a Coulomb branch expression for the 1d partition function.

It is straightforward to check the following asymptotics as $\mathsf{v}\to\pm i\infty$ with constant real part,
\begin{equation}\label{HS2decay}
  \begin{aligned}
    Z_{\RP^2}^{\text{vector}}\bigl(\mathsf{v},\mathsf{k}\bigr)
    & = O(\abs{\mathsf{v}}),
    \\
    Z_{\RP^2}^{\text{chiral}}\bigl(\mathsf{v}-R/2,\mathsf{k}\bigr)
    & = O\bigl(\abs{\mathsf{v}}^{(R-1)/2-\re\mathsf{v}}\bigr) .
  \end{aligned}
\end{equation}
At least when the different components of $\im\mathsf{v}$ are scaled by the same factor~$\lambda$, the vector multiplet contribution combines with the measure to give $\lambda^{\dim G}$ and the matter contribution compensates for it, provided the trial central charge~\eqref{trialcentralcharge} is positive.  This supports our earlier claim that the region $\abs{v}\to\infty$ does not contribute in~\eqref{ZRP2Higgs}.  It also suggests that the integral~\eqref{ZRP2Coul} converges.

\subsection{Seiberg-like dualities}

Higgs branch expressions are instrumental in proving that Seiberg dual theories have equal ${\rm S}^2$~partition functions.  Using our expressions we perform the same check for $\RP^2$~partition functions of one dual pair.

The electric theory is a $U(K)$ vector multiplet coupled to $N$ fundamental and $N$ antifundamental chiral multiplets.  We denote R-charges by $R_A$ and $\widetilde{R}_A$ for $1\leq A\leq N$.  The magnetic theory is a $U(N-K)$ vector multiplet coupled to $N$ fundamental and $N$ antifundamental chiral multiplets, themselves coupled by a cubic superpotential to $N^2$ neutral chiral multiplets (corresponding to mesons of the original theory).  These three sets of chiral multiplets have R-charges $(1-R_A)$, $(1-\widetilde{R}_B)$ and $(R_A+\widetilde{R}_B)$ for $1\leq A,B\leq N$, compatible with the fact that the cubic superpotential has total R-charge~$2$.

The $\RP^2$ partition function of the electric theory is
\begin{equation}
  \begin{aligned}
    Z_{\RP^2}(\zeta) & = \frac{1}{K!} \sum_{k\in\{0,1\}^K} \sum_{v_*\in\CC^K} e^{-2\pi\zeta\Tr v_*}\JKres_{v=v_*}{}_{\zeta} \Biggl(
    \prod_{i<j}^K Z_{\RP^2}^{\text{vector}}(v_i-v_j,k_i-k_j) \\
    & \qquad \times
    \prod_{i=1}^K \prod_{A=1}^N \Bigl( Z_{\RP^2}^{\text{chiral}}(v_i-R_A/2,k_i) Z_{\RP^2}^{\text{chiral}}(-v_i-\widetilde{R}_A/2,-k_i) \Bigr)
    \Biggr) \dd{^Kv}
  \end{aligned}
\end{equation}
with $Z_{\RP^2}^{\text{vector}}$ and $Z_{\RP^2}^{\text{chiral}}$ given in~\eqref{ZRP21loop}.  The poles to consider are at
\begin{equation}
  \begin{cases}
    v_j = R_A/2+l_j \text{ with } l_j\in\ZZ_{\geq 0} \text{ and } l_j - k_j \in 2\ZZ & \text{(chiral multiplet pole)}\\
    v_j \in v_i + k_i - k_j + 1 + 2\ZZ & \text{(vector multiplet pole)}
  \end{cases}
\end{equation}
where in the former case $1\leq A\leq N$ and in the latter case $v_i$ should be another component, itself set to $R_A/2+l_i$ or to some other $v_{i'}$ plus an integer, and so on.

As argued above, vector multiplet poles do not contribute.  Let us make the argument explicit in the case of $U(2)$.  If a vector multiplet pole is used then we have (up to exchanging $1\leftrightarrow 2$) $v_1=R_A/2+l_I$ and $v_2=v_1+\dots=R_A/2+l_2$ with $l_2\equiv k_1+(k_2-k_1+1)\equiv k_2+1\bmod 2$.  For $l_2<0$ the residue vanishes due to the $(2,A)$ factor in the chiral multiplet determinant, leaving only poles for which $l_2\geq 0$.  Denote such a pole by $(l_1,l_2,k_1,k_2)$.  Since shifting $k$'s by even integers does not change signs the residue there is the same as the residue at $(l_1,l_2,l_1,l_2+1)$.  That residue is antisymmetric in $l_1\leftrightarrow l_2$ because the arguments $(\mathsf{v}-\mathsf{k})$ in~\eqref{ZRP21loop} are invariant, the $1/\Gamma$ factors in $(1,B)$ and $(2,B)$ components of chiral multiplet determinants are interchanged, and the linear factor in the vector multiplet one-loop determinant changes sign.  We kept parities of $l_1-k_1$ and $l_2-k_2$ fixed so that the same components of one-loop determinants are singular to avoid having to track a sign due to the JK prescription.  In the general argument above, instead of using the pole at $(l_2,l_1,l_2,l_1+1)$ we had used its Weyl reflection $(l_1,l_2,l_1+1,l_2)$ because that contribution is easier to locate on general grounds.

Altogether, the enclosed poles are at $v_j=R_{A_j}/2+l_j$ for some choice of distinct flavors~$A_j$ and for some vorticities $l_j\in\ZZ_{\geq 0}$.  We denote $\mathcal{A}=\{A_1,\ldots,A_K\}$ and compute
\begin{equation}
  \label{RP2tovortex}
  \begin{aligned}
    Z_{\RP^2} =
    & \sum_{\mathcal{A}\subseteq\{1,\ldots,N\}} e^{-2\pi\zeta\sum_{A\in\mathcal{A}} R_A/2}
    \biggl( \prod_{A\in\mathcal{A}} \prod_{B\not\in\mathcal{A}}^N Z^{\text{chiral}}_{\RP^2}\Bigl(\frac{R_A-R_B}{2},0\Bigr) \\
    & \quad \times \prod_{A\in\mathcal{A}} \prod_{B=1}^N Z^{\text{chiral}}_{\RP^2}\Bigl(-\frac{R_A+\widetilde{R}_B}{2},0\Bigr) \biggr)
    Z_{\text{vortex}}^{(\mathcal{A})}\Bigl(R,\widetilde{R};x\Bigr)
  \end{aligned}
\end{equation}
where $x=(-1)^{N+K-1}e^{-2\pi\zeta}$ and
\begin{equation}
  Z_{\text{vortex}}^{\{A_j\}}(R,\widetilde{R};x)
  = \sum_{\{l_j\geq 0\}}
  \frac{\prod_{j=1}^K (x^{l_j}/l_j!) \prod_{B=1}^N \prod_{j=1}^K (\frac{\widetilde{R}_B}{2}+\frac{R_{A_j}}{2})_{l_j}}{
    \prod_{i\neq j}^K (-l_i-\frac{R_{A_i}}{2}+\frac{R_{A_j}}{2})_{l_j}
    \prod_{B\not\in\{A\}}^N \prod_{j=1}^K (1-\frac{R_B}{2}+\frac{R_{A_j}}{2})_{l_j}
  } .
\end{equation}
This vortex partition function is identical to the one appearing in the factorization of the ${\rm S}^2$~partition function, or in the hemisphere partition function.  One difference is that on $\RP^2$, $x=\pm e^{-2\pi\zeta}$ is real.  This can be traced back to the fact that the first Chern class $\frac{1}{2\pi}\int_{\RP^2}\Tr F$ vanishes so one cannot turn on a continuous theta term.

Using known results about vortex partition functions (found when comparing ${\rm S}^2$~partition functions of Seiberg dual
pairs) it is straightforward to get
\begin{equation}
  \begin{aligned}
    Z_{\RP^2,U(K),N}(R,\widetilde{R},\zeta)
    & =
    x^{-(N-K-\sum_AR_A)/2}
    \bigl(1-x\bigr)^{N-K-\sum_A(R_A+\widetilde{R}_A)/2} \\
    & \quad \times
    Z_{\RP^2,U(N-K),N}(1-R,1-\widetilde{R},\zeta)
    \prod_{A=1}^N \prod_{B=1}^N Z^{\text{chiral}}_{\RP^2}\Bigl(-\frac{R_A+\widetilde{R}_B}{2},0\Bigr) .
  \end{aligned}
\end{equation}
In other words the partition functions of Seiberg dual theories are equal up to powers of $x$ and $(1-x)$.  In the ${\rm S}^2$~case these were understood as regularization ambiguities and background FI parameters~\cite{Benini:2014mia}.

We must mention a subtle point that we haven't elucidated.  The BPS equations on $\RP^2$ (when the auxiliary field~$D$ is taken to be real) make $\star F$ be covariantly constant.  The authors of~\cite{Kim:2013ola} choose a gauge where $\star F$ is constant and deduce $F=0$, but this is only possible locally on~$\RP^2$.  Consider the double-cover of~$\RP^2$.  Since the Hodge star changes sign under the antipodal map, $\star F$ must take opposite values at antipodes, and $D(\star F)=0$ simply forces them to be conjugate.  For $G=U(2)$, an explicit class of examples for $B\in 2\ZZ+1$ (to simplify expressions) is
\begin{equation}
  F = \frac{B}{2} \begin{pmatrix}
    \cos\theta & e^{iB\varphi} \sin\theta \\
    e^{-iB\varphi}\sin\theta & -\cos\theta
  \end{pmatrix} \sin\theta \dd{\theta}\wedge\dd{\varphi} .
\end{equation}
In the region $0<\theta<\pi$ it derives from a potential
\begin{equation}
  A = \frac{i}{2} \begin{pmatrix} 0 & e^{iB\varphi} \\ -e^{-iB\varphi} & 0 \end{pmatrix} \dd{\theta} +
  \frac{B\sin\theta}{2} \begin{pmatrix} \sin\theta & -e^{iB\varphi} \cos\theta \\ -e^{-iB\varphi} \cos\theta & -\sin\theta \end{pmatrix} \dd{\varphi}
\end{equation}
which obeys $A_\theta(\pi-\theta,\pi+\varphi)=-A_\theta(\theta,\varphi)$ and $A_\varphi(\pi-\theta,\pi+\varphi)=A_\varphi(\theta,\varphi)$.

In principle one should take these nonabelian BPS configurations into account when localizing.  However, our check of Seiberg duality forbids them from contributing since we matched the $\RP^2$ partition functions of some nonabelian gauge theories to abelian ones.

\section{Other quotients of the ellipsoid}\label{app:quotients}

In this appendix we discuss what discrete quotients ${\rm S}^4_b/G$ can be probed using available 4d $\Nsusy=2$ localization results.  We use embedding coordinates $x\in\RR^5$ in which the squashed sphere is $x_0^2 + b^{-2} (x_1^2+x_2^2) +  b^2(x_3^2+x_4^2)=1$.  The round sphere ($b=1$) has more diverse quotients so this is what we focus on.  In the following, only the quotients involving products of cyclic groups are allowed for $b\neq 1$.

The relevant 4d theories are in general not conformal so $G$ must act by isometries, namely $G\subset\Lie{O}(5)$.  The localizing supercharge~$\Qcharge$ should be left invariant by the orbifold action, and in particular the poles should be fixed or exchanged (since they are fixed points of $\Qcharge^2$) hence $G\subset\ZZ_2\times\Lie{O}(4)$ where the $\ZZ_2$ is generated by the reflection $x_0\to -x_0$.  The two $\Lie{SU}(2)$ factors inside $\Lie{SO}(4)=(\Lie{SU}(2)_L\times\Lie{SU}(2)_R)/\ZZ_2$ act on the complex coordinates $(x_1+ix_2,x_3+ix_4)$ and $(x_1+ix_2,x_3-ix_4)$ respectively and their central elements $(-1)_L$ and $(-1)_R$ are identified since they both flip the sign of $(x_1,x_2,x_3,x_4)$.

Since $\Qcharge^2$ generates a $\Lie{U}(1)_L\subset\Lie{SU}(2)_L$, its commutant is $(\Lie{U}(1)_L\times\Lie{SU}(2)_R)/\ZZ_2$ and we deduce that $G\subset\ZZ_2\times(\ZZ_p\times\Gamma)/{\sim}$ for some discrete subgroups $\ZZ_p\subset\Lie{U}(1)_L$ and $\Gamma\subset\Lie{SU}(2)_R$.  Here, $\sim$ identifies $(-1)_L(-1)_R\sim 1$ if it belongs to the group.  The discrete subgroups $\Gamma\subset\Lie{SU}(2)$ are well-known, namely the odd cyclic groups $\ZZ_{2n+1}\subset\Lie{U}(1)$ and binary polyhedral groups.  The latter contain $\ZZ_2$ and have an ADE classification: $\Gamma/\ZZ_2\subset\Lie{SO}(3)$ are symmetry groups of pyramids, prisms, the tetrahedron, the octahedron, and the icosahedron.

Some of these quotients do not appear to preserve supersymmetry, even after turning on an R-symmetry current.

Since our analysis is not meant to be exhaustive let us concentrate on whether the quotient ${\rm S}^3/G$ of the equator preserves supersymmetry.  The round ${\rm S}^3$ is parametrized by unit-norm quaternions on which the two $\Lie{SU}(2)$ act by multiplication on the left/right respectively.  In the absence of background R-symmetry current, the round sphere admits two left-invariant and two right-invariant Killing spinors.

Thus, when $G$ is a (discrete) subgroup of $\Lie{SU}(2)_R$, the quotient ${\rm S}^3/G$ preserves two Killing spinors.  This case includes lens spaces $L(p,1)={\rm S}^3/\ZZ_p$, on which localization was performed: the one-loop determinants of 3d $\Nsusy=2$ multiplets on $L(p,1)$ are obtained by keeping only the $\ZZ_p$-invariant modes in the one-loop determinants on~${\rm S}^3$.  It also includes the quotients of ${\rm S}^3$ by dicyclic, binary tetrahedral, binary octahedral, and binary icosahedral groups, which have not been studied yet in the supersymmetric localization literature.  Through the 3d-3d correspondence such geometries might give rise to new topological theories in 3d.\footnote{We thank Mauricio Romo for this comment.}

For a more general quotient ${\rm S}^3/G$ one must turn on a $U(1)$ R-symmetry current, so that Killing spinors are not constant anymore (neither in the left-invariant nor the right-invariant frames).  Two of the Killing spinors are then only invariant under an $\Lie{O}(2)\subset\Lie{SU}(2)_R$, which contains dicyclic groups but not other binary polyhedral groups.  It should be possible to localize on quotients of ${\rm S}^3$ by subgroups of $(\ZZ_p\times\Gamma)/\sim$ when $\Gamma$ is cyclic or dicyclic.  This includes lens spaces $L(p,q)$ and more exotic manifolds and orbifolds.  However, we were unable to write an R-symmetry current such that Killing spinors are invariant both under a tetra-/octa-/icosahedral subgroup of $\Lie{SU}(2)_R$ and under a non-trivial subgroup of $\Lie{U}(1)_L$, so even the smooth quotients (spherical $3$-manifolds) among these do not seem to preserve supersymmetry.

Let us give a very short outline of the localization result on the quotient of ${\rm S}^3$ by a dicyclic subgroup of $\Lie{SU}(2)_R$.  The one-loop determinant of a free 3d $\Nsusy=2$ chiral multiplet of mass $\mu$ is obtained by keeping modes that are invariant under the dicyclic group
\begin{equation}
  Z_{\text{one-loop, free chiral multiplet}}^{{\rm S}^3/\text{dicyclic}}
  \sim \prod_{0\leq m\leq n,\,m-n\equiv 0\bmod p}\frac{(1+m+n-i\mu)}{(1+m+n+i\mu)}
\end{equation}
In a gauge theory, one must sum over flat gauge connections (labeled by conjugacy classes of morphisms from the dicyclic group to the gauge group).  The gauge connection affects which modes to keep in the chiral multiplet one-loop determinant, namely it changes the conditions $m-n\equiv 0$ and $m\leq n$.  An easy generalization is to give an R-charge to the chiral multiplet.  The one-loop determinant of off-diagonal components of the vector multiplet is then (essentially) one over that of a chiral multiplet of R-charge $0$ (because of the super-Higgs mechanism) hence equal to that of an adjoint of R-charge $2$ (because adjoints of R-charges summing to $2$ can be given a superpotential mass and can be integrated out).  There are sign ambiguities between different holonomy sectors, but they are no worse than with other methods of computing fermion determinants.  The common solution is to compare dual gauge/non-gauge theories and find what choice of signs makes partition functions equal.

It would be interesting to see what localizing on the quotients of ${\rm S}^3$ described here teaches us about global structures in 3d dualities.  It would also be interesting to perform the 4d localization on the corresponding quotients of ${\rm S}^4_b$ (with or without an action of $x_0\to-x_0$) and relate the results to various 2d CFTs.  This will involve instanton partition functions on orbifolds of~$\CC^2$ such as those studied in \cite{Belavin:2011pp, Belavin:2011tb,Belavin:2011sw,Bonelli:2012ny}.

\section{Conjectural hypermultiplet determinant}\label{app:rp4hyper}

We make here a conjecture for the one-loop determinant of a single 4d $\Nsusy=2$ hypermultiplet on the squashed projective space~$\RP^4_b$.  Our justification is quite schematic.  A proper treatment would require describing precisely how the~$\ZZ_2$ action on~${\rm S}^4_b$ acts on fermions, as we did for the vector multiplet in~\eqref{rp4identificationofvectormultiplet}.  As pointed out to us by Yuji Tachikawa, this issue is rather thornier for hypermultiplets~\cite{Tachikawa:2016xvs} as there are several choices of CP action on fermions in class~S theories.  It would be interesting to define the setup more precisely and give a first principle derivation of our proposal.

This appendix finds its logical place after we give the one-loop determinant~\eqref{S4-as-RP4} on~${\rm S}^4_b$, namely the one-loop determinant of a pair of hypermultiplets subject to holonomies $\pm 1$.  However, to emphasize that the main text does not rely on the conjecture made here, we moved it to the present appendix.  While the symmetry $\sigma\to -\sigma$ is broken by a boundary condition used as an intermediate step, it is restored in the final result.

The hypermultiplet one-loop determinant in~\cite{HS4} is the square root of the sphere determinant~\eqref{S4-as-RP4} so no 3d matter is needed in the decomposition analogous to~\eqref{S4bgluingvector}.  This would suggest to use that square root as our hypermultiplet determinant on~$\RP^4_b$.  However, the one-loop determinant computed in~\cite{HS4} was for mixed Dirichlet/Neumann boundary conditions, while the 3d deformation term treats all matter scalar fields on a same footing.  This suggests looking for a hypermultiplet determinant on~$\HS^4_b$ that would correspond to a different boundary condition.

The gluing~\eqref{S4bgluingvector} involved only a 3d $\Nsusy=2$ vector multiplet, even though the 3d restrictions of fields of the 4d vector multiplet belong to a 3d $\Nsusy=2$ vector and an adjoint chiral multiplet.  The restrictions of fields of the hypermultiplet belong to two chiral multiplets $q$, $\tilde{q}$ in conjugate representations, so in analogy we propose to use a hypermultiplet determinant that obeys a gluing similar to~\eqref{S4bgluingvector} with one of these chiral multiplets in 3d.  Namely,
\begin{equation}\label{oneloopHS4bforgluing}
  Z_{\HS^4_b,\text{for gluing}}^{\text{hypermultiplet}}(\sigma) = \Bigl( Z_{{\rm S}^4_b}^{\text{hypermultiplet}} \bigm/ Z_{{\rm S}^3_b}^{\text{chiral}} \Bigr)^{1/2}
  = \prod_{w\in\weights(\repr)} \Gamma_b(\tfrac{1}{2}(b+1/b) \pm \lb w,i\sigma\rb)
\end{equation}
where we used $Z_{{\rm S}^3_b}^{\text{chiral}}=\prod_wS_b(\tfrac{1}{2}(b+1/b)\mp\lb w,i\sigma\rb)$, the one-loop determinant\footnote{These 3d determinants are inverses of each other for the following reason.  Hypermultiplet scalars (denoted $q$ and $\tilde{q}^\dagger$) transform in a doublet of the $SU(2)$ R-symmetry in 4d hence have R-charges $+1$ and $-1$ under its Cartan torus, the $U(1)$ R-symmetry in 3d.  The normalization is fixed by giving R-charge~$1$ to supercharges.  Thus, the 3d $\Nsusy=2$ chiral multiplets $q$ and $\tilde{q}$ (without~$\dagger$) have R-charge~$1$ and transform in conjugate representations of~$G$ and of the flavor group.  Their 3d one-loop determinants must cancel, because one could turn on a superpotential mass term $W=\tilde{q}q$ which makes both massive.} for~$q$ or $\tilde{q}$.  We select the top sign here for definiteness.  We remind the reader that $S_b(x)=\Gamma_b(x)/\Gamma_b(b+1/b-x)$ and $\Upsilon(x)=1/\bigl(\Gamma_b(x)\Gamma_b(b+1/b-x)\bigr)$.

The Barnes Gamma function obeys a duplication formula of the following form\footnote{We would be interested in a reference.}
\begin{align}
  \Gamma_b\biggl(\frac{b+b^{-1}}{2}+ix\biggr)
  & = c(b) 2^{x^2/2} \prod_{\pm,\pm} \Gamma_b\biggl(\frac{b+b^{-1}}{2}\pm\frac{b}{4}\pm\frac{b^{-1}}{4}+\frac{ix}{2}\biggr) , \\
  S_b\biggl(\frac{b+b^{-1}}{2}+ix\biggr) & = \prod_{\pm,\pm} S_b\biggl(\frac{b+b^{-1}}{2}\pm\frac{b}{4}\pm\frac{b^{-1}}{4}+\frac{ix}{2}\biggr) ,
\end{align}
proven by checking that both sides obey the same shift relations under $x\to x+2b^{\pm 1}$.  This does not fix the constant $c(b)=c(1/b)$.  The chiral multiplet determinants on~${\rm S}^3_b/\ZZ_2$ contain two of these four factors, depending on the holonomy.  Combining with~\eqref{oneloopHS4bforgluing} and using the duplication formula we obtain an $\RP^4_b$ hypermultiplet determinant
\begin{equation}
  \prod_{w\in\weights(\repr)} \biggl[  c(b) 2^{\lb w,\sigma\rb^2/2} \prod_{s_1=\pm}\prod_{s_2=\pm} \Gamma_b\biggl(\frac{b+b^{-1}}{2}+s_1\frac{b}{4}+s_2\frac{b^{-1}}{4}\mp s_1s_2\frac{\lb w,i\sigma\rb}{2}\biggr) \biggr]
\end{equation}
where the $\mp$~depends on the holonomy, but also on whether we used $q$ or $\tilde{q}$ in our procedure.  This dependence (equivalently the lack of symmetry $\sigma\to-\sigma$) is inconsistent.  It is not restored by the sum over holonomies since there can be several hypermultiplets with different mass parameters.  In addition it seems strange that the even and odd holonomies would have so little difference.

A somewhat artificial resolution to both of these issues would be if the correct modes to keep in the ${\rm S}^3_b/\ZZ_2$ determinant are
\begin{equation}
  Z_{{\rm S}^3_b/\ZZ_2}^{\text{chiral}}
  = \prod_w \begin{cases}
    \displaystyle \frac{\Gamma_b(\tfrac{1}{4}(b+b^{-1})-\tfrac{1}{2}\lb w,i\sigma\rb) \Gamma_b(\tfrac{1}{4}(3b+3b^{-1})-\tfrac{1}{2}\lb w,i\sigma\rb)}{\Gamma_b(\tfrac{1}{4}(3b+b^{-1})+\tfrac{1}{2}\lb w,i\sigma\rb) \Gamma_b(\tfrac{1}{4}(b+3b^{-1})+\tfrac{1}{2}\lb w,i\sigma\rb)} & \text{(even holonomy),} \\[2ex]
    \displaystyle \frac{\Gamma_b(\tfrac{1}{4}(3b+b^{-1})-\tfrac{1}{2}\lb w,i\sigma\rb) \Gamma_b(\tfrac{1}{4}(b+3b^{-1})-\tfrac{1}{2}\lb w,i\sigma\rb)}{\Gamma_b(\tfrac{1}{4}(b+b^{-1})+\tfrac{1}{2}\lb w,i\sigma\rb) \Gamma_b(\tfrac{1}{4}(3b+3b^{-1})+\tfrac{1}{2}\lb w,i\sigma\rb)} & \text{(odd holonomy).}
  \end{cases}
\end{equation}
This leads to our conjecture for the $\RP^4_b$ hypermultiplet one-loop determinant:
\begin{equation}\label{RP4-hyper}
  Z_{\RP^4_b}^{\text{hypermultiplet}}
  =
  \prod_{w\in\weights(\repr)} \biggl[  c(b) 2^{\lb w,\sigma\rb^2/2} \prod_{\pm,\pm}
  \Gamma_b\biggl(\frac{b+b^{-1}}{2}\pm\frac{b+y^wb^{-1}}{4}\pm\frac{\lb w,i\sigma\rb}{2}\biggr) \biggr]
\end{equation}
where $y^w=+1$ for even holonomy and $y^w=-1$ for odd holonomy.
The product of determinants of hypermultiplets with even and odd holonomy in the same representation reproduces the determinant of a hypermultiplet on~${\rm S}^4_b$~\eqref{S4-as-RP4}.

\newpage

\begingroup\raggedright\endgroup

\end{document}